\documentclass[12pt,a4paper]{article}
\usepackage{geometry}                
\usepackage{amsmath}                
\usepackage{amsfonts}
\usepackage{amsmath}
\usepackage{color}
\usepackage{graphicx}
\usepackage{hyperref} 
\usepackage{wrapfig} 
\usepackage[T1]{fontenc}
\usepackage{bbm}
\definecolor{darkred}{RGB}{102,11,27}
\definecolor{lred}{RGB}{180,10,10}
\definecolor{lblue}{RGB}{5,5,160}
\definecolor{oldstuff}{RGB}{158,122,72}

\usepackage{multirow}

\usepackage[square,numbers,compress]{natbib}
\usepackage{bibentry}

\def\beq{\begin{equation}}
\def\eeq{\end{equation}}
\def\bea{\begin{eqnarray}}
\def\eea{\end{eqnarray}}
\def\Im{{\rm Im}}
\def\Re{{\rm Re}}
\def\OGroup{SO}
\DeclareMathOperator{\Tr}{Tr}

\def\V{{\bf N}}
\def\Vc{{\bf \bar N}}

\def\1{{\bf 1}}
\def\2{{\bf 2}}
\def\3{{\bf 3}}

\def\2c{{\bf \bar 2}}
\def\3c{{\bf \bar 3}}

\begin{document}
\pagestyle{plain}

	\makeatletter
	\@addtoreset{equation}{section}
	\makeatother
	\renewcommand{\theequation}{\thesection.\arabic{equation}}
	\pagestyle{empty}
\rightline{IFT-UAM/CSIC-22-150}
\rightline{NIKHEF 2022-024}
\rightline{ZMP-HH/22-22}
 
\vspace{1.2cm}

\begin{center}
{\large \bf
D-brane and F-theory Model Building
} 

\vskip 9 mm

Fernando Marchesano,${}^{1}$ Bert Schellekens${}^{2,3}$ and Timo Weigand${}^{4,5}$ 

\vskip 9 mm

\small ${}^{1}${\it Instituto de F\'isica Te\'orica UAM-CSIC, c/Nicolás Cabrera 13-15, 28049 Madrid, Spain} 

\vspace{2mm}

\small ${}^{2}${\it NIKHEF Theory Group, Kruislaan 409,
1098 SJ Amsterdam, The Netherlands}

\vspace{2mm}

\small ${}^{3}${\it Instituto de F\'isica Fundamental, CSIC, Serrano 123, Madrid 28006, Spain} 

\vspace{2mm}

\small ${}^{4}$ {\it II. Institut f\"ur Theoretische Physik, Universit\"at Hamburg, Luruper Chaussee 149,\\ 22607 Hamburg, Germany} 

\vspace{2mm}

\small ${}^{5}${\it Zentrum f\"ur Mathematische Physik, Universit\"at Hamburg, Bundesstrasse 55, \\ 20146 Hamburg, Germany  }   \\[3 mm]


\end{center}

\vskip 7mm

\begin{abstract}

We review recent progress in the construction of four-dimensional vacua of Type II string theory and F-theory which yield the Standard Model of particle physics (SM) or extensions thereof. In Type II orientifold compactifications the SM gauge group and chiral spectrum arise from the open string sector of the theory, namely from stacks of D-branes. The universal features of the chiral spectrum between various sets of D-branes allow for a general approach to build realistic models, which can be implemented in different setups. We describe the realisation of this strategy in Type II Calabi--Yau orientifold compactifications and Rational Conformal Field Theories, discussing the specific model building rules and features of each setting. The same  philosophy can be extended to F-theory constructions. These provide new model building possibilities, as they combine the localisation properties of D-branes with exceptional gauge groups and their representations. \\

\noindent This is an invited contribution to the {\it Handbook of Quantum Gravity} (Eds. C. Bambi, L. Modesto, and I. L. Shapiro, Springer 2023).

\end{abstract}

	\newpage
	\setcounter{page}{1}
	\pagestyle{plain}
	\renewcommand{\thefootnote}{\arabic{footnote}}
	\setcounter{footnote}{0}
	
	\tableofcontents





\section{D-branes and Orientifolds}
\label{s:general}

One of the aims of model building in string theory is to find string vacua whose spectrum of massless string excitations in four dimensions
resembles as  closely as possible the experimentally established and extremely successful Standard Model of particle physics (SM).
Embedding the SM within string theory as a consistent theory of quantum gravity is more than merely a proof of principle; it can be viewed as a first step in a more ambitious programme that hopes to {\it understand} some of the mysteries of particle physics from a string theoretic perspective.

The SM has a gauge group\footnote{Throughout this article, we will not distinguish between the gauge algebra and gauge group unless stated explicitly. In particular, we will not discern between $O(N)$ and $SO(N)$ D-brane groups, as the difference cannot be determined by the perturbative arguments that we use.}
\begin{equation}
SU(3) \times SU(2) \times U(1)
\end{equation}
with matter in the representation
\begin{equation}
\label{SM_Spectrum}
3 \times \left[({\bf 3},{\bf 2},\tfrac16)+({\bf \bar 3},{\bf 1},-\tfrac23)+ ({\bf \bar 3},{\bf 1},\tfrac13),+({\bf 1},{\bf 2} ,-\tfrac12)+
({\bf 1},{\bf 1},1)\right] \,. 
\end{equation}
We will refer to these multiplets as $Q, u^c, d^c, L$ and $e^+$, respectively.
They are left-handed fermions. Right-handed fermions in the same representations do not exist, and for this reason the spectrum is
called {\it chiral}. There may also exist non-chiral representations in nature. The most prominent candidates are singlets $(\1,\1,0)$, which could play the role
of right-handed  neutrinos. They are in fact highly desirable in neutrino physics, but their existence is still not established. Such singlets are non-chiral, because
their left- and right-handed components couple in the same way to  the Standard Model gauge group. This means in particular that a mass term can be written down
without breaking the Standard Model gauge group, {\it i.e.} without making use of the Standard Model Higgs mechanism. In general, many other non-chiral particles may
exist. They can be in non-trivial Standard Model representations, and a Dirac or Majorana mass term for these particles is allowed by the Standard Model symmetries.
We do not have any constraints on how large such a mass could be. However, a common assumption in nearly all attempts at string phenomenology is to allow such
non-chiral particles to exist in the massless spectrum, and to implicitly assume that by some unspecified mechanism they acquire a mass, lower than the string scale but
beyond the reach of current experiments.  Hence a main goal of string phenomenology is to find string solutions that reproduce the spectrum \eqref{SM_Spectrum} {\it chirally}. This is just a first step. If that is not possible, the whole idea is in serious doubt. 
A few string vacua have been identified in the literature where this spectrum is indeed realised exactly,
but even then the string solution may differ from the SM in many more details, like for instance the strength of the couplings.

The vast majority of the literature is about supersymmetric realisations of the Standard Model, and this is also what we implicitly assume here, unless stated otherwise. Non-supersymmetric realisations exist, but in general they have serious stability issues. However, the main feature we focus on here, which is the gauge group and the chiral fermion spectrum, is anyway the same.

Since 2011 we have known another particle in the ``light'' (in Planckian units) spectrum: the Higgs boson. In supersymmetric theories, the minimal way to accommodate it is to add two chiral supermultiplets to the massless spectrum, $H_u$ and $H_d$. These two form a non-chiral pair, in the representation 
$({\bf 1},{\bf 2} ,\tfrac12)+({\bf 1},{\bf 2} ,-\tfrac12)$, and therefore they can develop a mass term. Hence they are part of the set of non-chiral particles which string theory should reproduce. 

Historically, the goal of finding the chiral Standard Model spectrum within string theory was first achieved in the framework of compactifications of the ten-dimensional $E_8 \times E_8$ heterotic string, a theory of {\it closed} strings.  Compactification to four dimensions reduces the gauge group rather naturally to $E_6$ with chiral matter in the 78 dimension representation, and from there one could follow the well-known GUT 
path down to the Standard Model, usually via the intermediate groups $SO(10)$ and $SU(5)$. This approach was first considered in 1984 \cite{Candelas:1985en}. Although the ``second string revolution'' had been started in that same year by Green and Schwarz with a paper on {\it open} strings \cite{Green:1984sg}, that possibility was ignored for more than a decade. The ten-dimensional open string gauge group, $\OGroup(32)$, looked far less promising with regard to the SM, and open strings added an extra complication that most -- with the exception of a few courageous ones -- preferred to avoid. 

\subsection{D-branes and Chan-Paton Multiplicities}

This all changed with the discovery of D-branes. At the endpoints of open strings, boundary conditions must be imposed on the two-dimensional world-sheet fields. It had been known for a long time that
two kinds of boundary conditions were possible: Neumann and Dirichlet boundary conditions. If the former are imposed, the endpoints of the open string move through space-time at the speed of light. But if Dirichlet conditions are imposed, the endpoints of the string have a fixed space-time location. This implies the existence of a special point in space-time, which breaks translation invariance. While this option was immediately rejected by most people, it was realised around 1989 that one can impose Dirichlet boundary conditions in some directions of space-time, and Neumann conditions in others \cite{Dai:1989ua}. Then the open string endpoints are  fixed in some directions, but can move freely in others. These endpoints then sweep out a plane, or a membrane, which was called a Dirichlet brane or D-brane for short. The existence of such a  membrane does indeed break translation invariance in directions orthogonal to it, but not in directions parallel to it. It was then understood that our universe could live on top of such a brane, without any contradictions with translation invariance in our own space-time dimensions.

\begin{figure}[htb]
\begin{center}
\includegraphics[width=9cm, height=6cm]{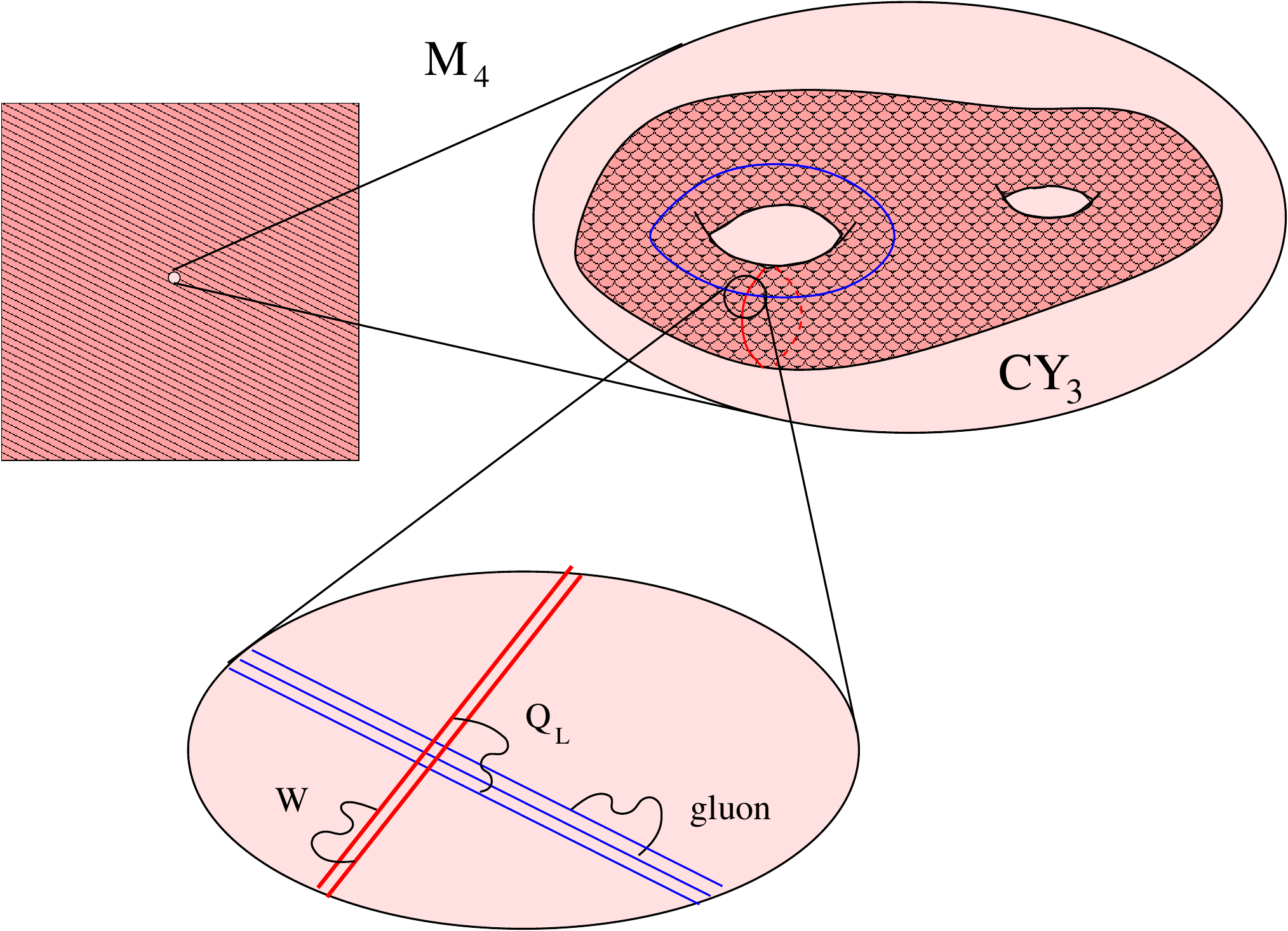}
\caption{Intersecting D-brane World scenario. Figure taken from \cite{Marchesano:2007de}.}
\label{interworld}
\end{center}
\end{figure}

This suggests a picture where we are living on a four-dimensional space-time membrane, embedded in the ten-dimensional space-time of string theory. There would be six uncompactified directions. But this cannot work, because gravity still detects all of space-time, and would therefore not exhibit the $\frac{1}{r^2}$ behaviour characteristic of Newtonian gravity. Hence the extra six dimensions must be compactified, but can remain relatively large as long as limits from fifth-force experiments are respected.\footnote{Alternatively, space in the extra six dimensions may be warped, rather than flat.} All interactions besides gravity are restricted to the brane, and hence impose no constraints on the extra dimensions. From here more involved scenarios can be formulated, because one may consider several higher dimensional branes on top of the four-dimensional space-time that also wrap cycles of the compact manifold, without constraining the strength of gravity. These D-branes can then intersect each other in the extra six dimensions, which is a mechanism to generate a 4d chiral spectrum \cite{Berkooz:1996km}. This has led to the name ``intersecting D-brane models''  \cite{Blumenhagen:2000wh,Aldazabal:2000dg,Aldazabal:2000cn}. See for instance \cite{Blumenhagen:2005mu,Blumenhagen:2006ci,Marchesano:2007de,Lust:2009kp,Ibanez:2012zz} for reviews on the subject, to which we refer for the vast original literature, and figure \ref{interworld} for a pictorial representation of the idea. 
\vspace{-5pt}

Since the early days of string theory, it has been understood that one could consistently assign multiplicities to the boundaries of open strings. These are called Chan-Paton multiplicities \cite{Paton:1969je}. The mode expansion of open strings always contains a massless vector boson, just as the mode expansion of closed strings always contains a massless rank-2 tensor field, the graviton. This massless vector boson behaves like a gauge boson. If there is a Chan-Paton multiplicity $N$, then there are in fact $N^2$ such gauge bosons, and by inspecting their interactions one can verify that they gauge a group $U(N)$. There can be many distinct D-branes in a theory, each defining a place for open strings to end on. For simplicity, one may think of them as D-branes wrapping different cycles on a compactification manifold. Each such brane $a$ can define a Chan-Paton multiplicity $N_a$. One may think of that multiplicity in terms of $N_a$ D-branes stacked on top of each other, and filling our four-dimensional space-time. These are called space-time filling branes. In this situation, an observer in this space-time sees a gauge group
\begin{equation}
U(N_a) \times U(N_b) \times U(N_c) \times \ldots
\end{equation}

\subsubsection{Oriented Strings: Groups and Representations}

Open strings with both ends on the same brane $a$ give rise to a gauge group $U(N_a)$. The matter produced by such open strings includes vector bosons in the adjoint representation of $U(N_a)$. This immediately suggests the possibility of open strings having their endpoints on {\it different} branes, say $a$ and $b$. It is clear that the physical particles produced by such strings must transform as a fundamental representation of $U(N_a)$ as well as that of $U(N_b)$. This is strongly suggested by the multiplicity $N_aN_b$ of these states, and can be verified by working out the scattering amplitudes. Hence what one obtains from such strings are particles in the {\it bi-fundamental} representation 
$({\bf N_a},{\bf N_b})$. The mass and spin of these particles does not follow from this argument alone; we will return to this later. 

The gauge group $U(N_a)$ has complex representations. Hence the multiplicity $N_a$ can correspond to the representation ${\bf N_a}$ or its complex conjugate, ${\bf \bar{N}_{a}}$. What determines which one of the two we get? The open strings we are considering here are actually {\it oriented}. This defines a sense of direction along the open string, or in other words, one can consistently draw an arrow along it. Hence the two endpoints are distinct. Now we assign the endpoint with an outgoing arrow to ${\bf N_a}$ and the one with an incoming arrow to ${\bf \bar{N}_{a}}$. This is the origin of particles in the adjoint representation: Open strings with both ends on the same brane produce a representation in the tensor product of ${\bf N_a}$ with ${\bf \bar {N}_a}$.

\subsubsection{An Oriented String Model}

\begin{wrapfigure}[14]{r}{0.45\textwidth}
\begin{center}
\includegraphics[width=2in]{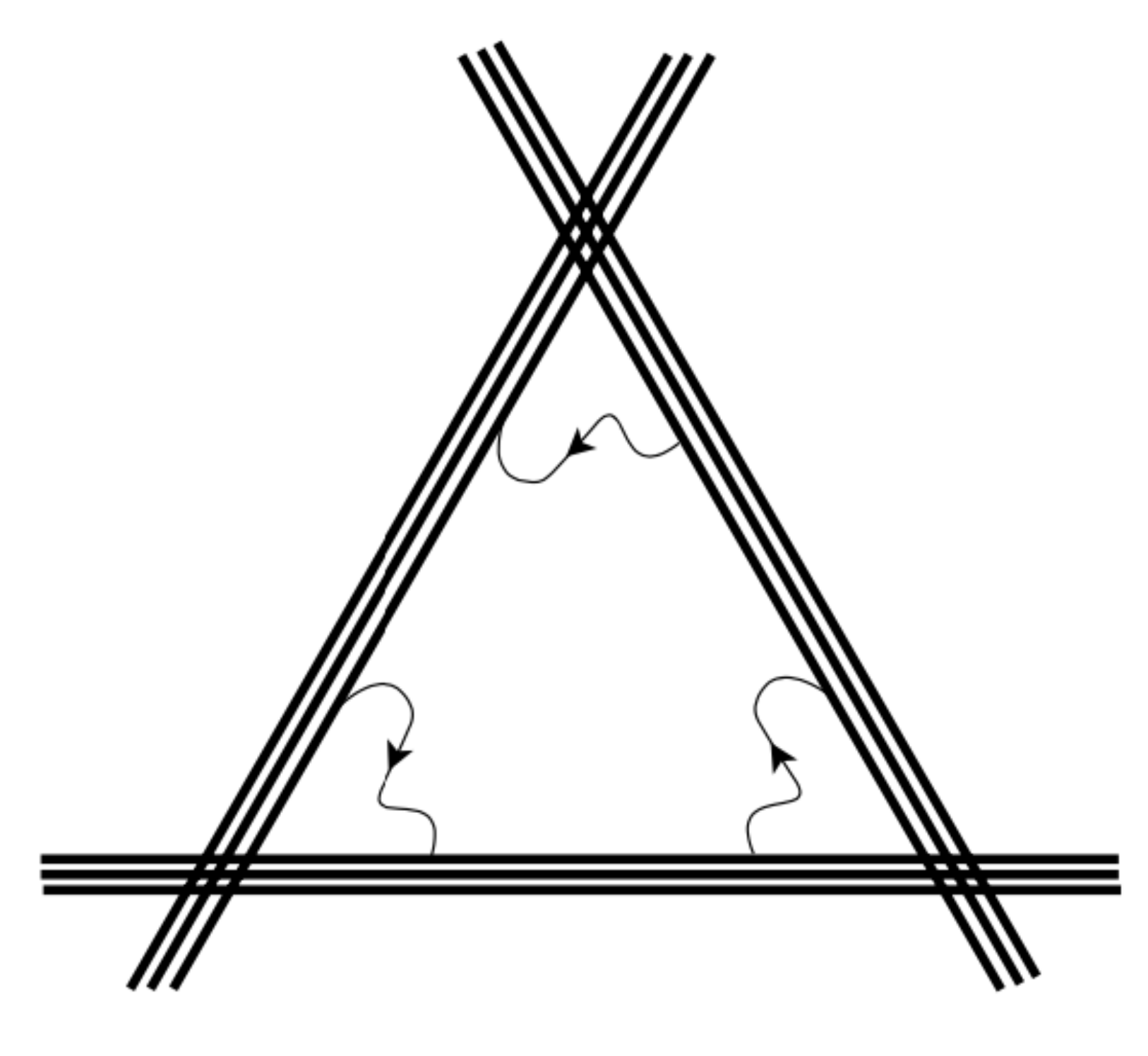}
\caption{\em Trinification, an example of an oriented D-brane model. \label{GUTTrinificationFigure}}
\end{center}
\end{wrapfigure}

As a warm-up exercise let us construct a simple brane configuration that will turn out to contain the Standard Model. 
Consider three stacks of $U(3)$ D-branes. Hence the gauge group is \begin{equation}
U(3)\times U(3) \times U(3) \,.
\end{equation}
Now connect each pair of stacks with an oriented string, such that each stack contains one start- and one endpoint of the oriented strings, as shown in figure \ref{GUTTrinificationFigure}. 

If these oriented strings have exactly three chiral (and hence massless) modes, the resulting  spectrum is
\begin{equation*}
3 \times \left[ ( {\bf 3},\3c,\1) + (\1,{\bf 3},\3c)+ (\3c,\1,{\bf 3})\right] \,.
\end{equation*}

This spectrum occurs naturally as a step in one of the symmetry breaking paths from $E_6$ Grand Unification to the Standard Model. There are 27 massless states per family; 10 of them occur as mutually chiral pairs and there are two right-handed neutrinos. To arrive at the Standard Model, one has to break the last two factors to $SU(2)\times U(1)$ in a suitable way. This model has plenty  of phenomenological issues, but there is a bigger problem we will have to deal with first. 

\vskip 1cm

\subsection{From Oriented String Models to Orientifolds}

\subsubsection{The Need for O-planes}\label{Oneed}

It turns out that in addition to branes another ingredient is always needed, at least in supersymmetric theories: unoriented strings. These are strings, open or closed, without a definite orientation. If an oriented open string, with endpoints {\it a} and {\it b}, traces out  a loop through space-time, the loop can only be closed by linking the {\it a} and {\it b} boundaries to themselves. The resulting string diagram is an annulus. But if a string is unoriented, the endpoints {\it a} and {\it b} are indistinguishable, so one can also link {\it b} to {\it a} when closing the loop. This results in a Moebius strip.  Analogously, for {\em orientable} closed strings the one-loop diagram is a torus, but in the case of {\em unorientable} closed strings there is an additional diagram, the Klein bottle. 

Constructions using unoriented strings are called ``orientifolds'' \cite{Sagnotti:1987tw,Horava:1989vt}, by analogy to orbifolds. The idea is that one uses world-sheet parity as an orbifold symmetry. For instance, the Klein bottle amplitude can be thought of as a closed string sweeping out a closed loop, inverting its orientation before closing the loop, see  \cite{Angelantonj:2002ct} for further details. 
Unorientable strings enter the discussion not only as a logical possibility providing additional model building options, but also because without them it is impossible to construct supersymmetric string vacua with space-time filling D-branes.

To understand why,  
note that D-branes carry charge under the higher-form gauge potentials in the Ramond-Ramond (RR) sector of the superstring theory.
Indeed, the worldvolume of a D$p$-brane enjoys an electric coupling of the form $S = \mu_p \int_{{\rm D}p{\rm -brane}} C_{p+1} +\ldots$ with $\mu_p$ the D$p$-brane charge.  
In compactifications of string theory to four dimensions, a D$p$-brane which extends along the visible 3+1 dimensions must fill a $(p-3)$-dimensional closed subspace -- a so-called $(p-3)$-cycle $\Pi_{(p-3)}$ --  on the six-dimensional compactification manifold $X_6$.  Due to their RR charge, the D$p$-branes act as a source for the $p$-form gauge potential along the $9-p$ dimensions normal to the D$p$-brane on the compact internal manifold. One can in fact characterise the charge under $C_{p+1}$ in terms of the homology class $[\Pi_{(p-3)}] \in H_{(p-3)}(X_6, \mathbbm{Z})$. This source is constrained by Gauss' law: The net charge on a compact space must vanish. In homological terms this amounts to requiring that a configuration of several D-branes satisfies 
\begin{equation}\label{tadpole}
\sum_a N_a [\Pi_{(p-3)}^a] = 0 ,
\end{equation}
where we have included the multiplicity of D-branes wrapping each internal cycle. If a D-brane couples to other RR fields due to non-trivial worldvolume fluxes or curvature corrections, the induced D-brane charge must be cancelled as well, a condition that can be formulated in a similar fashion to \eqref{tadpole}.

One can see that \eqref{tadpole} cannot be satisfied in a supersymmetric D-brane configuration unless additional objects with opposite charge and tension to D-branes are introduced. The reason is that two mutually BPS D-branes will add both their charge (as a sum of homology classes) and their tension (as a sum of positive numbers). Therefore, the total tension in a supersymmetric D-brane configuration is always a linear function of the D-brane total charge, and a vanishing-charge condition like \eqref{tadpole} cannot be compatible with only positive-tension objects.

From a 4d viewpoint, the problem can be understood in terms of the cancellation of the 4d dilaton tadpole. The 4d dilaton has positive couplings (tension) to all D-branes, and hence the one-point functions with a single dilaton external lines is a sum of positive terms. Hence  the 4d dilaton tadpole cannot cancel with only D-branes.

This requires the introduction of objects of negative charge and tension, the so-called orientifold $p$-planes, or O$p$-planes for short. 
As it turns out, such O$p$-planes are the fixed-point loci of involutions of the form $\Omega {\cal R} (-1)^N$, where $\Omega$ is the worldsheet parity operator that reverses the orientation of the string as described above and ${\cal R}$ acts as a geometric involution on the compactification space. Furthermore $(-1)^N$ is an operator that makes $\Omega {\cal R} (-1)^N$ square to the identity, and depends on the specific theory under consideration. In practice an O$p$-plane is specified by a submanifold or a sum of submanifolds $\Pi_{(p-3)}^{\rm O}$ of $X_6$ fixed by the geometric involution ${\cal R}$. 
Because this involution must be a symmetry of the compactification, if a D-brane internal worldvolume is not invariant under ${\cal R}$ there must be another D-brane located at $\Pi_{(p-3)}^{a'} = {\cal R} \Pi_{(p-3)}^a$, in order to identify their worldvolume theories. 
After O$p$-planes are introduced, the RR tadpole condition \eqref{tadpole} is modified to
\begin{equation}\label{tadpoleO}
\sum_a N_a \left( [\Pi_{(p-3)}^a] + [\Pi_{(p-3)}^{a'}]\right)  = Q_{{\rm O}p} [\Pi_{(p-3)}^{\rm O}] ,
\end{equation}
where $Q_{{\rm O}p} = 2^{p-4}$ is minus the relative charge of an O$p$-plane and a D$p$-brane wrapping the same $(p-3)$-cycle. Now there is no obstruction to building a supersymmetric D-brane configuration. The simplest one is to place all D-branes wrapping homological cycles to the O-planes:  $[\Pi_{(p-3)}^a] =[\Pi_{(p-3)}^{\rm O}]$ and $N_a =   Q_{{\rm O}p}/2$. 

To sum up, combining positive tension objects like D-branes in a supersymmetric fashion necessarily forces us to introduce negative-tension objects known as O-planes.\footnote{Non-supersymmetric setups have additional options, since one may introduce anti-branes.} These objects appear when one mods out an oriented configuration  by an orientation-reversal symmetry of the theory. From the Type II perspective they appear as non-dynamical objects of negative tension, whose microscopic nature can only be unveiled in a non-perturbative framework like F-theory.

\subsubsection{Unoriented Strings: Groups and Representations}

One direct consequence of the presence of O-planes is that new groups and representations appear. Intuitively, if a D-brane is invariant under the orientifold projection, the D-brane group becomes real: instead of $U(N)$ it becomes either $SO(N)$ or $USp(N)$ (with $N$ even). $U(N)$ groups are also still possible for some branes if their worldvolume is not fixed under the orientifold involution. One may think of this as two branes mapped to each other through the orientifold plane, a brane $a$ and its {orientifold image brane $a'$}. For a more precise discussion of brane groups in orientifold models see Section \ref{BraneGroups}.\footnote{{In the RCFT literature, it is more common to refer to the orientifold image brane $a'$ as the conjugate brane $a^c$. Both terms can be used interchangeably.}}

All open string states are bi-fundamentals of one or two brane groups. Hence in D-brane model building one must realise all Standard Model matter as bi-fundamental  representations. In the unoriented case, open strings can connect to branes passing through an O-plane, and reversing their orientation. This implies that a bi-fundamental between two unitary branes can be of the form $(\V,\Vc)$ in addition to $(\V,{\V})$. Here $\V$ denotes the fundamental (or vector) representation of $U(N), SO(N)$ or $USp(N)$.

Furthermore, the two endpoints of an open string can be a fundamental representation on the same brane. This allows the existence of rank-2 tensors. These tensors can be symmetric and anti-symmetric, and will be denoted ${\bf A}$ or ${\bf S}$ respectively. On unitary branes these tensors are complex representations of the brane group: one can have 
 ${\bf A}$, ${\bf \bar A}$, ${\bf S}$ and ${\bf \bar S}$.  Rank two tensors can also occur for real groups, and one can get adjoint representations of unitary groups if the open string endpoints are on a brane $a$ and its {orientifold image $a'$.}

Only complex representations give rise to chiral matter. This means that if a theory contains left-handed fermions in the representation $n(\V,\V)+{\overline n}({\Vc},{\Vc}) + m(\V,{\Vc})+{\overline m}({\Vc},\V)$, this is equivalent to $(n-{\overline n})(\V,\V)+ (m-{\overline m})(\V,{\Vc})$, up to non-chiral matter. Rank-2 tensors of real groups and adjoints of unitary groups are chirally irrelevant, and although they may appear in the massless spectrum of specific D-brane configurations, these are states that a priori are not protected against becoming massive via a number of effects. 

Despite this richer structure, it turns out that the 4d chiral spectrum obtained for D-branes in orientifold compactifications is quite universal. To describe it, one needs to define a chiral index $I_{ab}$ between two D-branes, that is a bilinear, anti-symmetric tensor of their D-brane charges (not counting their multiplicity $N_a$). The expression for $I_{ab}$ changes from one model-building setup to another, but it is always of topological nature. If the D-brane with charge $\Pi^b$ is not invariant under the orientifold action, there will be an orientifold image with charge $\Pi^{b'}$ and a corresponding index $I_{ab'}$. Finally, one can also extend this definition to include an index $I_{aO}$ between a D-brane and the O-plane content of the compactification. Once that this index has been defined, the chiral spectrum arising from the open string sector of the compactification reads as in Table \ref{specori}.

\begin{table}[htb]
\renewcommand{\arraystretch}{1.25}
\begin{tabular}{ll}
\hline
Non-Abelian gauge group & $\prod_a SU(N_a)$\\
Massless $U(1)$s &  $\sum_a c_a U(1)_a$ such that $\sum_a c_a ([\Pi^a] - [\Pi^{a'}]) = 0$  \\
Chiral fermions & $\sum_{a<b}\, I_{ab} (\V_a, \Vc_b) \, + \, I_{ab'} (\V_a, \V_b) $\\
& $\frac{1}{2} (I_{aa'} -  I_{aO}) {\bf S}_a\, + \, \frac{1}{2} (I_{aa'} +  I_{aO}) {\bf A}_a$\\
\hline
\end{tabular}
\caption{Chiral spectrum of a  D-brane orientifold compactification, in terms of the chiral index $I_{ab}$. For simplicity we are assuming that there are no D-branes with $\OGroup(N)$ or $USp(N)$ gauge group. $\Pi^a$ represents the charge of the D-brane. ${\bf S}_a$, ${\bf A}_a$ stand for the symmetric and antisymmetric representations of $U(N_a)$. }
\label{specori}
\end{table}

The fact that one has such a universal chiral spectrum for all perturbative orientifold models allows one to devise model-building strategies that are independent of their specific realisation, as will be discussed in the next section. Let us however stress that once one leaves the realm of perturbative constructions, new types of gauge groups, matter representations  and, consequently, model building possibilities arise. The reason is the appearance of non-perturbative bound states of strings which can have more than two endpoints, hence realising, for instance, higher tensor representations or spinor representations. 
This will be described in detail in the context of F-theory in Section \ref{sec_F-theory}.

\section{D-brane Model Building: Generalities} \label{sec_Generalities}

As it turns out, the Standard Model can be built very easily and naturally out of the limited set of bi-fundamentals and rank 2 tensors available already in perturbative D-brane models.
In this section we outline the systematics underlying the search for Standard Model like vacua in Type II orientifolds. This approach can then be applied both in geometric Type IIA or Type IIB orientifolds (see Section \ref{s:typeII})  and in conformal field theoretic models (Section \ref{s:RCFT}). Model building in the non-perturbative generalisation described by F-theory, in particular in the context of Grand Unified Theory (GUT) model building, is the topic of Section \ref{sec_F-theory}.

\subsection{Anomalies, Tadpoles and Axions}\label{ATA}

In any chiral model there is one important constraint to be taken into account: chiral anomaly cancellation. Anomalies cancel automatically in string theory, provided one satisfies all consistency conditions. The most important of these is in this context the cancellation of all RR-tadpoles. Note that there may also be NS-NS tadpoles. They automatically cancel in supersymmetric models that are free from RR-tadpoles. In non-supersymmetric setups  uncancelled NS-NS tadpoles  imply instabilities, which is a serious problem, but not an inconsistency.

The first step towards building the Standard Model consists of assembling a set of branes whose spectrum of chiral fermions is the same as that of the SM.  Unless one is extremely lucky, this set does not satisfy tadpole cancellation. In particular the NS-NS 4d dilaton tadpole can be oversaturated or undersaturated.  In the former case the total contribution of all branes plus the orientifold plane is positive. Then there is nothing one can do about this anymore. But if the total contribution is negative, one has the option of adding some additional branes to the configuration. This means that one chooses to assign a non-zero Chan-Paton multiplicity to some branes that are not part of the SM configuration. This must be done in such a way that no chiral particles are added to the spectrum: preferably no massless particles at all, or at least no chiral particles charged under the SM gauge group, dubbed {\em chiral exotics}. These additional branes are often referred to as a {\em hidden sector}.
Such a sector may have several other uses, such as breaking supersymmetry or providing dark matter.

\begin{table}[htb]
\renewcommand{\arraystretch}{1.5}
\begin{center}
\begin{tabular}{c|c|c|c|c|c|c}
$r$  & {\V} & ${\Vc}$ & {\bf S}  & ${\bf \bar S}$  & {\bf A} & ${\bf \bar A}$  \\
\hline
\hline
$D(r)$ & $N$ & $N$ & $\frac{N(N+1)}{2}$ & $\frac{N(N+1)}{2}$ & $\frac{N(N-1)}{2}$ & 
$\frac{N(N-1)}{2}$ \\ 
\hline
$Q(r)$ &  1 & -1 & 2 & -2 & 2 & -2 \\
\hline
$I_2(r)$ &  $1$ &  $1$ &  $N+2$ & $N+2$ & $N-2$ 
& $N-2$ \\
\hline
$I_3(r)$ & 1 & -1 & $N+4$ & $-N-4$ & $N-4$ & $-N+4$ \\
\end{tabular}
\caption{Properties of the fundamental (vector) {\V}, symmetric {\bf S}, and antisymmetric {\bf A} representations of $SU(N)$.  $D(r)$ is the dimension of the representation $r$, $Q(r)$ is the $U(1)$ charge of $r$ under the decomposition $U(N) = SU(N) \times U(1)$ and $I_2(r)$, $I_3(r)$ are the quadratic and cubic anomaly coefficients (also  known as Dynkin indices), respectively.}
\label{DynkinIndices}
\end{center}
\end{table}

\subsubsection{Non-abelian Anomalies}
Since finding hidden sectors can be very laborious, it helps to eliminate some SM configurations at an early stage.  This means first of all that all non-abelian anomalies must cancel. Since we are building the Standard Model, one may think that SM anomaly cancellation ensures this, but this is only partly true. Indeed, if one has realised the SM spectrum, anomaly cancellation ensures the absence of $SU(3)_{\rm QCD}$ anomalies. 

In  QFT, non-abelian anomaly cancellation is a condition on representations of $SU(N)$, $N\geq 3$. In particular, different representations of $SU(N)$ contribute to the $SU(N)$ cubic anomaly as their cubic anomaly coefficient, see Table \ref{DynkinIndices}, and 
the sum of their contributions must vanish. In QFT this must be imposed by hand, or else gauge invariance must be dropped.  In string theory all anomaly cancellations follow from some deeper consistency condition, such as modular invariance for closed strings.  In QFT, the limit $N \geq 3$ arises because the group $SU(2)$ has pseudo-real representations, so that ${\bf 2}$ is equivalent to $\2c$. Furthermore, $SU(1)$ is trivial. But in string theory $U(3)$, $U(2)$ and $U(1)$ brane stacks  are all on the same footing, and there is no reason to expect a lower limit on $N$. 

In open strings the anomaly cancellation condition was first derived by Bianchi and Morales \cite{BiMor}. As expected, they found 
that  non-abelian anomalies must cancel for $U(N)$ for all $N$, even if $N=2$ or $N=1$.  If a candidate SM configuration contains $U(2)$ or $U(1)$ branes the conditions must be checked, and if it not satisfied the configuration can never be realised. We refer to this class of anomalies as ``non-abelian'' anomalies using QFT terminology,  although $SU(2)$ is anomaly-free in QFT and $U(1)$ is abelian.

Let us see how the cubic non-abelian anomaly looks like in D-brane models. Using the spectrum of Table \ref{specori} and the anomaly coefficients of Table \ref{DynkinIndices} we find
\begin{equation}
  {\cal A}_{SU(N_a)^3}  =  \sum_{r\, {\rm irrep}} I_{3,a}(r) =  
\sum_{b} N_b \left(I_{ab} + I_{ab'} \right) - 4 I_{a O} ,
\label{cubic}
\end{equation}
which must vanish for any $SU(N), N \geq 3$, present in the model. In general one can see that the rhs vanishes when RR tadpole conditions are imposed, by using the appropriate generalisation of \eqref{tadpoleO} and bilinearity of the chiral index. As anticipated, this occurs even for $N_a <3$. The reason is that otherwise some other anomalies would be left uncancelled, namely mixed and abelian anomalies.

\subsubsection{Mixed and Abelian Anomalies}

Mixed and abelian anomalies are those that include $U(1)$ symmetries. Using again the content of Tables \ref{specori} and \ref{DynkinIndices} they read:
\begin{align}
   & {\cal A}_{U(1)_a-SU(N_b)^2} = \sum_{r\, {\rm irrep}} Q_a(r) I_{2,b} (r) =   \delta_{ab}  {\cal A}_{SU(N_a)^3}  +  N_a \left(I_{ab} + I_{ab'}\right) , \\
   & {\cal A}_{U(1)_a-U(1)_b^2}  = \sum_{r\, {\rm irrep}} Q_a(r) Q_b(r)^2 =
     \delta_{ab} N_a {\cal A}_{SU(N_a)^3}  + N_a N_b \left(I_{ab} + I_{ab'}\right) ,
\end{align}
where we assuming that the abelian factors arise from $U(N_a) = SU(N_a) \times U(1)_a$, and $r$ now runs over the irreps with respect to the groups involved in the anomaly.  Here, abusing notation, we have denoted by ${\cal A}_{SU(N_a)^3}$ the rhs of \eqref{cubic}, which is a well-defined quantity even for $N_a <3$.  It turns out that, in both cases, the term proportional to $I_{ab} + I_{ab'}$ is cancelled by a generalised Green--Schwarz  mechanism. However, the first term has to cancel due to the D-brane configuration. That is why, in practice, one needs to impose the condition ${\cal A}_{SU(N)^3} =0$ even for $N=1, 2$. 

The generalised Green--Schwarz  mechanism works by means of a mixing term of the longitudinal component of an abelian vector boson with an axion. If such term is present, the vector boson acquires a mass by absorbing the axion using the St\"uckelberg mechanism. This eliminates all vector bosons from the spectrum that couple to anomalous $U(1)$s, but the mechanism may also affect non-anomalous $U(1)$s. The result is that only the combinations of $U(1)$s that appear in Table \ref{specori} remain massless, and  this may be bad or good: one the one hand it may make the Standard Model $Y$-boson -- and hence the photon -- massive, but on the other hand it may eliminate undesirable non-anomalous $U(1)$s that occur in most models. Most frequently this occurs with vector bosons that involve $B-L$,  a non-anomalous symmetry in the Standard Model when three right-handed neutrinos  are added.

It is customary to distinguish local and global models. A local model has a Standard Model brane configuration and no non-abelian anomalies. Constructing it only requires knowledge of the chiral indices of the participating branes. If all tadpoles are cancelled and the photon remains massless one speaks of a global model. Note that checking the latter feature cannot be done in a local model, as it requires knowledge of the full set of available axions, which cannot be derived from the brane configuration alone.

\subsection{The Simplest Examples}\label{SimplestExamples}

In this section we will derive the simplest possible realisations of the Standard Model spectrum in terms of bi-fundamentals and rank-2 tensors. These are not explicit realisations, although we will indicate if such realisations are known. The steps towards an explicit realisation are: 
\begin{enumerate}
\item{Find a brane configuration that contains the Standard Model.}
\item{Check absence of cubic $U(N)$ anomalies, including $N=2$ and $N=1$.}
\item{Find a realisation of this configuration in terms of actual branes.}
\item{Check absence of a mass for the $Y$-boson.}
\item{Check tadpole cancellation, or cancel tadpoles by means of a hidden sector.}
\end{enumerate}
Here we will limit ourselves to discuss steps 1 and 2. 
The broadest exploration of explicit realisations was done in the context of RCFT Gepner models in \cite{ADKS}, where all five steps were considered. Models were collected if they passed step 4. Here we aim for maximal simplicity: the minimal number of branes, exact family repetition (so that we can focus on a single family), the minimal gauge group, and no superfluous non-chiral pairs. Since these guiding principles are debatable, we will make concessions if necessary. 

Perhaps one's first thought about realising the Standard Model with branes would be to take a $U(3)$ for QCD, a $U(2)$ or $USp(2)$ brane for the weak interactions and a $U(1)$ brane for the $Y$-charge\rlap.\footnote{Note that $USp(2)$ has the same Lie algebra as $SU(2)$. The same is true for $O(3)$, but open strings would give matter in the 3-dimensional vector representation of $O(3)$, which does not occur in the Standard Model.} 
However, associating the $Y$ charge with a separate $U(1)$ brane cannot work, because
the quark-doublet $({\bf 3},{\bf 2},\frac16)$ would then be a tri-fundamental, which do not occur in open string models. 

\subsubsection{The \texorpdfstring{$Y$}{Y}-charge}

Hence the $Y$-charge must involve the $U(1)$ factor of $U(3)$, the $U(1)$ factor of $U(2)$ (unless $USp(2)$ is used), and perhaps one or more additional branes.  
We label the branes as $a$ (for the QCD brane) $b$ (for the weak brane), and $c$, $d$, $\dots$ for any additional $U(1)$ branes. The $Y$-charge generator must then take the form 
\beq
Y=(x-\tfrac13) Q_a + (x - \tfrac12) Q_b  + \gamma Q_c + \delta Q_d + \ldots \,.
\eeq
We follow here the conventions of \cite{ADKS}, and assume that all branes are unitary and  $Q_i$ is the $U(1)$ generator of brane $i$. We normalise these generators so that a vector representation has charge $+1$. If the coefficient of any $Q_i$ vanishes, one can replace the corresponding brane by a real one with a symplectic or orthogonal group.

In the  following we will examine all possibilities for assigning quarks and anti-quarks, using the minimal number of branes, and for a single family.

\subsubsection{The Quark Doublet}

The coefficients of $Q_a$ and $Q_b$ ensure that the quark doublet gets the correct $Y$-charge, assuming  that the bi-fundamental we use for the quark doublet is $(\V,\Vc,0,\ldots)$.  Alternatively one could use $(\V,\V,0,\ldots)$.  This is just a convention, but it is useful to choose all multiplets as if we are in an orientable string theory, {\it i.e.} a fundamental representation on one end and an anti-fundamental on the other, until we do not have a choice anymore. At that point we know that we have reached a non-orientable configuration.  If $x=\tfrac12$ then $Q_b$ does not participate in $Y$, and we may use  $USp(2)$ instead of $U(2)$ for brane $b$. This choice makes the configuration non-orientable. Note that in orientable configurations the coefficients of $Q_i$  are not fully determined by the SM spectrum: if we shift all coefficients by the same amount $\Delta$, then $\Delta$ cancels out between the two open string ends.

\subsubsection{Anti-quarks}

Next we can try to assign the anti-quarks $u^c$ and $d^c$. The representation ${\bf \bar{3}}$ of the antiquark can be obtained in two ways: as a rank-2 anti-symmetric tensor, from an open string with both ends on the $U(3)$ brane, or as a $\Vc$ endpoint of an open string. In the latter case the other endpoint of that string has to end on another brane.  One then has the four options summarised in Table \ref{tablasm}.

\def\mystrut(#1,#2){\vrule height #1pt depth #2pt width 10pt}   

\begin{table}[htb] 
\renewcommand{\arraystretch}{1.35}
\begin{center}
\begin{tabular}{|c|c|c|c|c|c|c|c|}
\hline Class & $u^c$  &   $d^c$  &   $x$ & $\gamma$ & $\delta$  & L & $e^+$\\
\hline
\hline $(i)$
& $(\Vc,0,\V)$   &  $(\Vc,0,\Vc)$ &  $\tfrac12$    & $-\tfrac12$  & -- &   \parbox[t]{2cm}{$(0,\V,\V)$\\ ~~~~~ or\hfill  \\$(0,\Vc,\V)$} &  $(0,0,{\bf \bar S})$     \\
\hline $(ii)$  & $(\Vc,0,\V,0)$    &  $(\Vc,0,0,\V)$ &  $x$    & $x$  & $x-1$ &  \parbox[t]{2cm}{$(0,\V,\Vc,0)$\\ ~~~~~ or\hfill  \\$(0,\Vc,0,\V)$} & $(0,0,\V,\Vc)$\\
\hline $(iii)$ & $({\bf A},0,0)$ & $(\Vc,0,\V)$ &  0 & 0 & --  & $(0,\V,\V)$   & $(0,\bar{\bf A},0)$\\
\hline $(iv)$ & $(\Vc,0,\V)$   &  $({\bf A},0,0)$ &  $\tfrac12$   & $-\tfrac12$& --  & $(0,\V,\V)$    & $(0,0,\bar{\bf S})$\\
\hline
\end{tabular}
\end{center}
\vskip -.5cm
\caption{\small The four basic quark/anti-quark configurations for one family. Columns 2-5 define the class. The last two columns specify the possible completions to a full family, if no further branes are added, and for generic $x$ in case $(iv)$.}
\label{tablasm}
\end{table}
\begin{center}
\begin{figure}[ht!]
\includegraphics[width=5.6in]{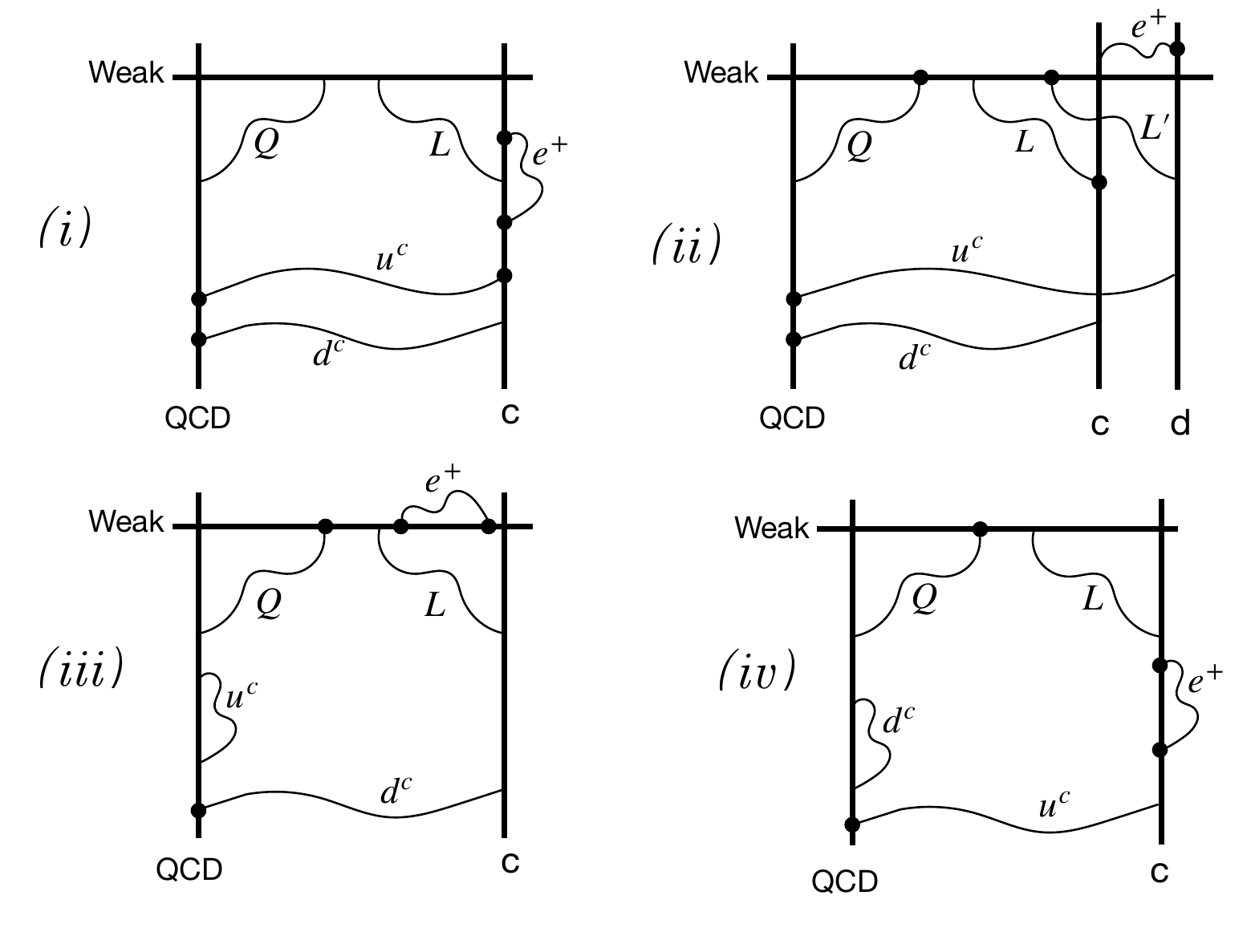}
\caption{The four basic classes from Table \ref{tablasm}. Dots indicate a coupling to the {orientifold image} brane. $L$ and $L'$ are two possible assignments of the lepton doublet. In case $(i)$ a weak $USp(2)$ group is assumed. \label{FourClasses}}
\end{figure}
\end{center}

At this point all configurations, except $(ii)$, are non-orientable. In the quark sector, all models in the literature necessarily have one of these for structures for a single family.
With three families one has the option to make different choices per family, as long as the coefficients of $Q_i$ match.   

\subsubsection{The Lepton Doublet}

Now we can try to assign the lepton doublet $({\bf 1},{\bf 2} ,-\tfrac12)$. If we do not add more branes, there is a unique choice in all cases except $(ii)$: 
a bi-fundamental $(0,\V,\V)$ between branes $b$ and $c$. In case $(ii)$ there are two options: Either $(0,\V,\Vc,0)$ or $(0,\Vc,0,\V)$. 
In both cases the configuration remains orientable. 

\subsubsection{The Left-handed Positron}

Finally we try to assign the charged lepton $({\bf 1}, {\bf 1},1)$. Being an $SU(2)$ singlet, it must either come from a string that does not end on brane $b$, or an 
anti-symmetric tensor on brane $b$. In   cases $(i)$ and $(iv)$ this fixes it uniquely to $(0,0,{\bf \bar S})$. In  case $(iii)$ the only option is
$(0,{\bf \bar A},0)$. In case $(ii)$ a natural choice is $(0,0,\V,\Vc)$. One could also use rank-2 tensors, but only for special values of $x$.

\subsubsection{Weak Interaction Anomalies}

The SM contains a quark doublet $({\bf 3}, {\bf 2},\frac16)$
and a lepton doublet $({\bf 1},{\bf 2} ,-\tfrac12)$ per family. The gauge group $SU(2)$ is anomaly free in quantum field theory. If it is realised as $USp(2)$ in string theory, this is also true. But if it is realised as $U(2)$ in string theory, there are anomalies not seen in field theory. We have to decide if the representation denoted ``${\bf 2}$'' is actual ${\bf 2}$ or $\2c$. If we do that in a single family it is immediately clear that there is no way to cancel the weak anomalies from the quark and the lepton doublets.  There are four ways out of this:
\begin{enumerate}
\item If $x=\tfrac12$ one may use $USp(2)$ instead of $U(2)$. Then there are no weak brane anomalies.
\item If $x=\tfrac12$ one may also use $(\V,\V)$ as a quark doublet. Then one may drop exact family repetition, to write the three doublet as $2(\V,\Vc)+(\V,\V)$.
\item One may add a rank-2 anti-symmetric anti-tensor. This contributes $+2$ to the weak anomaly, cancelling the contribution $-3$ of the quark doublet, and $+1$ of the lepton doublet. This adds a singlet chiral particle with charge $1-2x$ to the spectrum. For the most common values of $x$ this is a left-handed positron or a neutrino.
\item If there are two ways of realising a lepton doublet with opposite weak anomalies, one may add a non-chiral pair. For example suppose $(\V,X)$ and $(\Vc,Y)$ are both lepton doublets, where $X$ and $Y$ are some combination of brane representations. Now one may use a combination $2 \times (\V,X)+(\V,{\bar Y})$ to get three times the contribution of a single lepton doublet, but a net number of only one chiral lepton doublet. This adds a non-chiral pair to the spectrum with the features of a Higgsino pair $H_u+H_d$.
\end{enumerate}

\subsubsection{The Four Classes of Models} \label{sec:4classes}

\paragraph{Class {\em (i)}.} This class can be made anomaly free on the weak brane by means of mechanisms 1, 2 or 3. The $c$ brane anomaly cancellation is a bit awkward. The anomaly of the model as shown in the table would be -3. To cancel it we may add an anti-symmetric tensor ${\bf A}$ on brane $c$.  This has anomaly -3 and ground state dimension 0, and hence there are no massless states at all in this sector. Although this may seem a bit weird,  explicit realisations of such $U(1)$ branes were found in \cite{ADKS} in fully tadpole-free models. 

But there is a more appealing way. We can add a fourth brane with the same contribution to $Y$ as brane $c$, and realise the left-handed positron as $(0,0,\Vc,\Vc)$. Then we may add a string $(0,0,\V,\Vc)$, which has vanishing $Y$ charge, and connect the lepton doublet to brane $d$ instead of brane $c$. Now all anomalies on branes $c$ and $d$ cancel. Furthermore, if we use mechanism 1 or 2 on the weak brane we get a model entirely built out of bi-fundamentals. If we take the weak group as $USp(2)$ and assume mechanism 1, the full chiral spectrum is:\\
\begin{wrapfigure}[4]{l}{0.5\textwidth}
\begin{center}
\includegraphics[width=2.4in]{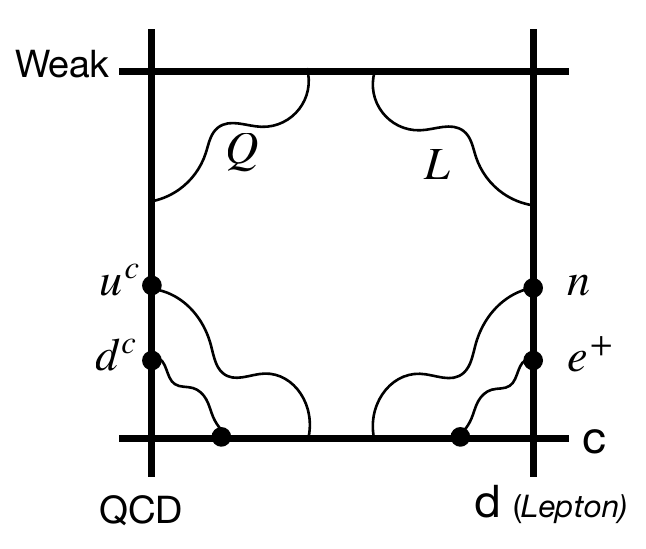}
\caption{The ``Madrid'' configuration, with a weak interaction group $USp(2)$. \label{MadridQuiver}}
\end{center}
\end{wrapfigure}
\begin{eqnarray*}\noindent
 & \  \quad \ \  \\
3& \times &(\V,\V,\ 0,0)  \quad\,   Q\\
3& \times &(\Vc,0,\V,0) \quad\ u^c\\
3& \times &(\Vc,0,\Vc,0) \quad\  d^c\\
3& \times &(0,\V,0,\V) \quad\   L\\
3& \times &(0,0,\Vc,\Vc) \quad\ e^+\\
3& \times &(0,0,\V,\Vc) \quad\  n\\
 & \  \quad \ \\  
 & \  \quad \ \\  
 & \  \quad \ \\  
 & \  \quad \ \\  
 & \  \quad \ \  
\label{MadridModel}
\end{eqnarray*}
\noindent This spectrum is shown in Figure \ref{MadridQuiver}. It is the most-studied class of brane models, first explored in detail in 
\cite{Ibanez:2001nd}. Numerous examples and variations have been found in subsequent papers.

A noteworthy feature is the presence of a baryon and a lepton brane: All quarks attach to brane $a$ and all leptons to brane $d$. In fact, branes $a$ and $d$ have the same intersections with branes $b$ and $c$. One may combine branes $a$ and $d$ into a $U(4)$ stack, and extend the $U(1)$ group on brane $2$ to $U(2)$ or $USp(2)$  to get a left-right symmetric model. Combining all this one obtains a $SU(4)\times SU(2)_L \times  SU(2)_R$ Pati--Salam model, see \cite{Cremades:2002qm,Cremades:2003qj} and Section \ref{s:typeII} for specific realisations of this idea.

The unitary factors of branes $a$ and $d$ are anomalous and therefore broken by axion mixing. They remain as global baryon number (B) and lepton number (L) symmetries. The linear combination $B-L$ is anomaly free, and the corresponding gauge boson may or may not become massive. This can only be decided by examining the full global model, not just the local configuration. In \cite{ADKS} global examples were found where the $B-L$ photon acquires a mass. Furthermore, examples were found with tadpole cancellation without any additional branes, both with a massless and a massive $B-L$ photon.


Bi-fundamentals between branes $b$ and $c$ are not used to build the fermion spectrum. But these states have precisely the right quantum numbers to be identified as supersymmetric Higgs multiplets $H_d$ and $H_u$. They have zero lepton number, so that potentially dangerous perturbative couplings such as $LH_u$ are automatically forbidden.

\paragraph{Class {\em (ii)}.} 

This class is characterised by having separate branes for the $u^c$ and $d^c$
endpoints. Models of this kind were first explored in {\cite{Antoniadis:2000ena,Antoniadis:2002qm}. However, weak anomaly cancellation was not considered in these papers. For arbitrary values of $x$, only mechanism 4 is available. Using that mechanism one arrives at a fully orientable brane configuration:
\begin{eqnarray*}
3& \times &(\V,\Vc,0,0) \quad Q \\
3& \times &(\Vc,0,\V,0)  \quad d^c\\
3& \times &(\Vc,0,0,\V) \quad u^v\\
6& \times &(0,\V,\Vc,0) \quad L\\
3& \times &(0,\V,0,\Vc) \quad L^c\\
3& \times &(0,0,\V,\Vc) \quad e^+\label{Oriented}
\end{eqnarray*}
There is a superfluous non-chiral pair $L+L^c$ per family, which has the quantum numbers of a Higgs pair, but is not distinguished from the lepton doublet by its quantum numbers. There is no left-handed anti-neutrino. 

The trinification model shown in fig. \ref{GUTTrinificationFigure} reduces to this model (plus additional non-chiral states) if one splits the second $U(3)$ to $U(2)\times U(1)$ and the third  to $U(1)^3$. 

Explicit realisations of this model have been found \cite{ADKS}, but no global realisations with tadpole cancellation. The models presented in \cite{Antoniadis:2002qm} are not explicit realisations, but hypothetical brane configurations with $x=0$ and $x=1$. Once $x$ is fixed to these values, it is possible to get left-handed positrons as anti-symmetric tensors. Note that in classes $(i)$, $(iii)$ and $(iv)$ the value of $x$  is either 0 or $\tfrac12$. 

\paragraph{Class {\em (iii)}.}

This class can be viewed as a brane realisation of an $SU(5)$ GUT model, with an extra $U(1)$. One takes an anti-symmetric tensor of $U(5)$ plus an anti-vector of $U(5)$, giving rise to the familiar ${\bf 10}+{\bf \bar 5}$. Now one may split the 5-stack into a 3-stack and a 2-stack. Physically, this may be realised by separating the two stacks by a certain amount in the compactified dimensions, or by using entirely different stacks with the same intersections. To construct a $U(5)$ GUT we need a vector of $U(5)$, which is an open string with one endpoint on the $U(5)$ stack and its other endpoint on another brane, labelled $c$. This brane does not contribute the $Y$-charge and hence the simplest choice is an $O(1)$ brane.
The explicit model in the table uses mechanism 3 to cancel the weak anomaly. This anomaly cancellation is inherited directly from the $SU(5)$ anomaly cancellation.

This class of models was first studied in \cite{Blumenhagen:2001te}, but without considering full tadpole cancellation. The latter problem was addressed in
 \cite{Cvetic:2002pj}, but only examples with chiral exotics ({\bf 15} of SU(5)) were found in this paper.
The exact model in the table, with a brane group $U(3)\times U(2)\times O(1)$ has been found frequently in the search of   \cite{ADKS} , and  there are even examples with full tadpole cancellation without any additional branes.  

We anticipate that $SU(5)$ GUTs based on D-branes suffer from the absence 
of a top quark Yukawa coupling at the perturbative level.
This problem is overcome in non-perturbative realisations of such models as described in Section \ref{sec_F-theory}.

\paragraph{Class {\em (iv)}.} This class can  be made anomaly-free on the weak brane by using mechanism 3. In this case the anti-symmetric tensor is a left-handed anti-neutrino. The $c$ brane anomaly cancels because the symmetric anti-tensor contributes -5, and the strings producing $d^c$ and $L$ have a contribution $3+2=5$.  

This class contains flipped $SU(5)$ models. In flipped $SU(5)$  $d^c$ (as opposed to $u^c$) is realised using an anti-symmetric tensor. Flipped $SU(5)$ requires an additional $U(1)$, and that $U(1)$ is realised here as a linear combination of the $U(1)$ from brane $c$ and the phase factor $U(1)$ of $U(5)$. 
Orientifold models of this kind were first studied in \cite{Chen:2005aba}, but also in the flipped $SU(5)$ case the first examples found had chiral exotics. 
Explicit examples of the spectrum shown in the table have been found in the search of \cite{ADKS}. There are even examples of full tadpole cancellation without a hidden sector. 


\subsection{Yukawa Couplings}\label{ss:yukis}

The quarks and leptons discussed above must all acquire a mass from a three-point coupling with a Higgs boson. The latter should
be present in the light spectrum. It must be a weak doublet, hence a bi-fundamental open string with one end on the weak brane. 
Let us call the brane with the other endpoint the Higgs brane.

In perturbative open string theories, three point couplings are represented by a disk diagram with three external lines. 
For these couplings to exist, the three fields must be bi-fundamentals between branes $(a,b)$, $(b,c)$ and $(c,a)$. If these branes are complex, the brane charges of each brane must cancel, in other words there must be an equal number of endpoints $\V$ and $\Vc$ on each brane. These charges, the phase factors of the brane group $U(N)$, are conserved in string perturbation theory, even though the corresponding $U(1)$s may acquire a mass via the St\"uckelberg mechanism. The latter fact means that there are non-perturbative effects, generated by instantons, that break those symmetries, but these symmetries remain as global symmetries in perturbation theory, analogously  to the baryon number in the Standard Model. 

\begin{wrapfigure}[16]{l}{0.41\textwidth}
\begin{center}
\includegraphics[width=2in]{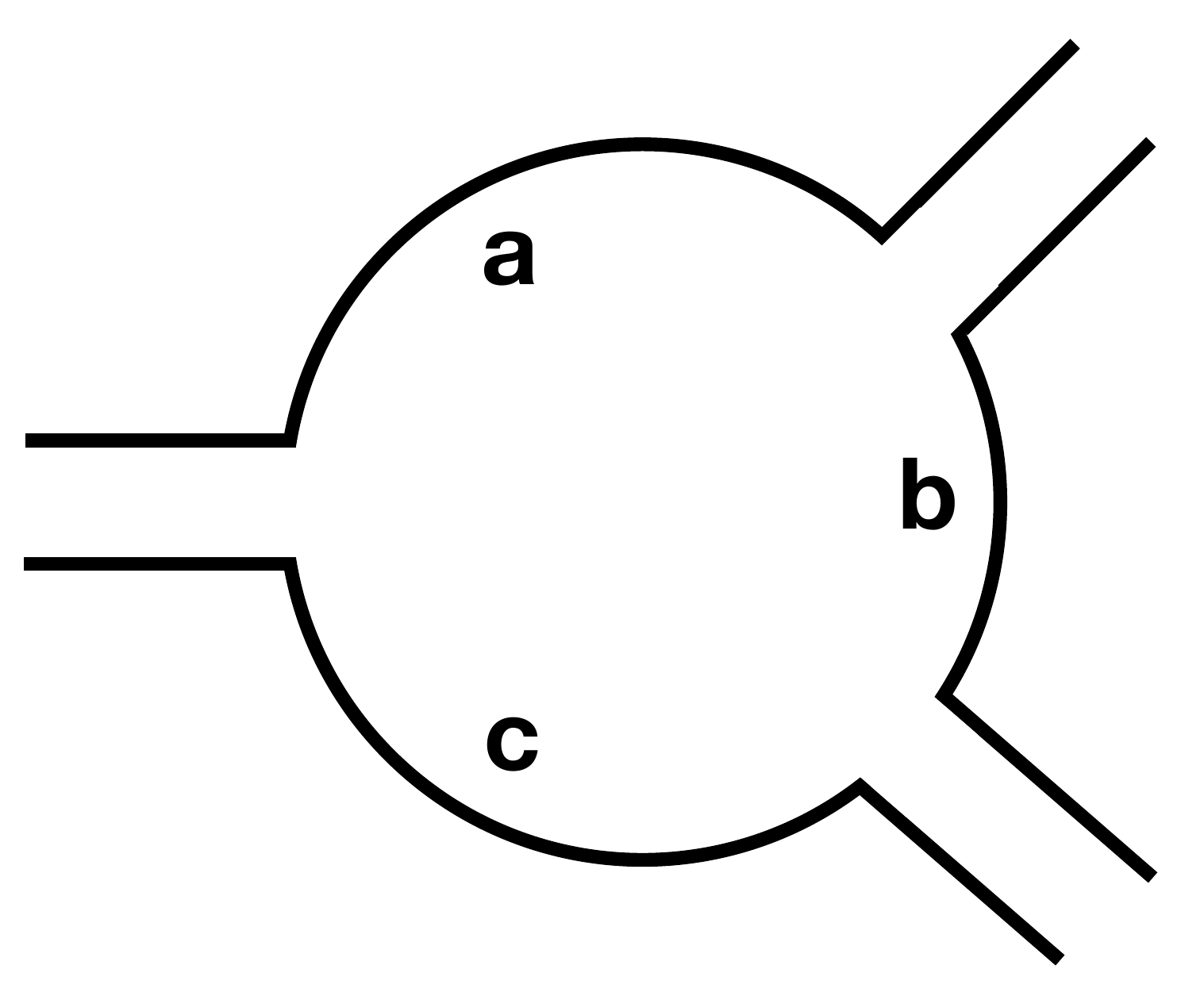}
\caption{Three-point open string coupling diagram. \label{ThreePoint}}
\end{center}
\end{wrapfigure}
Given this rule for three-point couplings, we can now compute the required Higgs representation for quarks by tensoring the quark doublet with each quark singlet, $u$ or $d$. In classes $(iii)$ and $(iv)$ we see immediately that respectively the up and down Yukawa couplings do not exist perturbatively,
because the quark singlets are realised as anti-symmetric tensors, so we must rely on three vectors of $U(3)$ coupling to a singlet. This is fine in $SU(3)$, but not possible in $U(3)$, as first pointed out in \cite{Blumenhagen:2001te}. One may try to generate the missing Yukawa couplings non-perturbatively using instantons \cite{Blumenhagen:2006xt,Ibanez:2006da,Florea:2006si,Haack:2006cy,Blumenhagen:2009qh}, or turn to F-theory models  where they arise more naturally, see Section \ref{FtheoryYukawas}.

In classes $(i)$ and $(ii)$ a perturbative Yukawa coupling is possible, if one assigns the Higgs to one of the multiplets denoted $L$ and $L'$ in Figure \ref{FourClasses}. Then there are a few more issues to worry about: weak brane and Higgs brane anomaly cancellation, and differences in weak $U(2)$ chirality for different families. In class $(i)$ this is most easily dealt with by choosing the Madrid configuration (see Figure \ref{MadridQuiver}) with $USp(2)$ as the weak brane group, 
and choosing brane $c$ as the Higgs brane. Now one can choose $(0,\V,\V,0)+(0,\V,\Vc,0)$ as the Higgs system. In class $(ii)$ the fully orientable configuration discussed above already comes with a Higgs pair per family, but one of the Higgses has the same quantum  numbers as the lepton doublet, which allows for undesirable couplings.

The discussion of lepton Yukawas goes along the same lines for the latter two models. Indeed, in the models of Figure \ref{MadridQuiver} quarks and leptons play a symmetric role. This is not true in the fully orientable model, which lacks a right-handed neutrino. However, it is not likely that neutrino masses are generated by the Standard Model Higgs mechanism. It is usually assumed that there is a Majorana mass component with a different, and necessarily non-perturbative,  origin, in order to understand the smallness of neutrino masses using the see-saw mechanism \cite{Blumenhagen:2006xt,Ibanez:2006da}.  This, as well as many other aspects of Yukawa couplings in orientifold models, is  beyond the scope of this chapter.

\section{Type II Orientifolds}
\label{s:typeII}

In this section we describe D-brane model building in specific Type II compactifications. Our framework will be Calabi--Yau (CY) three-fold orientifold compactifications,\footnote{Our discussion also applies to compactification backgrounds beyond Calabi--Yau metrics, like six-dimensional manifolds with $SU(3)$ or $SU(3) \times SU(3)$ structure, that feature a non-trivial warp factor and internal fluxes. We will however restrict ourselves to the CY case for simplicity. }  at large volume and weak string coupling. This regime is where most of the model building ideas have been developed in string theory,  because D-branes can be essentially treated as sub-manifolds in a compactification manifold $X_6$, and as a result most of the quantities that specify the resulting 4d EFT have a simple topological or geometric realisation. A large fraction of the intuition developed in this setup also applies to small-volume and strong-coupling compactifications, which will be dealt with in sections \ref{s:RCFT} and \ref{sec_F-theory}, respectively, and where also new model building features will arise.

\subsection{Type IIA Orientifolds}

Let us consider type IIA string theory on a background of the form $X_4 \times X_6$, where $X_6$ is a compact Calabi--Yau three-fold $X_6$, with K\"ahler two-form $J$ and holomorphic three-form $\Omega_3$. To this background we apply an orientifold quotient generated by $\Omega (-1)^{F_L}{\cal R}$, where $\Omega$ is the worldsheet parity reversal operator, ${F_L}$ is the space-time fermion number for the left-movers and ${\cal R}$ an anti-holomorphic involution of $X_6$ acting as ${\cal R} J=-J$, ${\cal R}\Omega_3 = - \overline{\Omega}_3$, respectively. The presence of $(-1)^{F_L}$ is important for the orientifold action to square to the identity. Performing the quotient has two main effects:

\begin{itemize}

\item[-] It reduces the supersymmetry on the gravity sector from 4d ${\cal N} =2$ to ${\cal N} =1$.

\item[-] It introduces a set of O-planes at the fixed loci of ${\cal R}$ which, as explained in Section \ref{s:general}, are necessary ingredients for D-brane model building. 

\end{itemize}

Since ${\cal R}$ is an anti-holomorphic involution, its fixed loci are given by a set of special Lagrangian three-cycles of $X_6$ that we collectively denote as $\Pi_{\rm O6}$, times the four non-compact dimensions $X_4$. Thus, we have a set of O6-planes on $X_4 \times \Pi_{\rm O6}$. In order to implement the strategy of Section \ref{sec_Generalities}, to this background we add a set of space-time filling  D-branes that give rise to the SM spectrum plus a hidden sector, and such that the RR tadpole cancellation conditions \eqref{tadpoleO} with $p=6$ are satisfied. The most natural option to build a vacuum is to consider D-branes that preserve the same supersymmetry as the bulk, because these are BPS objects that minimise their tension with respect to their RR charge, and then the cancellation of RR tadpoles implies that the bulk equations of motion are satisfied. If we focus on single D-branes, there are two types of objects that satisfy this condition. The first are D6-branes wrapped on a special Lagrangian three-cycle $\Pi_3$, satisfying \cite{Marino:1999af}
\begin{equation}
{\cal F} + i J = 0, \qquad \text{and} \qquad \Im \, \Omega_3 = 0 ,
\label{BPSD6}
\end{equation}
where 
\begin{equation} \label{calFdef}
{\cal F} = B + \frac{\ell_s^2}{2\pi} F
\end{equation}
is the gauge-invariant D-brane worldvolume field strength, with $\ell_s = 2\pi \sqrt{\alpha'}$ the string length. In all these expressions, bulk $p$-forms like $B$, $J$ and $\Omega_3$ are implicitly pulled-back to the D-brane worldvolume, in this case to the three-cycle $\Pi_3$. That is, the D6-brane wraps a special Lagrangian three-cycle with the same calibration phase as $\Pi_{\rm O6}$, along which it hosts a flat gauge bundle. These two conditions are referred to as F-flatness and D-flatness conditions in the literature, because if they are not met there will be an uncancelled D-term or F-term in the 4d EFT gauge sector, respectively. The second kind of object  are D8-branes wrapped on coisotropic five-cycles $\Xi$ of $X_6$ \cite{Kapustin:2001ij}, characterised by the BPS conditions 
\begin{equation}
\left({\cal F} + i J\right)^2 = 0, \qquad \text{and} \qquad \Im \, \Omega_3 \wedge {\cal F} = 0 ,
\end{equation}
which can again be interpreted as F-flatness and D-flatness conditions. While MSSM-like models have been built with coisotropic D8-branes \cite{Font:2006na} in the context of general CY geometries it is technically difficult to describe these objects, and so in the type IIA setting most of the activity has focused on building models based on D6-branes \cite{Blumenhagen:2005mu}. In the following we  describe the main features of such models.

\subsubsection{Intersecting D6-brane Models}

Type IIA orientifold models were one of the last frameworks to be explored in the D-brane model building literature, but they quickly gained a central place in our current description of this topic. The reason is that the formula for the chiral index $I_{ab}$ that appears in Table \ref{specori} is particularly simple, which gives us a lot of intuition about the chiral spectrum of these models. In particular, given $N_a$ D6-branes wrapping a three-cycle $\Pi_3^a \subset X_6$ and a second stack of $N_b$ D6-branes on $\Pi_3^b$, their chiral index reads
\begin{equation}
I_{ab} = [\Pi_3^a] \cdot [\Pi_3^b],
\label{inter}
\end{equation}
that is, the signed intersection number of the two three-cycles. More precisely, at each transverse intersection one finds a left-handed 4d ${\cal N} =1$ chiral multiplet in a bifundamental representation, which is either $(N_a, \bar{N}_b)$ or $(\bar{N}_a,N_b)$ depending on the sign of the intersection, and \eqref{inter} computes the net chiral spectrum in this sector. As anticipated, this is a topological invariant that only depends on the homology class of each three-cycle, or in other words of the RR charges of the branes. Finally, due to the orientifold symmetry, for each D6-brane stack wrapping $\Pi_3^a$ there is a similar number of D6-branes wrapping the orientifold image
\begin{equation}
\Pi_3^{a'} = {\cal R} \Pi_3^a .
\label{image}
\end{equation}
Including these three-cycles and their intersections and identifying them properly under the orientifold action one arrives at the spectrum of Table \ref{specori}. In particular, at transverse intersections between $\Pi_3^a$ and $\Pi_3^{a'}$ that are not on top of $\Pi_{\rm O6}$ we get adjoint ${\bf Adj} = {\bf S} + {\bf A}$ representations of $U(N_a)$, while for those on top of $\Pi_{\rm O6}$ one either gets a symmetric or an anti-symmetric representation. 

In this setup we can also specify those stacks of $N$ D6-branes that realise either an $\OGroup(N)$ or $USp(N)$ gauge group. As mentioned in Section \ref{s:general}, these are D-brane sectors that are left invariant under the orientifold action. More precisely they are realised by D6-branes on 3-cycles satisfying the property $\Pi_3^a = \Pi_3^{a'}$. There are essentially two kinds of such three-cycles: those that are left invariant point-wise and those that are only invariant as a set. Typically, the first kind gives rise to the real gauge group $\OGroup(N)$ and the second one, which requires an even number of D6-branes, to $USp(N)$ \cite{Marcus:1982fr}. 

The last piece of data needed to realise the content of Table \ref{specori} are those $U(1)$ factors that remain massless after the generalised Green--Schwarz mechanism and in particular all the $B\wedge F$ couplings have been taken into account. In the absence of an orientifold projection, these are the combinations $\sum_a c_a U(1)_a$ with $c_a \in \mathbb{Z}$ and such that the homology class $\sum_a c_a N_a [\Pi_3^a]$ is trivial in $H_3(X_3, \mathbb{Z})$. In the presence of O6-planes some of the RR fields mediating the Green--Schwarz mechanism are projected out, and only the weaker condition 
\begin{equation}
\sum_a c_a N_a \left([\Pi_3^a] - [\Pi_3^{a'}]\right) = 0
\label{masslessU1}
\end{equation}
needs to be imposed \cite{Camara:2011jg}. Those combinations that do not satisfy \eqref{masslessU1} acquire a mass via a St\"uckelberg mechanism, but they remain as perturbative global symmetries that are only broken by non-perturbative effects.  As discussed in Section \ref{ss:yukis}, they constrain the magnitude of those couplings that are not invariant under them, like certain Yukawa couplings, that can only be generated non-perturbatively \cite{Blumenhagen:2006xt,Ibanez:2006da,Florea:2006si,Haack:2006cy,Blumenhagen:2009qh}. Finally, if there are combinations of the form  \eqref{masslessU1} with g.c.d$\{c_a\} =1$ which are $2k$ multiples of some non-trivial element of $H_3(X_3, \mathbb{Z})$, the corresponding massive $U(1)$ contains a $\mathbb{Z}_k$ subgroup that is an exact gauge symmetry.  This will prevent the appearance of certain couplings even at the non-perturbative level \cite{Berasaluce-Gonzalez:2011gos}. 

With these ingredients one may already start discussing explicit examples of intersecting D6-brane models that realise the model building philosophy of Section \ref{sec_Generalities}. In general, the topological data that one needs are the lattice $H_3(X_6, \mathbbm{Z})$, the action of the involution ${\cal R}$ on it, the class $[\Pi_{\rm O6}]$ as well as the intersection product \eqref{inter}. Further data that are important for D6-brane model building are those classes $[\Pi] \in H_3(X_6, \mathbbm{Z})$ with $\Im \int_{\Pi} \Omega_3  = 0$ that contain special Lagrangian representatives. Determining them is  the hardest part of the problem, and hence a large fraction of type IIA orientifold model building is performed in simple geometries like toroidal orbifolds. 

\subsubsection{A Simple Model}

Let us illustrate the general strategy of Section \ref{sec_Generalities} in a simple MSSM-like model. We focus on building a Class $(i)$  model in the classification of Section \ref{SimplestExamples}, following \cite{Marchesano:2004yq,Marchesano:2004xz}. The first step is to specify the four sets of D6-branes that host the MSSM-like spectrum, which we do as in Table \ref{guay}. We consider two stacks of D6-branes ($b$ and $c$) invariant under the orientifold projection and such that the gauge group for each of them is $USp(2) \simeq SU(2)$. For the remaining two stacks ($a$ and $d$) we choose $\Pi_3^a$ and $\Pi_3^d$ to lie in the same homology class, up to a torsion element in $H_3(X_6, \mathbb{Z})$. 
\begin{table}[htb]
\begin{center}
\begin{tabular}{lccccc}
\hline
D6-brane content & & $3 \Pi_3^a$ & $\Pi_3^b$ & $\Pi_3^c$ & $\Pi_{d}$\\
Gauge group & & $SU(3) \times U(1)_a$ & $USp(2)$ & $USp(2)$ & $U(1)_d$ \\
\hline
\end{tabular}
\caption{\small Left-right model of intersecting D6-branes.  Here $[\Pi_3^a] - [\Pi_3^d] \in {\rm Tor} H_3(X_6, \mathbb{Z})$.}
\label{guay}
\end{center}
\end{table}
This implies that they will have the same intersection number with any other three-cycle, and that $U(1)_{B-L} = \frac{1}{3}U(1)_a - U(1)_d$  will remain massless. With these choices we only need to specify four intersection numbers, $I_{ab} = - I_{ac} = 3$ and $I_{aa'} = I_{aO} =0$, in order to reproduce the model of Figure \ref{MadridQuiver}, or more precisely a left-right extension of the SM gauge group. If in addition $\Pi_3^a = \Pi_3^d$, one arrives at a Pati--Salam model, whose spectrum is specified in the upper part of Table \ref{Yspectrum}. This case is particularly simple to realise in a concrete CY geometry because three-cycles hosting $USp$ gauge groups automatically satisfy the supersymmetry condition \eqref{BPSD6}, and so it only remains to verify that the cycles $\Pi_3^a = \Pi_3^d$ also preserve supersymmetry at some point in the CY moduli space. 

To proceed with the construction of the model one must specify the CY geometry. As noted, the simplest choices correspond to toroidal orbifolds and, in this case, to the orbifold $X_6 = (T^2 \times T^2 \times T^2) / \mathbb{Z}_2 \times \mathbb{Z}_2$  with cohomology $(h^{1,1},h^{2,1}) = (51,3)$, whose O6-plane quotient and D6-brane model building rules were worked out in \cite{Cvetic:2001tj,Cvetic:2001nr}. The simplicity of this geometry stems from the fact that the twisted sector only contains collapsed two-cycles, and so $H_3(X_6, \mathbbm{Z})$ and its intersection product is essentially that of $T^2 \times T^2 \times T^2$. To specify a three-cycle class one must provide a one-cycle class on each $T^2$ factor, that is specify the following wrapping numbers
\begin{equation}
[\Pi_3^a] = (n_a^1, m_a^1) \times (n_a^2, m_a^2) \times (n_a^3, m_a^3), \qquad n_a^i, m_a^i \in \mathbbm{Z} ,
\label{3cycleT6}
\end{equation}
and then the intersection number between two three-cycles is given by
\begin{equation}
I_{ab} = [\Pi_3^a]\cdot[\Pi_3^b]\, =\, (n_a^1m_b^1-m_a^1n_b^1) \times (n_a^2m_b^2-m_a^2n_b^2) \times (n_a^3m_b^3-m_a^3n_b^3).
\label{interT6}
\end{equation}
In a toroidal orientifold geometry such three-cycles must be accompanied by their images under the orientifold group. In the case at hand the $ \mathbb{Z}_2 \times \mathbb{Z}_2$ orbifold group the wrapping numbers \eqref{3cycleT6} are mapped to themselves, and the smallest or fractional D-brane objects correspond to two copies of \eqref{3cycleT6} in certain locations. The orientifold image is dictated by the anti-holomorphic involution ${\cal R} : z^i \to \bar{z}^i$, where $z^i$ is the complex coordinate on the $i^{\rm th}$ $T^2$. That this involution is a symmetry of the internal metric restricts the moduli space of complex structures, with one possibility being that each $T^2$ is rectangular. In that case, which we will assume in the following, the orientifold image of \eqref{3cycleT6} is
\begin{equation}
[\Pi_3^{a'}] = (n_a^1, -m_a^1) \times (n_a^2, -m_a^2) \times (n_a^3, -m_a^3), 
\label{3cycleT6m}
\end{equation}
and the O6-plane class is given by
\begin{align}
[\Pi_{\rm O6}] = & 4 \Big[ (1,0)  \times(1,0)  \times(1,0) +  (1,0) \times (0,1)  \times(0,-1) \\ & \quad +   (0,1)  \times (1,0)  \times(0,-1) +   (0,1)  \times(0,-1)  \times(1,0) \Big] ,
\label{3cycleori}
\end{align}
in fractional three-cycle units. A stack of D6-branes with the wrapping numbers of one of the components of \eqref{3cycleori} hosts a $USp$ gauge group. Because such wrapping numbers are invariant under the full orientifold group, one only needs two D6-branes in the covering space to host a $USp (2)$ gauge group.

With these ingredients one may already build an explicit model, by providing the wrapping number content of Table \ref{Ymodel}.
\begin{table}[htb]
\begin{center}
\begin{tabular}{ccccc}
$N_\alpha$  & & $(n_\alpha^{1},m_\alpha^{1})$  &  $(n_\alpha^{2},m_\alpha^{2})$
&  $(n_\alpha^{3},m_\alpha^{3})$ \\
\hline
\hline $N_a = 3 + 1$ & & $(1,0)$ & $(3,1)$ & $(3,-1)$  \\
$N_b=1$ & & $(0,1)$ &  $ (1,0)$  & $(0,-1)$ \\
$N_c=1$ & & $(0,1)$ &  $(0,-1)$  & $(1,0)$  \\
\hline 
$N_{h_1}= 1$ & & $(-2,1)$  & $(-3,1)$ & $(-4,1)$ \\
$N_{h_2}= 1$ & & $(-2,1)$ & $(-4,1)$ & $(-3,1)$ \\
$N_f = 20$ & & $(1,0)$ &  $(1,0)$  & $(1,0)$  \\
\end{tabular}
\end{center}
\caption{Explicit $(T^2 \times T^2 \times T^2)/\mathbb{Z}_2 \times \mathbb{Z}_2$ orientifold model  realising the MSSM-like model of Table \ref{guay}. D6-brane multiplicities are given in fractional three-cycle units. \label{Ymodel}}
\end{table}
Notice that the upper part of the table realises the intersection numbers needed for the Pati--Salam spectrum, and one finds in addition a minimal Higgs sector with a non-vanishing $\mu$-term, see Table \ref{Yspectrum}. The lower part of table  \ref{Ymodel} is the one needed to satisfy the RR tadpole cancellation conditions \eqref{tadpoleO}.  In principle this sector introduces an additional gauge group $U(1)' \times USp(40)$, with $U(1)' = \frac{1}{3}U(1)_a + 2[U(1)_{h_1} - U(1)_{h_2}]$, that we would like to treat as a hidden sector of the theory. However, the $U(1)$ branes $h_1$ and $h_2$ intersect with those of the Pati--Salam sector, generating chiral exotics that naively prevent us from doing so, see  \cite{Marchesano:2004yq,Marchesano:2004xz} for the resulting spectrum.

Nevertheless, one of the main advantages of the model building strategy outlined in Section \ref{sec_Generalities} is that this kind of chiral exotics is easily avoidable. Indeed, because the particle content of the Pati--Salam sector cancels all cubic $SU(N)$ anomalies by itself (including those for $N=2$ and $N=1$, which in this particular case are trivial) the chiral exotics that arise from additional D-brane sectors appear as vector-like from the viewpoint of the visible sector gauge group. In practice, this means that there is some direction in moduli space that allows us to get rid of such chiral exotics, as it is the case in this example. More precisely, by moving in the complex structure moduli space of the the first $T^2$ one can induce a tachyon in the bifundamental of $U(1)_{h_1} \times U(1)_{h_2}$ via a D-term potential, see below. Condensation of this tachyon corresponds to the D-brane recombination process $h_1 + h_2' \to h$, which does not affect the low energy gauge group, but greatly simplifies the chiral spectrum. \begin{table}[htb]
\begin{center}
\begin{tabular}{ccc}
Sector  &  Matter & $SU(4) \times SU(2) \times SU(2) \times U(1)' \times USp(40)$  \\
\hline
\hline 
(ab) & $F_L$ & 3(4,2,1)$_{1/3}$  \\
(ac) & $F_R$ & 3(4,1,2)$_{-1/3}$  \\
(bc) & $H$ &  (1,2,2)$_0$ \\
\hline 
(bh) & & 2(1,2,1)$_2$ \\
(ch) & & 2(1,1,2)$_{-2}$  \\
\end{tabular}
\end{center}
\caption{Spectrum of the model of Table \ref{Ymodel} after the D6-brane recombination $h_1 + h_2' \to h$, with the subindex indicating the charge under $U(1)' = \frac{1}{3} U(1)_a + 2U(1)_h$. There is no chiral matter under the gauge group $USp(40)$, which is a hidden sector. \label{Yspectrum}}
\end{table}
The final result is given in Table \ref{Yspectrum}, where one can see that the lower part of the spectrum is not chiral under the Pati--Salam gauge group. Given the simplicity of this construction this is quite an encouraging result, which one may hope to improve by exploring more general setups. Indeed, applying the same approach to more involved toroidal orbifolds, general Calabi--Yau manifolds and RCFT models yields even more realistic models.

\subsubsection{Further EFT Features}

Besides a reasonable chiral spectrum, a realistic model must display a set of couplings and a non-chiral light spectrum that are compatible with the MSSM or extensions thereof. In the sequel we outline the computation of these additional features in the context of intersecting D6-brane models. 

The simplest quantity to consider is the gauge kinetic function associated with each stack of D6-brane. It reads
\begin{equation}
f_{a}^{\rm D6} = \frac{1}{2\pi\ell_s^3} \int_{\Pi_3^a} e^{-\phi} \Re\, \Omega_3 + i C_3 ,
\end{equation}
from where one can compute the gauge couplings of the non-abelian and abelian gauge factors. In the latter case there will generically be a kinetic mixing with bulk $U(1)$ gauge symmetries, if present \cite{Marchesano:2014bia}. 

One may also consider the presence of light or massless non-chiral particles in the D-brane sector, such as  D-brane moduli that appear as ${\cal N}=1$ multiplets in the adjoint. A stack of $N$ D6-branes wrapping a special Lagrangian three-cycle $\Pi_3$ has $b_1(\Pi)$ deformations of its embedding that preserve the special Lagrangian condition, as it follows from McLean's theorem. These are complexified by the same amount of Wilson lines, giving rise to $b_1(\Pi)$ ${\cal N} =1$ chiral multiplets in the adjoint representation of $SU(N)$. Aiming to build models without adjoints leads us to either consider three-cycles with $b_1(\Pi)=0$ or D6-branes with deformations that are fixed by an F-term potential. There are two sources for the latter. The first source are elements of $H_1(\Pi, \mathbb Z)$ dual to two-cycles of $\Pi_3$ that are non-trivial in $H_2(X_6,\mathbb{Z})$ \cite{Marchesano:2014iea}. The second one is the superpotential generated by disc worldsheet instantons ending on one-cycles of $\Pi_3$ \cite{Kachru:2000ih}. While this second source in principle affects all adjoint fields, it is generically expected to give them exponentially suppressed masses in the large volume regime in which we are working. In general, D-brane adjoints  redefine the 4d holomorphic variables that arise from bulk moduli upon dimensional reduction. In this case D6-brane moduli redefine the CY complex structure moduli see e.g. \cite{Grimm:2011dx,Kerstan:2011dy,Carta:2016ynn}, and their mass spectrum should be treated in the broader framework of moduli stabilisation. Such complex structure moduli enter the D-flatness condition for D6-branes, and can induce Fayet-Iliopoulos terms which  break supersymmetry and may trigger D6-brane recombination \cite{Kachru:1999vj}.

Besides adjoint masses, worldsheet instantons with the topology of a disc are a source for mass terms for vector-like pairs that arise from the transverse intersections of a pair of D6-branes, as illustrated in Figure \ref{ws}, as well as for Yukawa couplings \cite{Aldazabal:2000cn}. 
\begin{figure}[htb]
\includegraphics[width=70mm,height=40mm]{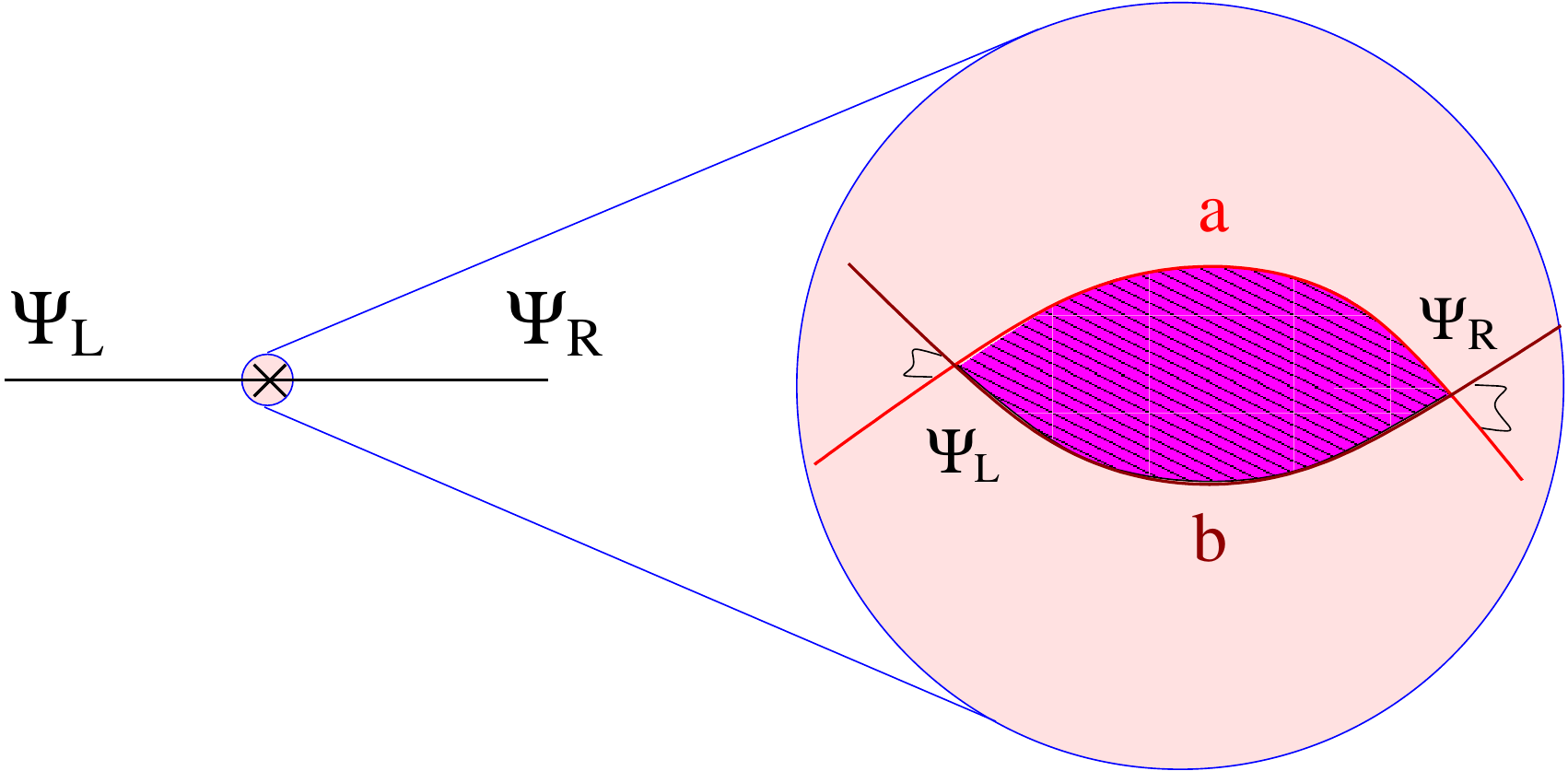}~a)
\hfil
\includegraphics[width=70mm,height=40mm]{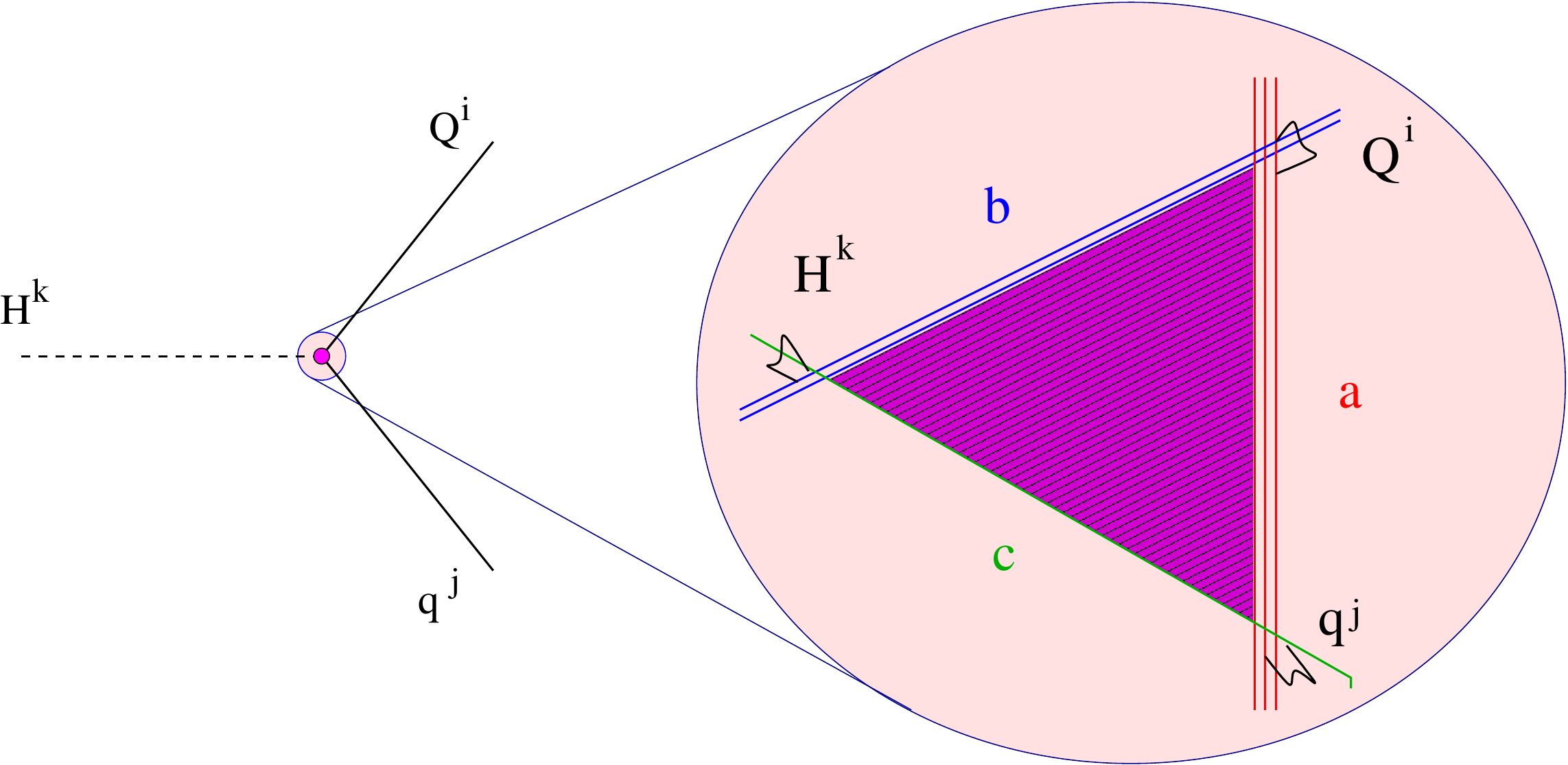}~b)
\caption{Worldsheet instantons as generators of {\em a)} masses for vector-like pairs and {\em b)} Yukawa couplings. Figures taken from \cite{Marchesano:2007de}.}
\label{ws}
\end{figure}
These two quantities have a direct interpretation in terms of Kontsevich's homological mirror symmetry conjecture, and their computation is a rich mathematical subject of research.  While they are difficult to compute in general, in simple examples like toroidal orbifold models one can perform the computation quite explicitly, showing for instance that the Pati--Salam model described above leads to Yukawa couplings of rank one \cite{Cremades:2003qj}. 

What worldsheet instantons cannot generate are couplings that are forbidden by D-brane $U(1)$ symmetries that become massive due to a St\"uckelberg mechanism. In that case, couplings should be generated by D-brane instantons, which in this case are D2-brane instantons wrapping special Lagrangian three-cycles of $X_6$ \cite{Blumenhagen:2009qh}. In typical models such couplings include right-handed neutrino masses and Yukawas forbidden by global $U(1)$ symmetries, as described in Section \ref{sec_Generalities}. Moreover, if any of these couplings is forbidden by the discrete gauge symmetries that are remnants of the massive $U(1)$s they will not be generated even at the non-perturbative level. The model-building challenge then resides in using this structure to forbid unwanted couplings and obtain those that are necessary phenomenologically, with the appropriate magnitude given by their suppression factors.

\subsection{Type IIB Orientifolds} \label{subsec_IIBorientifolds}

We now turn to Type IIB string theory on the 10d background $X_4 \times X_6$, where again $X_6$ is taken to be a compact Calabi--Yau three-fold $X_6$, with K\"ahler two-form $J$ and holomorphic three-form $\Omega_3$. There are two different kinds of orientifold projections which are compatible with a large compactification volume:
\begin{align}\nonumber
{\rm O3/O7 \, projection}: & \quad \Omega  (-1)^{F_L}{\cal R} \quad \text{such that} \quad {\cal R} J = J, \quad {\cal R}\Omega_3 = -\Omega_3 , \\
{\rm O5/O9  \, projection}: & \quad \Omega  {\cal R} \qquad \qquad \text{such that} \quad {\cal R} J = J, \quad {\cal R}\Omega_3 = \Omega_3 ,
\nonumber
\end{align} 
where ${\cal R}$ is now a holomorphic involution of $X_6$. As in the Type IIA case, this projection reduces the bulk supersymmetry from 4d ${\cal N} =2$ to ${\cal N} =1$ and introduces a series of O-planes at the fixed loci of ${\cal R}$. The difference is that these fixed loci are even-dimensional submanifolds of $X_6$. In the first projection they are given by points or holomorphic four-cycles, leading to O3 and/or O7-planes, while the second projection leaves invariant curves or the whole of $X_6$. These holomorphic involutions are much better understood than their anti-holomorphic counterparts, which has resulted in the construction of models in geometries beyond toroidal orbifolds. The same remark applies to the space-time D-branes that host gauge interactions in these models, which now consist of D3, D5, D7 and D9-branes wrapping internal even-dimensional cycles $\Pi_{p-3}$ of $X_6$. The BPS conditions for a single D-brane  of this kind read
\begin{eqnarray}
\label{FIIB}
\Pi_{p-3} \, \text{holomorphic}, \quad {\cal F}^{(2,0)} =  0, & \quad & \text{F-flatness} , \\
\label{DIIB}
\Im\, e^{-i\theta} e^{{\cal F} + i J} {\sqrt{\hat{A}_{\Pi_{p-3}}}}   =  0, & \quad & \text{D-flatness} ,
\end{eqnarray}
where $\theta = 0$ for the O3/O7-projection and $\theta=\pi/2$ in the O5/O9 projection, and $\hat{A}$ is the A-roof genus of the tangent bundle of $\Pi_{p-3}$, encoding part of the D-brane curvature couplings. As a non-trivial worldvolume field-strength is allowed, these objects must typically be understood as  D3/D5/D7/D9 bound states \cite{Douglas:1995bn} or, from a more mathematical perspective, as coherent sheaves \cite{Douglas:2000gi,Aspinwall:2004jr}. These abelian BPS conditions can be generalised to non-abelian D-brane configurations that are allowed for stacks of several D-branes, featuring a non-abelian field strength ${\cal F}$ and/or a non-abelian D-brane worldvolume embedding. 

Non-abelian D-brane configurations have mostly played a role in models with D9-branes, due to their analogy with heterotic compactifications. Due to this similarity, they will not be discussed here, nor will be models with O5/O9 projection. Instead, we will focus on models with O3/O7 projection, which display a set of features that are very representative of D-brane model building: 

\begin{itemize}

\item[-] The SM gauge group or its extension can be localised in a patch of the compact manifold $X_6$. This simplifies its construction, which can be done in two steps: {\it i)} first building a model in a local patch and {\it ii)} embedding it into a compact manifold, see Figure \ref{bottomsup}. This procedure is known as the bottom-up approach to model building \cite{Aldazabal:2000sa}, and it allows one to distinguish between those physical features that are only sensitive to local data and those that depend on global data of the construction. 

\item[-] Due to this localisation, and the fact that 4d gauge couplings are controlled by the internal volume of D-branes, these models naturally realise the idea of gauge coupling unification. They in addition permit to decouple the strength of gauge and gravitational interactions, even to the extent to implement  the  large extra dimension scenario \cite{Arkani-Hamed:1998jmv,Antoniadis:1998ig}.

\item[-] One can easily combine these models with additional ingredients that implement mechanisms for moduli stabilisation at large volume, like background  fluxes \cite{Douglas:2006es}, in order to build more sophisticated models, see e.g. \cite{Quevedo:2014xia}.

\end{itemize}
 
In the following we will discuss two classes of models that illustrate how these features are realised: D-branes at singularities and intersecting D7-branes. The latter can be thought of as a particular case of the F-theory constructions of Section \ref{sec_F-theory}, which also incorporate these attractive features. 

\begin{figure}[htb]
 \begin{center}
\includegraphics[width=13cm, height=5cm]{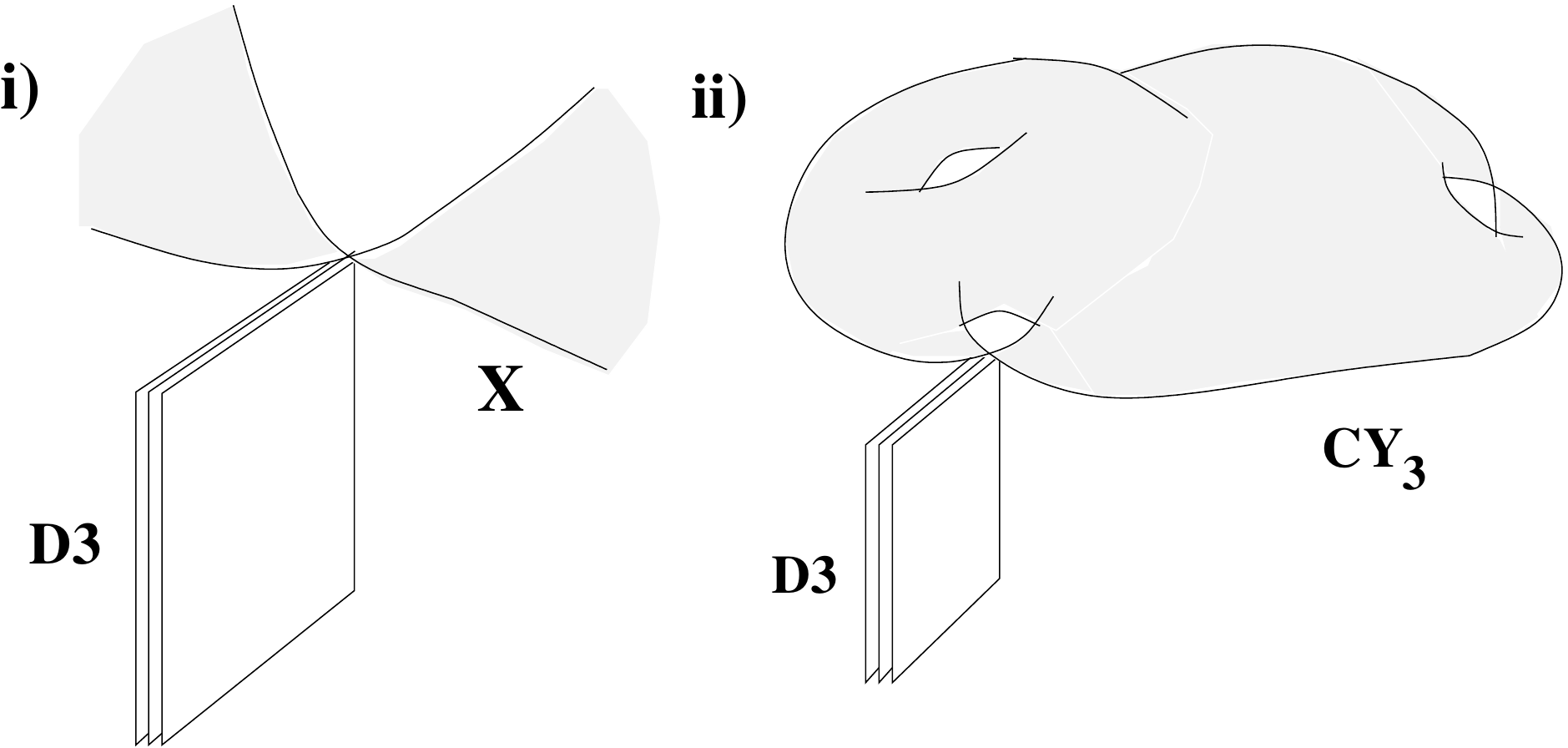}
\caption{Two-step procedure for building $4d$ models based on D3-branes at singularities. Figure taken from \cite{Aldazabal:2000sa}.}
\label{bottomsup}
 \end{center}
\end{figure}

\subsubsection{D-branes at Singularities}

Strictly speaking, D-branes at singularities do not correspond to large volume models that can be treated in the 10d supergravity regime. They are engineered in neighbourhoods of the compact manifold $X_6$ that display a singular geometry, obtained for instance from collapsing some of its cycles. Nevertheless, one can analyse this system by directly quantising open strings in a such a singular geometry, following the techniques initiated in \cite{Douglas:1996sw,Douglas:1997de}, and then embed the resulting gauge sector into a fully-fledged compactification, along the lines of the bottom-up philosophy described above. 

The simplest example of this class of models is given by D3-branes at orbifold singularities. A stack of $N$ D3-branes in this flat space yields a 4d ${\cal N} =4$ $U(N)$ gauge theory and therefore a non-chiral gauge sector. The non-chiral nature of the gauge sector persists if the D3-brane is placed at any smooth point of a compactification manifold $X_6$, since the effect of curvature and background fluxes can at best lead to a massive deformation of 4d ${\cal N} =4$ SYM. The only way to obtain a chiral spectrum is to place the D3-brane on top of a singular geometry, like the one obtained by an orbifold quotient of the form $\mathbbm{C}^3/\Gamma$ with fixed points. Let us for simplicity consider the cyclic orbifold group $\Gamma = \mathbbm{Z}_k$ generated by an action on $\mathbbm{C}^3$ of the form
\begin{equation}
(z^1, z^2, z^3) \mapsto \left( e^{2\pi i \frac{a_1}{k}} z^1, e^{2\pi i \frac{a_2}{k}} z^2, e^{2\pi i \frac{a_3}{k}} z^3\right) , \qquad a_i \in \mathbb{Z} .
\label{Zkorb}
\end{equation}
In order for spinors to be well-defined in this background one needs to require that $\sum_i a_i \in 2\mathbb{Z}$. Then one can embed $\Gamma$ into $SU(4)$ as ${\rm diag} (e^{2\pi i \frac{b_0}{k}}, e^{2\pi i \frac{b_1}{k}}, e^{2\pi i \frac{b_2}{k}},e^{2\pi i \frac{b_3}{k}})$ with $b_i \in \mathbb{Z}$ and $\sum_i b_i = 0 \mod k$, such that $a_1 = b_2 + b_3$, $a_2 = b_3+ b_1$ and $a_3 = b_1 + b_2$, and quantise closed strings in this background \cite{Dixon:1985jw}. To embed this singularity into a CY geometry one must impose local $SU(3)$ holonomy, which amounts to $a_1 + a_2 + a_3 = 0 \mod k$. Then one can assume $b_0 = 0$ and $b_i = -a_i$. 

Placing a stack of $N$ D3-branes at the fixed point of \eqref{Zkorb} yields at gauge sector that is an orbifold projection of the initial 4d ${\cal N} =4$ $U(N)$ gauge theory. The result depends on how the orbifold generator acts on the D3-brane Chan-Paton degrees of freedom, which is specified by an element of $U(N)$ of the form
\begin{equation}
\gamma =  {\rm diag}\, \left( \mathbbm{1}_{N_0}, \omega \mathbbm{1}_{N_1}, \dots , \omega^{k-1} \mathbbm{1}_{N_{k-1}} \right) , \qquad \omega = e^{\frac{2\pi i}{k}} ,  
\label{gammaD3}
\end{equation}
with $\sum_{a=0}^{k-1} N_a = N$. The 4d ${\cal N} =4$ vector multiplet gets projected out to those Chan-Paton degrees of freedom $\lambda$ invariant under the adjoint action $\lambda \mapsto \gamma \lambda \gamma^{-1}$, while for the three adjoint chiral multiplets $\Phi^i$ only the modes invariant under  $\lambda \mapsto  e^{-2\pi i \frac{a_i}{k}} \gamma \lambda \gamma^{-1}$ survive. This results in the following spectrum:
\begin{align}\label{gaugesingu}
\text{Vector multiplet}: & \quad  \prod_{a=0}^{k-1} U(N_a) , \\
\text{Chiral multiplets}: & \quad   \sum_{a=0}^{k-1} \left[(\V_a, \Vc_{a+a_1}) + (\V_a, \Vc_{a+a_2}) +  (\V_a, \Vc_{a+a_3}) \right],
\nonumber
\end{align} 
with a set of Yukawa couplings that arise from truncation of the parent ${\cal N} =4$ superpotential $W = \Tr ( \Phi^1 [\Phi^2,\Phi^3])$. Notice that this is a particular case of the general spectrum of Table \ref{specori}, with the chiral index $I_{ab}$ determined by the orbifold twists $a_i$, and without the presence of those representations that arise due to the orientifold projection. Indeed, while  orientifold planes are a necessary ingredient of the global construction, a  singularity can be located at a point $p \in X_6$ away from any O-plane. Then the orientifold projection simply requires that there is an identical singularity with similar D-brane content located at ${\cal R}p$. 

This chiral spectrum has a limited capacity of family replication, which occurs when two or more orbifold twists $a_i$ are equal mod $k$. This sets an upper bound of three families and makes the orbifold $\mathbbm{Z}_3$ with twist $\{a_i\} = (1,1,-2)$ particularly attractive \cite{Aldazabal:2000sa,Maharana:2012tu}. One may take $N_0 = 3$, $N_1=2$ and $N_2=1$, yielding a SM gauge group with hypercharge $U(1)_Y = U(1)_0/3 +  U(1)_1/2 + U(1)_2$ and a partial SM chiral spectrum. The spectrum can then be completed by considering a stack of D7-branes going through the orbifold singularity and with a non-trivial action of the orbifold group on their Chan-Paton degrees of freedom \cite{Aldazabal:2000sa}.

Besides the particularities of each model, there is a series of general features common to all of them that are worth mentioning:

\begin{itemize}

\item[-] These models directly realise gauge coupling unification at the compactification scale because the coupling of all the gauge groups in \eqref{gaugesingu} is given by the 4d dilaton. When introducing D7-branes these may carry their own gauge group, but their gauge couplings are suppressed with respect to the local ones by the volume of the four-cycle wrapped by the D7-brane, and so from the viewpoint of the local model they are treated as flavour branes. In general, the role of gauge groups coming from D7-branes can only be determined upon the global completion of the local model. 

\item[-] There is always an anomaly-free massless $U(1)$ combination given by 
\begin{equation}
U(1)_{\rm diag} = \sum_{a=0}^{k-1} \frac{U(1)_a}{N_a}  ,  
\label{U1diag}
\end{equation}
which plays the role of hypercharge in the $\mathbbm{C}^3/\mathbbm{Z}_3$ model discussed above. Other $U(1)$ symmetries are typically anomalous and acquire a mass of the order of the string scale via a Green--Schwarz mechanism \cite{Ibanez:1998qp}.

\item[-] The gauge and chiral content of these local models can be encoded in a quiver diagram, in which each node represents a gauge group and a set of  arrows connecting them the bifundamental representations. A D3-brane on a given node is dubbed fractional D3-brane, and the set of D3-branes that add up to the regular representation in \eqref{gammaD3} is identified as a bulk D3-brane that can be separated from the singularity and probe the bulk of $X_6$. One can also extend the diagram to incorporate flavour D7-branes and their associated chiral content, see \cite{Uranga:2000ck} for a short review.

\item[-] When choosing \eqref{gammaD3}, or more generally the D-branes in the quiver, one has to observe the local RR tadpole conditions. These form a subset of the whole set of tadpole conditions of the compactification which is only sensitive to fractional D3-brane and D7-brane charges. As usual satisfying these conditions implies that the non-abelian and mixed anomalies cancel \cite{Aldazabal:2000sa}. 

\end{itemize}

Additionally, this simple orbifold setup can be generalised in a number of ways, which give rise to more and more sophisticated models, and which we briefly summarise in the following:

\begin{itemize}

\item[-] One may study more general orbifold groups $\Gamma$ such as $\mathbb{Z}_k \times \mathbb{Z}_m$, or non-abelian subgroups of $SU(3)$ like for instance $\Gamma = \Delta_{27}$, which also features family triplication \cite{Aldazabal:2000sa}.

\item[-] One may consider non-orbifold toric singularities such as conifold or del Pezzo singularities. In this case the gauge theory data are more efficiently encoded in a so-called dimer diagram \cite{Feng:2000mi}, which is a tiling of $T^2$.  While these models provide more flexibility, certain features like the upper bound on three families remain generic \cite{Krippendorf:2010hj}.

\item[-] One may explore orientifolded singularities. In particular certain orientifolded del Pezzo singularities give rise to realistic SM spectra without the need of D7-branes \cite{Cicoli:2021dhg}, using the full spectrum of Table \ref{specori}. Note that this result is in agreement with the general model building philosophy of Section \ref{sec_Generalities}, in the sense that one can build a realistic spectrum that avoids the necessity of additional D-brane sectors by cancelling all anomalies locally. 

\end{itemize}

Finally, it is worth mentioning the efforts to perform systematic embeddings of local models of D-branes at singularities into global compactifications, mostly using toric geometry techniques, see e.g.  \cite{Cascales:2003wn,Balasubramanian:2012wd,Cicoli:2012vw}. 

\subsubsection{Intersecting D7-branes}

Models of D-branes at singularities are a particular class of a broader set, which in the generic regime without collapsed cycles correspond to models of intersecting D7-branes. In this case the basic object is a D7-brane wrapping a holomorphic four-cycle $\Pi_4 \subset X_4$, threaded by a non-trivial worldvolume field strength ${\cal F}$ defined as in \eqref{calFdef}. The full spectrum between two D7-branes wrapping 4-cycles $\Pi_4^a$ and $\Pi_4^b$ must be computed in terms of {\it Ext} groups \cite{Katz:2002gh}, but in order to determine the chiral index of Table \ref{specori} one may use the Riemann-Roch-Hirzebruch theorem to arrive at the expression \cite{Blumenhagen:2008zz}:
\begin{equation} 
I_{ab} = \int_{X_6} [\Pi_4^a] \wedge [\Pi_4^b] \wedge \left(c_1(F_a) - c_1(F_b)\right)  \,.  
\label{chiralIIB}
\end{equation}
Here $[\Pi_4^a]$ is the two-form Poincar\'e dual to the divisor class of $\Pi_4^a$, and $c_1(F_a)$ is the first Chern class of the quantised piece of the worldvolume flux ${\cal F}$ threading $\Pi_4^a$, viewed as an element of $H^2(X_6, \mathbbm{Z})$. This formula extends to orientifold images by using that $(\Pi_4^{a'}, F_{a'}) = ({\cal R} \Pi_4^a, -{\cal R}F_{a})$, and to the orientifold as $I_{aO} =  2 \int_{X_6} [\Pi_4^a] \wedge [\Pi_{\rm O7}] \wedge c_1(F_a)$. With these expressions at hand, the strategy to build models works quite similarly to the Type IIA case, and for instance one may reproduce the Pati--Salam sector of Table \ref{Yspectrum} by using three stacks of intersecting branes \cite{Marchesano:2004yq,Marchesano:2004xz}. 

A natural arena for D7-brane model building is in the context of local models, following the general philosophy outlined above. The gauge group is now localised on a (fluxed) non-trivial four-cycle $\Pi_4$ which can be collapsed by moving in moduli space, and that may host chiral matter either via self-intersection or via the intersection with flavour D7-branes. Contracting such a four-cycle leads to a singularity of the type described above with the wrapped D7-branes becoming fractional D3-branes \cite{Diaconescu:1999dt}. While this shows that the two classes of Type IIB models under discussion are secretly similar to each other, in practice the model building possibilities are quite different. The reason is that the spectrum of BPS D-branes at the singular point and at volumes large compared to the string scale are quite different. In this sense both classes of models should be considered separately.

Still, they have a number of similarities. If the volume of the contractible four-cycle $\Pi_4$ is large enough, one may achieve an approximate gauge coupling unification. This is because the gauge coupling constant associated to a D7-brane gauge group is set by its volume and its worldvolume flux, and in the regime of dilute flux densities the former is the dominant contribution. This suggests GUT model building as an obvious model building option, and in particular $SU(5)$ GUTs, with a {\bf 10} representation coming from intersection with an O7-plane \cite{Blumenhagen:2008zz}. These models suffer from problems similar to the ones mentioned in Section \ref{ss:yukis}, in the sense that the massive $U(1)$ symmetry within $U(5)$ forbids the generation of the top Yukawa coupling ${\bf 10} \, {\bf 10} \, {\bf 5}_H$ at the perturbative level. This motivates applying the same  model building philosophy but in the more general context of F-theory GUTs, to be discussed in Section \ref{sec_F-theory}. Given the similarities between F-theory and D7-brane GUTs, we will refrain from further discussing  the latter, and refer the reader to \cite{Blumenhagen:2008zz,Maharana:2012tu} for details.

\section{Rational Conformal Field Theories}

\label{s:RCFT}

A conformal field theory (CFT) is a field theory with invariance under the conformal group. Here we are only interested  in two-dimensional CFT's living on the worldsheet of strings. The properties (such as spectra and correlation functions) of such a theory can be described in terms of the algebra of conformal field theory, the Virasoro algebra. 
This algebra is characterised by a number $c$ called the central charge. For $c<1$ the representation theory of the Virasoro algebra is well-known \cite{Friedan:1983xq}.  It has a discrete series of unitary representations for certain rational values of $c$, accumulating at $c=1$. For each $c$ in the series there is a finite number of representations, characterised by a conformal weight denoted as $h$. Conformal field theories with a finite number of representations are called rational conformal field theories, or RCFT for short. 

Conformal field theory encompasses all world-sheet descriptions of perturbative string theory, including geometric orbifolds and orientifolds, like the ones discussed in the previous section. But in those cases the geometric language provides the more powerful description. Using RCFTs will allow us to go into uncharted territory not easily accessible geometrically. 

In addition to the Virasoro algebra, generated by currents of spin 2, other algebras may be present. They may be generated by currents of spin 1 (affine Lie algebras, often called Kac-Moody algebras), higher spin (called $W$ algebras), spin-$\tfrac12$ (free fermions) and spin-$\tfrac32$ (superconformal algebras). As a general rule, the extra symmetry makes representations larger, and reduces their number. In particular, the number of representations may become finite, producing an RCFT. The set of generators, acting on either the right- or left-moving modes of the theory, is called the {\it chiral algebra}.

\subsection{Gepner Models}

Compactified closed superstring theories are often described geometrically, as strings propagating in a space with $D$ flat dimensions, and $10-D$ dimensions rolled up on a torus, an orbifold or a Calabi--Yau manifold. But there is an alternative description  in terms of a suitable CFT on the worldsheet. 

To characterise the compactified sector of a superstring with four uncompactified dimensions we need superconformal field theories with a total central charge of 9, and two supersymmetries.  This is because the uncompactified theory is defined by world-sheet fields $X^I$ and $\Psi^I$, $I=1,\ldots, 6$. The free world-sheet theory has ${\cal N}=2$ supersymmetry, and its conformal field theory has a total central charge of $6+3=9$ (each word-sheet boson contributes 1, each fermion $\tfrac12$). The compactified  theory has to mimic these properties to preserve the consistency of the theory. In short, we need a $c=9$, ${\cal N}=2$ superconformal field theory.
 
 The ${\cal N}=2$ superconformal field theories also have a discrete series, this time accumulating at $c=3$. The values of $c$ are
\beq
c=\frac{3k}{k+2} \,.
\eeq
These ${\cal N}=2$ minimal models are superconformal RCFTs, but they do not have the required central charge of 9. This can be solved by ``tensoring" several copies of them, in such a way that the sum of their central charges is 9.
There are 168 solutions $(k_1,\ldots,k_n)$ to the equations
\beq
\sum_i \frac{3k_i}{k_i+2}=9 \,.
\eeq
One needs 4 to 9 copies, for example $(3,8,8,8)$ or $(1,1,1,1,1,1,1,1,1)$. This method was first used by D. Gepner \cite{Gepner:1987vz} to construct compactified heterotic strings, and hence these tensor products are often referred to as Gepner models. Here we are using them to describe the compactified sector of a Type II superstring. 

Gepner models are by no means the only way to construct superconformal RCFTs with $c=9$, but they are the only ones that are both non-trivial and mathematically under control, so that relevant quantities are computable. Therefore they present  an excellent theoretical laboratory  for investigating both closed and open string models. 

\subsection{Modular Invariant Partition Functions}\label{MIPF}

After choosing an RCFT, a second important choice to be made is that of the modular invariant partition function (MIPF), which enters the discussion 
when the world-sheet diagram is a torus. In closed string theories this diagram encodes the closed
string spectrum, in terms of a partition function 
\begin{equation}\label{PartF}
\begin{split}
 \int {\cal D} X e^{-S_E(X)}&={\rm Tr}\ e^{2\pi i \tau (L_0-1)}e^{-2\pi i \bar\tau (\bar L_0-1)} \,,
\end{split}
 \end{equation}
 where the left-hand side denotes -- symbolically -- the path integral. 
The complex parameter $\tau$ denotes the modular parameter of the torus, which describes its shape. 
  By expanding in powers of 
\begin{equation}
q=e^{2\pi i \tau} \ \ \hbox{and} \ \ \bar q=e^{-2\pi i \bar \tau} 
 \end{equation}
 one can read off the multiplicities of the closed string spectrum. 
 Only states with the same power of $q$ and $\bar q$ are physical.

The right-hand side of (\ref{PartF}) can be expanded in terms of characters of the (extended) conformal field theory,
 \beq
 \chi_i={\rm Tr}\mid_i\ e^{2\pi i \tau (L_0-1)} \,,
 \eeq
 where the trace is over all states in the representation built on ground state $i$ by the action of the Virasoro generators $L_n$ and the generators of some extension
 of the Virasoro algebra. The characters have the following expansion
\beq
\chi_i(\tau) = q^{h_i-c/24}\sum_n d_n q^n \,,
\eeq
where $d_n$ are non-negative integers.

The number of such representations will in general be infinite, but if the extension of the symmetry algebra is large enough we have an RCFT, and then the number is finite.  The character label 0 corresponds to the vacuum representation. It contains the vacuum and all states obtained from it by the action of the Virasoro generators and all generators of the extended algebra (if any), modulo states of zero norm (null states).

 The torus partition function has the following character expansion:
\begin{equation}
{\rm Tr}\ e^{2\pi i \tau (L_0-1)}e^{-2\pi i \bar\tau (\bar L_0-1)} = \sum_{ij} Z_{ij} \chi_i(\tau) \bar\chi_j(\bar \tau) \,.
 \end{equation}
Here the coefficients $Z_{ij}$ must be non-negative integers, subject to the constraint of modular invariance. This constraint is a consequence of the fact that there are infinitely many parametrisations of the torus that must all give the same result. These reparametrisations are generated by discrete transformations of the parameter $\tau$:
\bea
T: &\tau \rightarrow \tau+1 \,,\\
S: &\tau \rightarrow -\frac{1}{\tau} \,. \label{STrans}
\eea
These transformations are represented as matrices $S$ and $T$ on the set of characters $\chi_i$, and the conditions for modular invariance are therefore
\beq
\left[S,Z\right]=\left[T,Z\right]=0 \,.
\eeq
Furthermore $Z_{00}$, the multiplicity of the vacuum state,
must be equal to 1. In closed string theory this has the consequence that there is precisely one graviton in the spectrum.

There has been a lot of work on finding solutions to the conditions for $Z_{ij}$, but this has been completed only for a few extended algebras. Notable examples are the unextended Virasoro algebra with $c<1$ as discussed above, and the $SU(2)$ Kac-Moody algebra \cite{Cappelli:1986hf}. 
Furthermore there are known classes of general solutions that are valid for any RCFT: the charge conjugation invariant $Z_{ij}=C_{ij}$, the diagonal invariant $Z_{ij}=\delta_{ij}$, conformal embeddings \cite{Schellekens:1986mb, Bais:1986zs} and simple current invariants \cite{Schellekens:1989am}. Here $C_{ij}$ is a bijection that corresponds to charge conjugation in the world-sheet  theory, acting on the ground states. 
It is known that the diagonal invariant is always a solution to the conditions of modular invariance at one loop, but does not always define a consistent CFT 
\cite{Davydow}. On the other hand, the charge conjugation invariant always defines a consistent CFT, and should be viewed as the canonical definition of the theory. We refer to this case as ``C-diagonal".

\subsection{Fusion Rules and Simple Currents}
The fusion rules of an RCFT indicate how many couplings exist when two representations $[i]$ and $[j]$ are combined. It can be formally written as
\beq
[i] \times [j] = \sum_k N_{ij}^{~~k}[k]\ ,
\eeq
where $[i]$ denotes an (extended) CFT representation. 
A simple current \cite{Schellekens:1989am, Intriligator:1989zw} is a special representation $J$ with the particular feature that just one
term (labelled $Ji$) exists on the right-hand side:
\beq
[J] \times [i] = [Ji] \,.
\eeq
The set of representations generated by the action of $J$ on $[i]$ is called the {\it orbit} of $[i]$. 

The reason simple currents enter the story is that they allow us to construct a large number of modular invariant partition functions. This works roughly as follows.
The action of the simple currents on themselves defines an abelian discrete group. Take any subgroup of that discrete group. Now take a set of generators of that subgroup. On that basis of generators one defines a matrix $X$. This is a matrix of rational numbers \cite{Kreuzer:1993tf} computed from a simple equation, which occasionally has no solution, but generally has a number of solutions that grows exponentially with the size of the basis. We will omit the details here, but the main point is that for any subgroup of the simple current group one has many modular invariants.

The simple current group of ${\cal N}=2$ minimal model is fairly large. For $k$ odd it is 
$\mathbb{Z}_{4k}$ and for $k$ even it is $\mathbb{Z}_2 \times \mathbb{Z}_{2k}$. But much more importantly, in a tensor product one gets a product of all these groups. For example, the combination $(3,8,8,8)$ yields $\mathbb{Z}_{12}\times (\mathbb{Z}_2)^4 \times (\mathbb{Z}_{16})^4$. These discrete groups have a huge number of subgroups, and hence for every Gepner model we get a huge number of MIPFs. This boosts the number of available Gepner models from 168 (for just the C-diagonal MIPF) to 5403.


\def\ket#1{|#1\rangle}
\def\bra#1{\langle#1|}
\def\Zbf{{\bf Z}}
\newcommand{\be}{\begin{equation}}
\newcommand{\ee}{\end{equation}}
\subsection{Open String CFT}

To use RCFTs for open string model building one needs a description of ends of open strings. In a worldsheet description they sweep out worldsheets with boundaries.
  The most general worldsheet for oriented {\it closed} strings is a Riemann surface of arbitrary genus $g$, which is a torus with $g$ handles. One can attach tubes to act as external closed strings. To get all {\it open} string diagrams, one can make holes in those surfaces with the topology of a circle.  
  Finally, one can add strips to the edges of the holes to act as external open strings.  
  
  In section (\ref{Oneed}) we have seen that O-planes are needed in order to build consistent open string models (at least in the supersymmetric case). Just as D-branes
are described by means of boundaries of the surface, O-planes are described by means of another topological feature, 
 a {\it crosscap}. A crosscap is added to a surface
 by making a hole, as  above, but identifying the opposite points of the boundary circle to each other in an orientation reversing way. Hence an ant crawling on one side of the circle finds itself on the other side after crossing the crosscap. One can add more than one crosscap, but not all resulting surfaces are topologically distinct. A sphere with two holes is topologically an annulus; a sphere with one  hole and a crosscap is equivalent to a Moebius strip, and a sphere with two crosscaps is a  Klein bottle. 
 
 Now we have to determine the behaviour of the CFT near the edges of the surface or in the prasence of a crosscap

\subsubsection{Boundary and Crosscap States}

Any surface with boundaries or crosscaps has a double cover which only has handles, and on which
one defines a closed, oriented conformal field theory. 
This CFT is the starting point for constructions of open (and 
unoriented) strings, which were referred to as ``open descendants'' of
the closed string theories in \cite{Sagnotti:1996eb}. 
The
presence of boundaries and crosscaps is described by 
boundary and crosscap ``state'', which are not really states themselves,
but in fact non-normalisable linear combinations of states in
the closed string Hilbert space.

Here we will assume that the entire chiral algebra
remains unbroken at the boundary or by a crosscap, ignoring the interesting possibility of breaking part of the closed string symmetries.
 The condition that 
a symmetry is not broken by a boundary or a crosscap
is
\be 
(W_n + (-1)^{h_W}\tilde W_n) \ket{B}=0 \ ,\ \ \ \ \ \ \
 (W_n + (-1)^{h_W+n} \tilde W_n) \ket{C}=0 \,,
\ee
where $W_n$ is a mode of a chiral current, $\tilde W_n$
a mode of an anti-chiral current and $h_W$ its conformal weight; $\ket B$ a boundary state
and $\ket C$ a crosscap state. A basis for the solutions to
these conditions is formed by the Ishibashi states \cite{Ishibashi:1988kg}
\be \label{IshiState} \ket{B_i} = \sum_I \ket{I}_i \otimes U_B \ket{I}_{i^c} \ ,\ \ \ \ \ \ \ 
\ket{C_i}= \sum_I \ket{I}_i \otimes U_C \ket{I}_{i^c} \ .
\ee
Here the $i$ labels a representation of the chiral algebra and
$i^c$ its charge conjugate. The sum is over all states in the
representation, and $U_B$ and $U_C$ are operators satisfying
\be 
\tilde W_n U_B = (-1)^{h_W} U_B \tilde W_n\ ,\ \ \ \ \ \ \ 
\tilde W_n U_C = (-1)^{h_W+n} U_C \tilde W_n \,.
\ee
Any boundary state must be a linear combination of
these Ishibashi states, {\it i.e.}
\be
\ket{B_a} = \sum_i B_{ia} \ket{B_i}\ ,\ \ \ \ \ \ \
\ket{C} = \sum_i \Gamma_{i} \ket{C_i}  \,.
\ee 
It turns out that in general one can allow for several boundary
states, labelled by a boundary label $a$, but for only one
crosscap state for a given theory.  For a given  MIPF $Z_{ij}$ more than one crosscap state may exist, but one cannot mix them.   

A choice of a set of boundary labels $a$, and a set of coefficients
$B_{ai}$ and $\Gamma_i$ form part of the data that define an
open string CFT. 
Although more is required to specify all correlation functions
on arbitrary surfaces, this information is sufficient 
to compute the one-loop diagrams without external lines that
contribute to the open and closed string partition functions. 
The relevant string diagrams are
those with vanishing Euler number. From these diagrams we can  compute the spectrum of the theory. 

\subsubsection{Orientifold Partition Functions}

In the presence of boundaries and crosscaps there are four topologically distinct surfaces with vanishing
Euler number: 
the torus, the Klein bottle, the annulus and the M\"obius strip. These contributions can be expanded in (bi)linears of characters:
\bea \label{OpenStringPart}
{\cal T} = \sum_{ij} Z_{ij} \chi_i(\tau) \chi_{j}^*(\tau)\,, & &  \;\;\;  \;{\cal K} = \sum_i K^i \chi_i(2it) \,, \\
{\cal A} = \sum_{ab} N_{a} N_{b} A^{i}_{ab} \chi_i(it/2) \,,
& & \;\;\;  \; {\cal M} = \sum_{a} N_{a} M^i_{a} e^{-i\pi(h_i-c/24)}{\chi}_i\left(\tfrac12(1+it)\right)\,. \;\;\; 
\eea
Here $\tau$ is the  modular parameter of the torus, as before, and $t={\rm Im}\tau$. As discussed in section \ref{MIPF}, the torus defines the oriented closed string
partition function. Likewise, the sum of the torus and the Klein bottle defines the unoriented closed string partition function. The annulus is an open string loop, and defines.  
together with the M\"obius strip, the open string partition function. The Klein bottle amplitude and the M\"obius strip act as an orientifold projection on the closed and open string spectrum respectively.

\subsubsection{Channel Transformations}

The diagrams that describe the spectrum are computed in the {\it transverse} channel, in which
closed strings propagate between two boundaries, a boundary
and a crosscap, or two crosscaps. In Figure \ref{ChannelTrans} this is illustrated for the simplest case: a diagram of closed strings propagation from boundary $a$ to boundary $b$ is transformed to a diagram where open strings with endpoints $a$ and $b$ propagate in a closed loop. This is called the {\it direct} channel. By the rules of string perturbation theory this is the same diagram, but with two different parametrisations. To transform from one parametrisation to another one has to interchange the world-sheet space and time directions. This can be done by means of the transformation $\tau \rightarrow -\frac{1}{\tau}$, which acts on the characters as a matrix $S$. 
\begin{figure}[h]
\begin{center}
\includegraphics[width=4.3in]{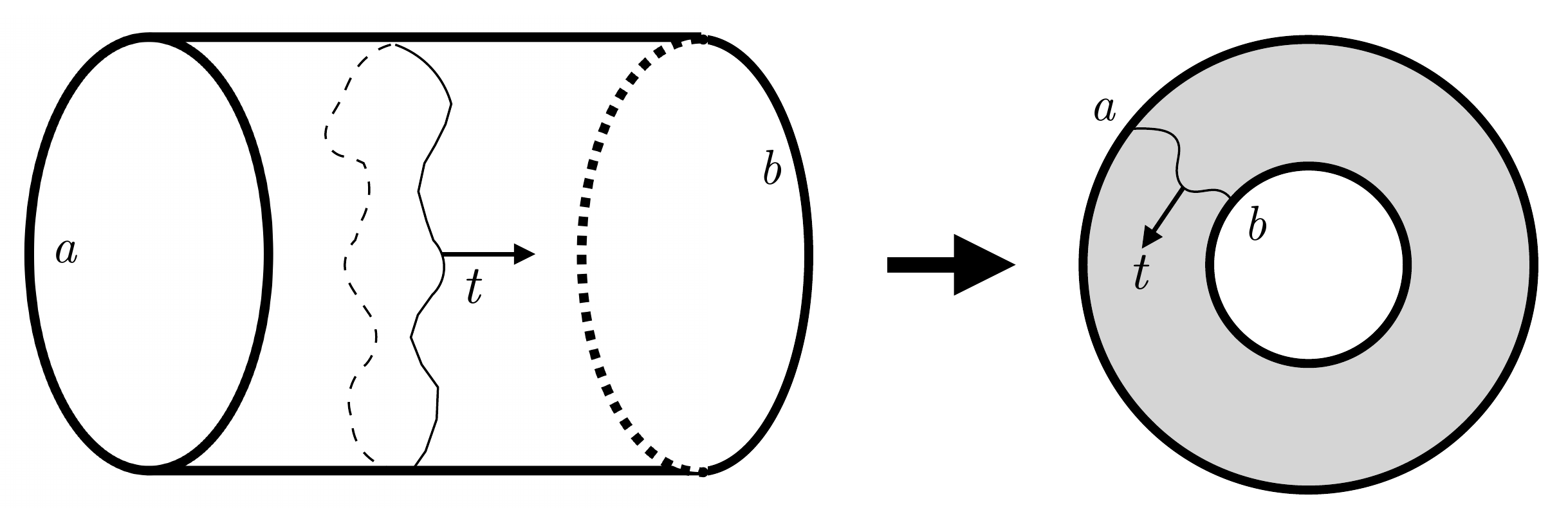}
\caption{Channel transformation.\label{ChannelTrans}}
\end{center}
\end{figure}
The analogous figure for non-orientable surfaces is harder to draw, but the direct channel for closed strings propagating between a boundary and a crosscap is a Moebius strip, and the direct channel for propagation between two crosscaps is a Klein bottle. The required transformations are also different. Here we will just give the result. 
The details and many references may be found in  \cite{Angelantonj:2002ct}.

The three basic transverse channel amplitudes are obtained by sandwiching the closed string propagators between boundary and crosscap states:
\begin{eqnarray}
&\hbox{Transverse Annulus:~~~~~~~~~~} &N_a N_b \bra{B^c_a}e^{i\tau H}\ket{B_b}\, ,
 \\ 
&\hbox{Transverse Moebius strip:~~~} &N_a \left[\bra{B^c_a}e^{i\tau H}
\ket{C}+\bra{C^c}e^{i\tau H}\ket{B_a}\right]\, , \\ 
&\hbox{Transverse Klein bottle:~~~~~} &N_a \bra{C^c}e^{i\tau H}\ket{C} \,.
\end{eqnarray}
Here $H$ is the closed string Hamiltonian: 
$H=2\pi(L_0+\tilde L_0 - c/12)$, and $\tau$ is a real number
representing the length of the cylinder. The subscript ``$c$'' indicates
that a CPT conjugate state is to be used.  
The integers $N_a$ are the
Chan-Paton multiplicities. 
One can express these amplitudes in terms of characters of the
representation $i$. 
By means of a transformation of the parameter $\tau$ one
can then compute the corresponding amplitudes in the direct channel
(the open and closed string loop channels). In the case of the
Klein bottle and the annulus this transformation acts on the characters
as the modular transformation matrix $S$, whereas in the case of
the Moebius strip one uses the matrix $P=\sqrt{T}ST^2S\sqrt{T}$, with
$\sqrt{T}$ defined as $\exp{i\pi(L_0-c/24})$. Then one arrives at
the following expressions: 
\def\half{\tfrac12}
\begin{eqnarray}
&\hbox{Direct Annulus:~~~~~~~~~~}\hfill &\half N_a N_b A^i_{~ab} \chi_i(\half\tau) \, ,
 \\ 
&\hbox{Direct Moebius strip:~~~}\hfill &\half N_a M^i_a \hat\chi_i(\half+\half \tau)\, , \\ 
&\hbox{Direct Klein bottle:~~~~~}\hfill &\half K^i \chi_i(2\tau)  \,.
\end{eqnarray}
Here $\hat\chi_i  \equiv T^{-\half}\chi_i $, and the parameter
$\tau$ is purely imaginary. The coefficients are
\bea A^i_{~ab}&= \sum_m S^i_{~m} B_{ma} B_{mb}\, ,  \label{AnnulusCoeff}\\
M^i_{~a}&= \sum_m P^i_{~m} B_{ma} \Gamma_{m}\, ,  \\
K^i&= \sum_m S^i_{~m} \Gamma_{m}\Gamma_{m}\  \,.
\eea

\subsubsection{Integrality Conditions}

To interpret these expressions in terms of state counting it is clearly important that all the relevant coefficients be non-negative integers. Indeed, in all cases discussed here 
$Z_{ij}$ and $A^i_{ab}$ are explicitly non-negative integers, and $K^i$ and $M^i_a$ are integers. But that is not sufficient, because the actual state multiplicities are sums and differences of these numbers.   Note first of all that the argument of the Klein bottle term, $2it$, coincides precisely with the terms in the expansion of 
$\chi_i(\tau) \chi_{j}^*(\tau)$ for the physical states: those
with the same powers of $q$ and $\bar q$. Hence this term may alter the multiplicity of the physical states. It is the total multiplicity that must be a non-negative integer.
The vacuum 
representation, $i=0$, has $Z_{00}=1$ and $K_0=1$. As stated before, this representation produces the graviton, which should have multiplicity 1. Hence we find that we must
 divide the entire partition function by 2 to get the right multiplicity. To get correct multiplicities for all the other closed string states, we need
$$K^i=Z_{ii}\ {\rm mod}\ 2\ \hbox{ and} \ |K_i |\leq Z_{ii}\ ,$$
which is indeed satisfied in all known case. 
Hence the multiplicity of the   closed string states is given by 
$$ \frac12 (Z_{ii}+K_i)\,.$$
Note that it is possible for closed string states to be ``projected out'', {\it i.e.} removed from the spectrum, if $K_i=-Z_{ii}$.

The second line in (\ref{OpenStringPart}) gives rise to open string states, projected by the Moebius amplitude. In this case the coefficients satisfy, in all known theories
$$M^i_a=A^i_{aa} \ {\rm mod}\ 2\hbox{ and}\ |M^i_a| \leq A^i_{aa} \,.$$
The natural interpretation is that diagonal states (open strings between the same boundaries) are projected to obtain the following multiplicities 
$$ \frac12 (N_a N_a A^i_{aa}+N_a M^a_i) \,,$$
whereas the off-diagonal terms are not affected by the Moebius amplitude. In the simplest case, $A^i_{aa}=1$ and $M^a_i=\pm 1$, one gets dimensions
of symmetric and anti-symmetric tensors. Further work is needed to demonstrate that these particles do indeed couple in that manner. 

\subsubsection{Gauge Groups}\label{BraneGroups}

The gauge bosons come from the identity character. Its massless state is a space-time vector. In the closed string a left-right combination of two vectors gives the graviton, the dilaton and the $B_{\mu\nu}$ Kalb-Ramond field. In the open string sector, the vector comes with multiplicity
$$\frac12 ( A^0_{aa} N_a N_a + M^0_a N_a) \,. $$
It turns out that the matrix $A^0_{ab}$ in the space of boundaries is a bijection: it has the property that for any {\it a} there is precisely one label $b$ with  $A^0_{ab}=1$, and $A^0_{ab}=0$ for all other $b$. The label $b$ for which $A^0_{ab}=1$ is called the complex conjugate boundary, and is denoted $a^c$. If $a=a^c$ the boundary is called self-conjugate or real. In that case,
$M^0_a=\pm 1$. The number of gauge bosons is either $\frac12 N_a (N_a + 1)$, suggesting that the gauge group is $USp(N_a)$, or $\frac12 N_a (N_a - 1)$, suggesting that it is $SO(N_a)$. Further studies of the amplitudes are necessary to verify that this interpretation is correct, because from these arguments we only obtained multiplicities. Note that in the symplectic case we must have $N_a$ even. 

For complex boundaries, those with $a\not=a^c$, we get a vector boson multiplicity $\frac12 N_a N_a \left(A^0_{aa^c}+A^0_{a^ca}\right)=(N_a)^2$, since $M^0_a=0$. This
suggests a group $U(N_a)$, which is indeed correct. Hence now we can get, in principle, the same brane configurations as in the oriented case, but with the additional option of symmetric and anti-symmetric tensors.
If $b$ is not equal to $a$ or $a^c$ we get a bi-fundamental.

\subsubsection{Completeness, Integrality and Sewing Constraints}

Now we have to determine the coefficients $B_{ia}$ and $\Gamma_{i}$. Indeed, the first task is to determine the set of labels $a$, which determine the set of boundaries, and hence, in geometric language, the number of D-branes at our disposal. 
The answer is provided by a conjecture called the completeness condition for boundaries, formulated in \cite{Pradisi:1996yd}. It states essentially that $B_{ia}$ is an invertible matrix, so the number of labels $a$ is equal to the number of labels $i$. The latter number can be inferred from (\ref{IshiState}). Note that this definition pairs states $i$ with their charge conjugates $i^c$. The closed string states that propagate in the transverse channel are those with $Z_{ii^c} \not =0$, and they have multiplicity $Z_{ii^c}$. Hence the number of labels $i$, also known as Ishibashi labels, is equal to $\sum_i Z_{ii^c}={\rm Tr} ZC$. By the completeness conjecture the number of labels $a$ is the same. It follows that the boundary coefficients $B_{ia}$ depend on the choice of partition function $Z_{ij}$ of the closed string theory. 

The boundary and crosscap coefficients are subject to the integrality conditions described above, but more importantly by {\it sewing constraints}. These constraints follow from the requirement that different ways to build a Riemann surface from three-point functions by ``sewing'' must yield the same answer. See \cite{Sonoda:1988mf} for a discussion of these constraints  for closed strings, \cite{Cardy:1991tv} and \cite{Lewellen:1991tb} for open strings and \cite{Fioravanti:1993hf} for unoriented strings. 

\subsubsection{A Solution for C-diagonal Theories}

The first solution to these constraints in the oriented case was found by Cardy \cite{Cardy:1989ir}. He considered the C-diagonal modular invariant, and gave the following formula for the boundary coefficients:
\be\label{CardyB}
B_{ia} = \frac{S_{ia}}{\sqrt{S_{0i}}} \,,
\ee
where $S$ is the matrix defined in (\ref{STrans}).   Note that this explicitly satisfies the completeness condition, because $S$ is a square matrix. 
Substituting this in (\ref{AnnulusCoeff}) we get
\be
A^i_{~ab} = \sum_m \frac{S^i_{~m} S_{ma}S_{mb}}{S_{m0}} = N^i_{~ab}\ ,
\ee
where in the last step we used the Verlinde formula for fusion multiplicities  \cite{Verlinde:1988sn}. This shows that also the integrality conditions are satisfied, because fusion coefficients are integer.
In fact, the purpose of Cardy's work was to understand the Verlinde formula, but while doing so he proposed  and initiated the field of boundary CFT. This was picked up almost immediately \cite{Bianchi:1990yu} by the Tor Vergata (Rome II) group, who did a lot of pioneering work in this area.

They also considered 
unoriented surfaces, a subject not treated by Cardy. After some earlier work, in 
\cite{Pradisi:1995qy} they proposed the following formula for the crosscap coefficients for the case $Z=C$:
\be
\Gamma_i=\frac{P_{0i}}{\sqrt{S_{0i}}}\ ,
\ee
which uses the $P$-matrix introduced above.

\subsubsection{Simple Current Results}

Now attempts started to obtain similar results for general simple current modular invariants. This culminated about a decade later in a completely general formula \cite{Fuchs:2000gv} for all simple current MIPFs. This paper contains references to all the earlier partial results obtained by various groups.  The formula for boundary coefficients is
\be B_{(i,J)[j,\psi]} =\sqrt{\frac{|{\cal G}|} {|{\cal S}_i|| {\cal C}_j|} }
{ \frac{\alpha(J)S^J_{ij} }{ \sqrt{S_{0i}}}} \psi(J)\ , \ee
and the crosscap formula is
\be \Gamma_{(i,J)}={\frac{1}{ | {\cal G} |}} \sum_{K \in {\cal G}} \eta(K)  P_{Ki}\delta^{J0} \,. \ee
We only present these results here to give a flavour of what is involved. The simple currents form a subgroup ${\cal G}$ under fusion. Multiplicities $Z_{ij} > 1$ occur when a field $i$ is a fixed point of the simple current action: $Ji=i$. The simple currents that fix $i$ for a subgroup of ${\cal G}$ called the stabiliser of $i$ and denoted ${\cal S}_i$. Consequently, the Ishibashi labels $(i,J)$ are equipped  with a degeneracy label $J \in {\cal S}_i$, and $i$ is a label such that $Z_{ii^c}=1$. The boundary states are labelled by orbit representatives $i$ (that is, one representative of each orbit $(i,Ji,J^2i,\ldots)$), and their degeneracy is governed by the size of another discrete group, ${\cal C}_j \subset {\cal S}_j$. The discrete group characters $\psi(J)$ are used as degeneracy labels for the boundaries. The matrices $S^J$ are modular transformation matrices of a ``fixed point CFT'', an algebraic structure associated with the fixed points, intuitively introduced in  \cite{Schellekens:1989uf} and more rigorously defined  in \cite{Fuchs:1995zr}. For further details, such as the precise definition of ${\cal C}_j$, the phase $\alpha(J)$, and the signs $\eta(J)$ we refer the reader to \cite{Fuchs:2000gv}. 

The boundary coefficients are uniquely defined as soon as the torus partition function $Z_{ij}$ is known, but for the crosscap there is a variety of possibilities. First of all there is a ``Klein bottle current'' $K$, and secondly the signs $\eta(J)$ must satisfy a condition that may have several solutions.  
Taking into account these choices for different crosscaps increases the total number of distinct, unoriented Gepner CFT models to 49304. However only 33012 of them have non-zero tension. The remaining ones are not usable for building supersymmetric orientifold models.

\subsection{RCFT Model Building}

We are now ready to apply this machinery to model building. In the situation of interest, the relevant CFT is the usual world-sheet theory of the superstring (or non-supersymmetric fermionic string) in four flat dimensions, combined with a non-trivial superconformal CFT to describe the six ``compactified'' dimensions. The characters are 
products of superstring characters and internal characters. The former determine the space-time properties of the string excitations, in particular spin and chirality, whereas the latter contribute to the counting of states. The gauge representations of the physical states can be read off from the Chan-Paton labels of the string state under consideration.

Now we start with a choice of an RCFT (in practice always a Gepner model), a choice of a MIPF, a choice of a crosscap, and a choice of three or four boundary labels $(a,b,c)$ and $d$, depending on the configuration we want to realise, as discussed in Section \ref{SimplestExamples}.   Usually $a$ is taken to be the QCD label and $b$ is the weak label. We can already  make sure that $N_a=3$ and $N_b=2$ do not oversaturate the dilaton tadpole. Now
we compute the annulus coefficients $A^i_{ab}$, $A^i_{ac}$, $A^i_{bc}$ etc. for any two 
chosen labels and check if the chiral intersections (as defined in Table \ref{specori}) match the required spectrum. 

In case of success, we now check if the putative $Y$ boson remains massless after taking into account axion mixing. This requires checking all axions in the full closed string spectrum. At the same time we may check if any other $U(1)$ vector bosons in the spectrum acquire a mass. If they do not, this is a phenomenological issue which we will have to deal with later because it is not solved at the level  of the RCFT.

The next step is to cancel all tadpoles by finding a suitable hidden sector, as explained in Section \ref{ATA}. This can be very time-consuming, because it requires considering all subsets of all branes not used in the Standard Model configuration. In practice, this can usually not be done exhaustively.

There is still one more check to be made: the absence of global anomalies, as discussed in \cite{Uranga:2000xp}.

The first work along these lines appeared in \cite{Angelantonj:1996mw}. In this paper six-dimensional theories were studied, and examples with chiral spectra were found. In \cite{Blumenhagen:2004cg}, building on \cite{Blumenhagen:2003su}, the first chiral spectra were found in four dimensions. Then in \cite{Dijkstra:2004ym,Dijkstra:2004cc} a general search was undertaken for the Madrid configuration shown in Figure \ref{MadridQuiver} and some of its variations. 

In \cite{ADKS} a different approach was taken. Rather than searching for specific pre-selected brane configurations, these authors searched for any combinations of brane labels that yields the Standard Model, in a rather generous definition of the latter. 
This includes enlarged gauge groups, non-chiral pairs of quarks and leptons as chiral brane matter ({\it i.e.} chiral matter that becomes non-chiral if only the SM group is considered) and gauged flavour symmetries. The result essentially demonstrated that anything one could theoretically propose as a Standard Model configuration will likely  be realisable if one has a large enough scope of brane models to start with. 

One may think that this scope can be extended significantly by moving to other RCFTs, but in practice this is not easy. For most RCFTs we simply do not have all the relevant data available. It is often easy to get the spectrum modulo integers, but a lot harder to get the exact spectrum, as required. Furthermore any RCFT building blocks other than ${\cal N}=2$ minimal models (one example are the Kazama-Suzuki models \cite{Kazama:1988qp}) have much smaller simple current groups, and hence far fewer MIPFs. There might exist vast numbers of exceptional (not simple current related) MIPFs in some cases, but a general formalism to compute their boundary and crosscap coefficients is not available. Free fermion orientifolds have been considered, but are a far less fertile area \cite{Kiritsis:2008mu}. So it seems that Gepner models are in a sense the optimal possibility.

\section{F-theory Model Building} \label{sec_F-theory}

F-theory was introduced in \cite{Vafa:1996xn} as a non-perturbative formulation of Type IIB compactifications with 7-branes.
Its importance for model building is owed to its generality, including the fact that it allows for the construction of gauge sectors which enjoy an embedding into the exceptional Lie group $E_8$. This property makes it a particularly natural framework to study Grand Unified Theories (GUTs) in string theory and distinguishes it from its perturbative cousins discussed in the previous sections, which are based on gauge groups $U(N)$, $\OGroup(N)$ and $USp(N)$.
Hence, F-theory combines the attractive features of model building with D-branes - the localisation of gauge degrees of freedom on branes, which in principle invite a local approach to model building within a certain realm of questions - with the appearance of exceptional gauge symmetry as in heterotic $E_8 \times E_8$ string theory, which bears its fruit in the context of GUT model building. The goal of this programme is
to solve some of the outstanding model building challenges faced by four-dimensional SUSY GUTs in the higher-dimensional brane-world framework provided by F-theory.

In addition, F-theory offers a formulation of D-branes in terms of the geometry of so-called elliptic (or more generally genus-one) fibrations; many involved questions of brane dynamics are hence translated into entirely geometric questions and oftentimes have a clear answer in algebraic or arithmetic geometry.

General introductions to F-theory are provided for instance in \cite{Denef:2008wq,Heckman:2010bq,Weigand:2010wm,Taylor:2011wt,Weigand:2018rez,Cvetic:2018bni}, to which we refer for details and the original references. In the sequel, after briefly presenting some of the technical foundations of F-theory, we focus on its role for particle physics model building.

\subsection{From \texorpdfstring{$[p,q]$}{[p,q]} 7-branes to Exceptional Gauge Algebras}

The starting point for F-theory is Type IIB string theory. 
Type IIB string theory contains in its massless spectrum higher-form Ramond-Ramond (RR) gauge potentials of even degree, $C_{2p}$, with $p=0,\ldots, 4$. 
The 2-form $C_2$ couples electrically to a D1-brane, which is a string-like soliton of tension $T_{\rm D1} = \frac{2 \pi}{\ell_s^2} \frac{1}{g_s}$ in perturbative string theory. The Neveu-Schwarz 2-form potential $B_2$, on the other hand, couples electrically to the fundamental, or F1-, string of tension $T_{\rm F1} = \frac{2 \pi}{\ell_s^2}$. These two types of strings can form BPS bound states: A bound state of $p$ F1-strings and $q$
 D1-branes is called a $(p,q)$ string, and it exists as a BPS bound state for $p$ and $q$ co-prime integers. 
Type IIB string theory enjoys a weak-strong coupling duality, which maps an F1- or $(1,0)$-string into a D1- or $(0,1)$-string, and vice versa. 
 The theory is believed to be invariant, at the non-perturbative level, under an $SL(2, \mathbb Z)$ duality transformation, which in particular acts on the axio-dilaton $\tau = C_0 + \frac{i}{g_s}$ and the two-form potentials as
\bea \label{SL2Z}
\tau \to \frac{a \tau +b}{c \tau +d} \,, \qquad \left(\begin{matrix} C_2 \\B_2 \end{matrix}\right) \to  M  \left(\begin{matrix} C_2 \\B_2 \end{matrix}\right) \,, \qquad  M = \left(\begin{matrix} a  & b \\ c  & d \end{matrix}\right) \in SL(2, \mathbb Z)  \,.
\eea

Of central interest for the formulation of F-theory are the Type IIB 7-branes. 
A D7-brane, often called $[1,0]$-brane in the sequel, is, by definition, a 7-brane on which an F1-string can end; $SL(2,\mathbb Z)$ duality implies that there must then exist also another type of 7-brane, a so-called $[0,1]$-brane, on which a D1-string can end.
More generally one defines a $[p,q]$ 7-brane as a 7-brane on which a $(p,q)$-string can end.

The fact that a 7-brane is an object of real codimension two in ten dimensions implies a rather severe backreaction on the supergravity background due to its tension and RR charge. 
If one considers a single D7-brane and introduces the complex coordinate $z = x^8 + i x^9$ in the complex plane spanned by the two directions normal to the brane, the axio-dilaton $\tau$ acquires a varying profile which close to the brane at $z=0$ can be approximated as
\bea \label{tauprofile}
\tau(z) = \tau_0 + \frac{1}{2 \pi i} {\rm log}\left(\frac{z}{z_0}\right) \,.
\eea
Due to the logarithmic branch cut, $\tau(z)$ undergoes a shift $\tau \to \tau +1$ as one encircles the D7-brane at $z=0$. This multi-valuedness is consistent because it is accompanied by a
an $SL(2, \mathbb Z)$ duality transformation on all the Type IIB fields, which merely accounts for a change of duality frame.  
The $SL(2, \mathbb Z)$ monodromy induced in this way by a single $[p,q]$ 7-brane can be represented by the action of a monodromy matrix of the form
\bea
M_{[p,q]} =  \left(\begin{matrix}  1 - p q & p^2 \\ - q^2 & 1 + p q     \end{matrix}\right) \,.
\eea
Two $[p,q]$ 7-branes are called mutually non-local if their monodromy matrices cannot be brought into the same form by an $SL(2, \mathbb Z)$ transformation.

Not all types of mutually non-local $[p,q]$ 7-branes can be consistently placed on top of each other to form a supersymmetric brane stack without leading to a drastic degeneration of the theory such as an effective decompactification. Let us first consider 7-branes in flat space $\mathbb R^{1,9}$. In order to describe the allowed configurations of coincident $[p,q]$ 7-branes in 10 dimensions, it suffices to consider three types of mutually non-local $[p,q]$ 7-branes. A possible such generating system consists of the following three brane types:
\bea
A : [p,q] = [1,0] \,, \qquad B : [p,q] = [3,-1] \,,  \qquad C : [p,q] = [1,-1]   \,.
\eea
In the chosen duality frame, an $A$-type brane corresponds to a perturbative D7-brane. 
The O7-plane from Type IIB orientifolds is realised as a $BC$-brane system. Consistently the monodromy matrix $M_{BC} = M_B M_C $ acts as multiplication with $(-1)$ on a $(1,0)$ string, which is interpreted as worldsheet parity. In this way one recovers the perturbative gauge groups $SU(N)$ (omitting the diagonal $U(1)$ factor) by a configuration of $N$ $A$-type branes, while $\OGroup(N)$ groups correspond to a system of $N$ $A$-type branes on top of a $BC$ brane system, i.e. on an O7-plane.\footnote{For $N < 4$, non-perturbative effects lead to a dynamical separation of the branes.} 
The most important observation, however, is that in addition to these perturbative gauge groups, 
also the exceptional gauge groups $E_N$ for $N=6,7,8$ can be obtained in flat space from brane configurations of the form $A^{N-1} B C C$. 

The emergence of exceptional gauge algebras is rooted in the possible $[p,q]$-strings stretched between the mutually non-local branes in the configuration. These include, in the present setup, bound states of $[p,q]$-strings with three endpoints - so-called multi-pronged strings - which end on three mutually non-local branes, rather than just two branes of the same type as in perturbative setups. 
A careful analysis of the possible multi-pronged strings \cite{Gaberdiel:1997ud} for the configuration $A^{N-1} B C C$
indeed identifies all the roots of the exceptional series $E_N$, $N=6,7,8$.

The easiest way to classify the allowed configurations of coincident 7-branes is by passing to the geometrised description of F-theory as pioneered in \cite{Vafa:1996xn,Morrison:1996na,Morrison:1996pp}. 
The behaviour of the axio-dilaton $\tau$ under an $SL(2, \mathbb Z)$ transformation, \eqref{SL2Z}, is reminiscent of the transformation of the modular parameter of an elliptic curve under its modular group. This motivates interpreting $\tau$ as the complex structure parameter of an elliptic curve $\mathbb E_{\tau}$, which varies holomorphically in the directions normal to the 7-brane according to \eqref{tauprofile}. This structure defines an elliptic, or more generally a genus-one \cite{Braun:2014oya}\footnote{The difference is that an elliptic fibration  necessarily has a section. In the sequel we will, for simplicity, use the term elliptic fibration for both constructions.}, fibration over the directions normal to the branes.

The simplest manifestation of this idea is to consider compactifications of F-theory to eight dimensions with general $[p,q]$ 7-branes filling the uncompactified dimensions $\mathbb R^{1,7}$. 
Due to the backreation of the 7-branes,
the two compact normal directions must be curved to form a projective sphere $\mathbb P^1$.
The variation of the axio-dilaton over $\mathbb P^1$ can be identified with the variation of the complex structure parameter of an elliptic curve fibered over $\mathbb P^1$. Supersymmetry requires that the elliptic fibration defined in this fashion is a Calabi--Yau manifold, and in this case an elliptically fibered K3 surface.
The physical compactification space from ten to eight dimensions, $\mathbb P^1$, represents the base of this elliptic fibration. Note that this physical compactification space is {\it not} Calabi--Yau, but rather carries positive curvature as a result of the backreaction.
The $[p,q]$ 7-branes sit at special points on the base $\mathbb P^1$ over which 
 the complex structure $\tau$ of the elliptic fiber $\mathbb E_\tau$ degenerates to reflect the singular behaviour of the axio-dilaton profile \eqref{tauprofile} for $z \to 0$. The monodromies around the location of a $[p,q]$ 7-brane have a direct geometric interpretation as monodromies that transform a local basis of one-cycles on the elliptic fiber into one another as one transports it around the 7-brane location. 

The possible types of monodromies that can consistently occur in this manner in an effectively eight-dimensional compactification have been classified geometrically by Kodaira and N\'eron \cite{Kodaira2,Kodaira3,Neron} and are in one-to-one correspondence with A-D-E type singularities of the elliptic fiber.
By interpreting a monodromy as a product of monodromy matrices of different $[p,q]$ 7-branes one can translate this geometric classification into a classification of allowed stacks of coincident 7-branes on $\mathbb P^1$.

 The classification starts from the notion of a Weierstrass model for the elliptic fibration:
 An elliptic curve $\mathbb E_\tau$ with modular parameter $\tau$ can be represented as the hypersurface 
 \bea
 P_W := y^2 - (x^3 + f x z^4 + g z^6)  = 0
 \eea
  in the weighted projected space $\mathbb P_{2,3,1}$ with homogenous coordinates $[x : y : z]$.
Here $f$ and $g$ are complex parameters which determine the complex structure or modular parameter $\tau$ as
\bea
j(\tau) = \frac{4 (24 f)^3}{\Delta}  \,, \qquad \Delta = 4 f^3 + 27 g^2 \,,
\eea
where $j(\tau)$ is the $SL(2, \mathbb Z)$ invariant Jacobi $j$-function. 
When the discriminant $\Delta$ vanishes,  the elliptic curve $\mathbb E_\tau$ degenerates.
This description is promoted to a Weierstrass model for the elliptically fibered K3 surface by allowing $f$ and $g$  --  and hence $\tau$ -- to vary suitably over the base $\mathbb P^1$.\footnote{Since $\mathbb P^1$ is compact, $f$ and $g$ cannot be globally defined functions, but must rather represent sections of a certain line bundle, whose degree is fixed uniquely by the requirement that the elliptic fibration be Calabi--Yau. In fact, $f$ and $g$ must be sections of $\bar K_{\mathbb P1}^4$ and $\bar K_{\mathbb P1}^6$, respectively, where $\bar K_{\mathbb P1} = {\cal O}_{\mathbb P^1}(2)$ is the anti-canonical bundle on $\mathbb P^1$.}
The possible types of singularities in the elliptic fiber are classified as in Table \ref{Kodtable}.

\begin{table}\begin{center}
\begin{tabular}{|c|c|c|c|c|c|l|c|}
\hline
type & ${\rm ord}(f)$ &  ${\rm ord}(g)$ &  ${\rm ord}(\Delta)$ & sing.       & $G$    \\ \hline \hline 
  $I_0$ &        $ \geq 0 $            &                $\geq 0$      &                      $0$         &   $  -$                         &   $ -$ \\ \hline
  $I_1$ &        $0$             &                $0$      &                      $1$         &     $-  $                      &   $ -$   \\ \hline 
  $II$ &          $ \geq1$             &                $1$      &                      $2$        &      $- $                       &    $-$  \\ \hline 
   $III$ &        $1$             &                $\geq 2$      &                      $3$         &$A_1$                  &   ${SU}(2)$      \\ \hline 
   $IV$ &        $\geq 2$             &                $2$      &                      $4$         &$A_2$                  &   ${SU}(3)$      \\ \hline 
   $I_m$ &        $0$             &                $0$      &                      $m$         &$A_{m-1}$                  &   ${SU}(m)$      \\ \hline 
   $I_{m-6}^\ast$ &        $2$             &                $3$      &                      $m$         &$D_{m-2}$                  &   ${SO}(2m-4)$      \\ \hline 
    $IV^\ast$ &        $\geq 3$             &                $4$      &                      $8$         &$E_6$                  &   ${E}_6$      \\ \hline 
      $III^\ast$ &        $3$             &                $\geq 5$      &                      $9$         &$E_7$                  &   ${E}_7$      \\ \hline 
      $II^\ast$ &        $\geq 4$             &                $5$      &                      $10$         &$E_8$                  &   ${E}_8$      \\ \hline  \hline 
     \text{non-min} &        $\geq 4$             &                $\geq 6$      &                      $\geq 12$         &                  &        \\ \hline 
\end{tabular}
\caption{Kodaira table for singular fibers of an elliptic K3 \cite{Kodaira2,Kodaira3,Neron}. The non-minimal enhancements in the last row can lie at finite or infinite distance \cite{Lee:2021qkx}; in the latter case, one can associate a loop algebra to the singularity.   \label{Kodtable}}
\end{center}
\end{table}

\subsection{F-theory on Elliptic Four-folds} \label{subsec_Fon4folds}

This geometric classification can be extended to higher-dimensional elliptic fibrations. Relevant for compactifications of F-theory to four dimensions are
elliptic Calabi--Yau fourfolds $Y_4$. These automatically preserve 4d ${\cal N}=1$ supersymmetry at the geometric level. From the Type IIB perspective, the base $B_3$ of the elliptic fibration represents the physical compactification space, while the additional two directions along the torus fiber 
keep track of the type and location of the 7-branes. In this sense, $B_3$ takes the role of the compactification space $X_6$ of Section \ref{subsec_IIBorientifolds}, including the effect of the orientifold action.\footnote{The subscript in $B_3$ refers to its complex dimension.} As was the case for the base of an elliptic K3 surface, $B_3$ has positive curvature due to the 7-brane backreaction.

Such four-dimensional F-theory compactifications depend on two types of data - the geometry of the elliptic fibration and additional gauge backgrounds.

\subsubsection{Geometric Data: Gauge Group, Matter, and Couplings}

The nature of F-theory as a theory of intersecting 7-branes is reflected in the singularity  
structure of the elliptic fiber of $Y_4$ along strata of codimension one, two and three on the base $B_3$:

\begin{enumerate}
\item {\it Non-abelian gauge algebras from codimension-one singularities:}
The discriminant locus $\Delta$ on $B_3$ is defined as the vanishing locus of the combination $4 f^3 + 27 g^2$ of the sections $f$ and $g$ that enter the definition of the Weierstrass model.
 It represents a holomorphic four-cycle on $B_3$ which is identified  with the cycle wrapped by the 7-branes.
In general,  the 4-cycle class of $\Delta$ can be decomposed as
\bea
[\Delta] = \sum_i a_i [\Sigma_i] + [\Delta']  \,,
\eea
where the holomorphic four-cycle $\Sigma_i$ carries a stack of 7-branes associated with a vanishing order $a_i$ in Table \ref{Kodtable}, while 
$ [\Delta']$ is the remaining piece of the discriminant with vanishing order $1$.
The non-abelian gauge algebra $G_i$ associated with $\Sigma_i$ can be read off from Table \ref{Kodtable}, where in addition one has to take into account monodromies \cite{Tate,Bershadsky:1996nh,Grassi:2011hq} along 
$\Sigma_i$ which can lead to smaller gauge algebras than on K3, including all non-simply laced ones.
As in the Type IIB orientifold context, the inverse gauge coupling squared associated with $G_i$ is set by the volume of the wrapped cycle $\Sigma_i$.
\item {\it Localised matter from codimension-two singularities:}
Over special curves $C_{\bf R}$, extra charged massless matter multiplets reside which transform in some representation ${\bf R}$ of the gauge algebra.
These curves $C_{\bf R}$ correspond to the intersection loci of the 7-branes, including possible self-intersections.
The charged massless matter fields arise from open $(p,q)$-strings stretched between the intersecting 7-branes. 
One can systematically identify the matter curves by searching for enhancements in the vanishing orders of the Weierstrass model data, which signal an enhancement in the singularity
structure of the elliptic fiber \cite{Katz:1996xe,Grassi:2011hq}.

\item {\it Yukawa couplings from codimension-three singularities:}
At the intersection of two or more matter curves, Yukawa interactions between the matter multiplets are localised \cite{Beasley:2008dc,Beasley:2008kw,Donagi:2008ca,Hayashi:2008ba,Donagi:2008kj}.
The strength of the couplings depends on the overlap of the wavefunctions of the participating massless modes. 
The physics rationale behind these couplings is completely analogous to the Type II orientifold setting (see Section \ref{ss:yukis}).
Geometrically, to each such Yukawa point one can associate a higher singularity type in the fiber, and the resulting Yukawa couplings follow from the group theoretically allowed triple couplings within the associated higher symmetry group.
Note that the higher singularity types characterising the singularity enhancements over curves and points do not correspond to gauge algebras in the four-dimensional effective action. 
\end{enumerate}

The discussion so far has focused on the non-abelian part of the gauge algebra and its charged matter and their Yukawa type couplings. This data can, to a certain extent, be described already locally by analysing the gauge theory along a stack of 7-branes together with additional matter on curves where other branes intersect the brane stack. 
The local approach amounts to zooming into the neighbourhood of one of the components $\Sigma_i$ of the discriminant.
This is analogous to the local approach described in the perturbative context in Section \ref{subsec_IIBorientifolds}.  The benefits of such a local approach in F-theory have been advocated in particular in and following \cite{Beasley:2008dc,Beasley:2008kw,Donagi:2008ca,Donagi:2008kj}.

Understanding non-Cartan abelian gauge symmetries, on the other hand,  requires going beyond a local analysis of a given brane stack. As in the Type II orientifolds reviewed in Section \ref{s:typeII}, non-Cartan abelian gauge symmetries depend on global, rather than local, data: The diagonal abelian gauge symmetries associated with the $U(N)$ gauge groups can acquire a St\"uckelberg mass, and only certain linear combinations of $U(1)$s from different brane stacks remain as massless $U(1)$s. Similarly, the existence of a massless non-Cartan $U(1)$ depends on global geometric properties of the elliptic fibration in F-theory.

The underlying construction builds on the definition of F-theory via duality with eleven-dimensional M-theory.
Compactification of M-theory on the same elliptic fourfold $Y_4$ gives a theory in three dimensions, which is identified with the compactification of F-theory on $Y_4$ times an additional circle $S^1_{\rm M}$ (see e.g. the reviews \cite{Denef:2008wq,Weigand:2018rez} for details and references). In particular, this approach admits a detailed derivation of the F-theory effective action, as explained in \cite{Grimm:2010ks}.
To treat the M-theory compactification on $Y_4$ in supergravity, one must resolve the 
 singularities in the elliptic fiber of $Y_4$ over $\Delta$. Let us denote the four-fold after the resolution as $\hat Y_4$.
At the abelian level, the gauge potentials in M-theory arise from the expansion of the M-theory three-form in terms of harmonic 2-forms of $\hat Y_4$,
\bea
C_3 = \sum_{\alpha = 1}^{h^{1,1}(\hat Y_4)} A_\alpha \wedge \omega_\alpha + \ldots
\eea
for $\omega_\alpha$ a basis of $H^{1,1}(\hat Y_4)$. 
Not all of these three-dimensional gauge potentials correspond to gauge fields in the four-dimensional F-theory. It turns out that only two types of 2-forms give rise to a gauge field in three dimensions which is associated with a  7-brane gauge field in F-theory:

First, the resolution process from $Y_4$ to $\hat Y_4$ induces a set of exceptional divisors $E_{\alpha_i}$ on $\hat Y_4$ which are fibered by rational curves over the four-cycles $\Sigma_i$. Here $\alpha_i = 1, \ldots, {\rm rk}(G_{i})$ runs over the generators of the Cartan subalgebra of  the Lie algebra $G_i$.  
The associated gauge potentials are the abelian Cartan gauge fields associated with the non-abelian gauge sector \cite{Aspinwall:2000kf}.

Non-Cartan U(1) gauge potentials on the other hand require a different source of harmonic 2-forms: These are provided by so-called rational sections
 of $\hat Y_4$ \cite{Morrison:1996pp,Aspinwall:1998xj,Aspinwall:2000kf}. A rational section embeds the base $B_3$ into $\hat Y_4$ as a divisor, and the different ways of doing so gives additional independent dual harmonic 2-form classes. The set of rational sections $S_a$ forms a finitely generated abelian group, the Mordell--Weil group ${\rm MW}(\hat Y_4)$, whose free part is in one-to-one correspondence with non-Cartan $U(1)_a$ gauge fields in F-theory. By taking a certain linear combination of the
two-forms associated with the extra section, the zero-section and other 2-form classes pulled back from $B_3$ one obtains
an element $\sigma(S_a) \in H^{1,1}(\hat Y_4)$ (called the image of the Shioda map) with the property that 
\be
C_3 = \sum_{a=1}^{{\rm rk}({\rm MW}(\hat Y_4))}A_a \wedge \sigma(S_a) + \ldots
\ee
 yields the non-Cartan gauge potentials $A_a$ \cite{Grimm:2010ez,Park:2011ji,Morrison:2012ei}.
 More details and a guide to the vast literature on this topic are provided in the reviews \cite{Weigand:2018rez,Cvetic:2018bni}.

 A subtle point concerns that fate of the diagonal $U(1)$ gauge groups which play such an important role for model building in Type II orientifolds as detailed in Sections \ref{sec_Generalities} and \ref{s:typeII}.
For instance, in Type IIB orientifolds, the diagonal $U(1)$
gauge field in the $U(N)$ group can acquire a St\"uckelberg mass even before the effect of gauge fluxes is taken into account. 
We will refer to such $U(1)$s as geometrically massive.
Depending on the geometric details, the $U(1)$
gauge group is broken to a $\mathbb Z_k$ gauge group \cite{Berasaluce-Gonzalez:2011gos}. If $k \geq 2$, this manifests itself in a corresponding selection rule on the allowed couplings, while for $k=1$ no
such selection rules survive and the geometrically massive $U(1)$ cannot be detected at the level of the effective action.
In F-theory, a geometrically massive $U(1)$ is realised directly in terms of its remnant discrete $\mathbb Z_k$ subgroup. Such discrete symmetries are associated with multi-sections which occur on genus-one fibrations not possessing a zero-section \cite{Braun:2014oya}. On the other hand, if a geometrically massive $U(1)$
does not leave behind a $\mathbb Z_k$ symmetry for $k \geq 2$, it only manifests itself at worst in certain $\mathbb Q$-factorial terminal singularities over curves in the base \cite{Braun:2014nva,Arras:2016evy}. 

Finally, the global structure of the 7-brane gauge group, as opposed to the gauge algebra, is encoded in the torsional part of the Mordell--Weil group \cite{Aspinwall:1998xj,Mayrhofer:2014opa} in F-theory, with additional subtleties appearing in presence of abelian gauge algebra factors \cite{Cvetic:2017epq,Cvetic:2018bni}.\footnote{As throughout this article, we will mostly not distinguish between the gauge group and the algebra in the sequel.}

Apart from the gauge group supported on 7-branes, four-dimensional F-theory  compactifications exhibit a gauge sector on spacetime-filling D3-branes as well as an abelian gauge sector in the Type IIB Ramond-Ramond sector \cite{Grimm:2010ks,Greiner:2017ery}. These extra gauge sectors cannot support chiral charged matter.

 \subsubsection{Gauge Backgrounds: Stückelberg Terms, Matter Multiplicities}

In addition to this purely geometric structure, the gauge background affects both the gauge algebra and the matter spectrum. This is in complete analogy to the effect of gauge background in Type IIB orientifolds, see Section \ref{subsec_IIBorientifolds}.
For simplicity we will only consider abelian gauge backgrounds. Part of the information is encoded in the background value of the gauge {\it field strengths} along the compactified dimensions.
In the language of M-theory, this corresponds to a background for the M-theory field strength $G_4 = d C_3$,
\bea \label{G4gauge}
G_4 = \sum_{\alpha_i=1}^{{\rm rk}(G_i)} F_{\alpha_i} \wedge [E_{\alpha_i}] +  \sum_{a=1}^{{\rm rk}({\rm MW}(\hat Y_4))} F_{a} \wedge [\sigma(S_a)] \,.
\eea
Here $F_{\alpha_i}$ and $F_a$ take values in $H^{1,1}(B_3)$ and parametrise the internal part of the Cartan and non-Cartan $U(1)$ field strengths, respectively.
The flux must satisfy the following constraints:
\begin{itemize}
\item[-]
The flux background is subject to a D3-brane tadpole equation, which can be elegantly written as \cite{Sethi:1996es}
\bea
n_{\rm D3} + \frac{1}{2} \int_{\hat Y_4} G_4 \wedge G_4 = \chi(\hat Y_4) \,.
\eea
The Euler characteristic of $\hat Y_4$, $\chi(\hat Y_4)$, accounts for curvature contributions to the D3-brane tadpole on the 7-branes, and
$n_{\rm D3}$ is the number of spacetime-filling D3-branes. 
\item[-]
The flux must obey the Freed-Witten quantisation condition $G_4 + \frac{1}{2} c_2(\hat Y_4) \in H^{2}(\hat Y_4, \mathbb Z)$, where $c_2$ represents the second Chern class \cite{Witten:1996md,Collinucci:2010gz,Collinucci:2012as}. 
Whenever $\frac{1}{2} c_2(\hat Y_4)$ is not by itself an integer class,
this condition implies that the gauge background must be non-vanishing in a consistent compactification. 
\item[-]
The flux induces a D-term supersymmetry condition involving the K\"ahler moduli, which is satisfied if
\begin{equation}
\int_{B_3} J \wedge F_A \wedge \pi_\ast(\omega_A \wedge \omega_B) = 0  \,, \qquad \quad \omega_A \in \{ [E_{\alpha_i}], [\sigma(S_a)]\}\,.
\end{equation}
This is essentially the condition \eqref{DIIB}. The F-term condition \eqref{FIIB} translates into the condition $G_4 \in H^{2,2}_{\rm vert}$, the primary vertical subspace \cite{Greene:1993vm}, which
is automatically fulfilled
for the choice \eqref{G4gauge}.

\end{itemize}

The gauge background must also be specified at the level of the underlying three-form potential in M-theory rather than merely the field strength. The gauge background is fully specified by an element of the D\'eligne cohomology group \cite{Curio:1998bva}, which can partly be parametrised by the Chow group of $\hat Y_4$ \cite{Bies:2014sra,Bies:2017fam}.

The gauge background plays at least three different roles in the model building context in F-theory:

\begin{itemize}

\item[-]
A Cartan gauge background breaks the non-abelian gauge algebra to a subgroup involving abelian gauge factors. This effect will be discussed in more detail in the next section.

\item[-]
Cartan and non-Cartan gauge fluxes both in general induce St\"uckelberg mass terms for the abelian gauge symmetries (along with a D-term potential for the 
K\"ahler moduli).
If one collectively denotes by $\omega_A$ the two-forms $[E_{\alpha_i}]$ and $\sigma(S_a)$, then the St\"uckelberg mass matrix $M_{AB}$ for the $U(1)_A$ gauge potentials
is proportional to
\bea \label{Stmassmatrix}
M_{AB} \propto \int_{\hat Y_4} G_4 \wedge \omega_A \wedge \omega_B \,.
\eea
For Standard Model constructions, it must therefore be checked whether the hypercharge $U(1)_Y$ remains massless in presence of gauge backgrounds.

\item[-]
The gauge background determines the multiplicities of the massless chiral superfields charged under the gauge algebra. In particular, the chiral index can be expressed as an integral of the form \cite{Donagi:2008kj,Braun:2011zm,Marsano:2011hv,Krause:2011xj,Grimm:2011fx}
\bea \label{chi-def}
\chi({\mathbf R}) = n_{\bf R} - n_{\bf \bar R} = \int_{S_{\mathbf R}} G_4 \,,
\eea
where $S_{\mathbf R}$ is a complex surface on $\hat Y_4$ which can be attributed to every representation of the gauge group as  
detailed in \cite{Weigand:2018rez} and references therein. This is the F-theory version of the expression \eqref{chiralIIB}.
The amount of vector-like matter, i.e. the multiplicities $n_{\bf R}$  and $n_{\bf \bar R}$ rather than merely their difference, is likewise encoded in the gauge background, but is sensitive to the finer information contained in the D\'eligne cohomology group
\cite{Bies:2014sra,Bies:2017fam}.

\end{itemize}

\subsection{Standard Model Constructions}

There are two different classes of Standard Model realisations in F-theory:

\subsubsection{Direct Standard Model Constructions } 

The first approach is to directly engineer, in the geometry of the elliptic four-fold, Standard Model quivers of a form similar to the ones reviewed in Section \ref{sec_Generalities}.
In such constructions, the gauge background controls the multiplicities of the massless charged matter fields and provides a mass term
for additional abelian gauge fields other than hypercharge, if present. According to our discussion of the previous section, a geometric realisation of a gauge algebra of the form
\bea 
G = SU(3) \times SU(2) \times  \prod_a U(1)_a
\eea
requires 
\begin{itemize}
\item 
a fibral singularity of either Kodaira Type $I_3$ or Type IV without monodromy along a divisor $\Sigma_{SU(3)}$,
\item
a fibral singularity of Kodaira Type $I_2$ or Type III, or of Type IV with monodromy, along a divisor $\Sigma_{SU(2)}$, 
\item
additional $U(1)_a$ gauge group factors as a consequence of extra rational sections of the fibration.
\end{itemize}
Note that in direct Standard Model constructions which cannot be unhiggsed to a theory with a GUT group, $\Sigma_{SU(3)}$ and $\Sigma_{SU(2)}$ lie in different $B_3$ homology classes. 

One linear combination of the abelian group factors must correspond to hypercharge $U(1)_Y$, while the orthogonal linear combinations must acquire a St\"uckelberg mass
by a suitable choice of gauge background. 
This approach to Standard Model building is comparable in spirit to the constructions in perturbative Type II orientifolds even though the F-theory framework is considerably more general and includes brane configurations which are not realisable in perturbative constructions. In particular, the engineering of the $SU(3)$ and $SU(2)$ factors via Kodaira fibers of Type IV or Type III has no perturbative analogue as it involves mutually non-local 7-brane stacks. 

The first realisation of this approach in \cite{Lin:2014qga,Lin:2016vus} specialises a fibration with Mordell--Weil group of rank two constructed in \cite{Borchmann:2013hta,Cvetic:2013nia} to accommodate $SU(3) \times SU(2) \times U(1)_a \times U(1)_b$ (with the non-abelian part realised via $I_3$ and $I_2$ fibers), where in realistic constructions one linear combination of the abelian factors must be massive through the gauge background.
Instead of an additional massive $U(1)$, dimension four proton decay is prohibited in the construction of \cite{Cvetic:2018ryq} by a discrete symmetry group $\mathbb Z_2$ (matter parity).
In \cite{Cvetic:2015txa}, one of the toric elliptic fibers dubbed $F_{11}$ in \cite{Klevers:2014bqa} automatically encodes  
$(SU(3) \times SU(2) \times U(1))/\mathbb Z_6$ without additional symmetries to protect proton decay; this construction admits a plethora of chiral three-general models \cite{Cvetic:2019gnh,Bies:2021nje}. 
A systematic analysis of $(SU(3) \times SU(2) \times U(1))/\mathbb Z_6$ models (containing these  as special subcases) has been undertaken in \cite{Taylor:2019wnm,Raghuram:2019efb,Jefferson:2022yya}. 
Non-perturbative Standard Model with Kodaira fibers of Type IV and III are investigated in \cite{Grassi:2014zxa}; in such scenarios, for instance the $SU(3)$ group can be non-higgsable \cite{Morrison:2012np,Morrison:2014lca}, while the incorporation of the abelian factors is to date less well understood.

\subsubsection{GUT Constructions: Georgi-Glashow \texorpdfstring{$SU(5)$}{} GUTs}\label{FtheoryYukawas}

The second approach is via Grand Unified Theories (GUTs) and was initiated in \cite{Beasley:2008dc,Beasley:2008kw,Donagi:2008ca,Donagi:2008kj}. This approach makes full use of the non-perturbative nature of the F-theory construction  because unlike perturbative Type II orientifolds \cite{Blumenhagen:2007zk,Blumenhagen:2008zz}, F-theory models admit a natural engineering also of the top quark Yukawa couplings with order one coefficients \cite{Beasley:2008dc,Beasley:2008kw,Donagi:2008ca,Hayashi:2008ba,Donagi:2008kj}. 
The general idea is to geometrically engineer a GUT group $G$ containing $SU(3) \times SU(2) \times U(1)_Y$ and to break $G$ to the latter by a suitable choice of gauge background.


In the context of F-theory, GUT groups $G=SU(5)$ (beginning with \cite{Beasley:2008dc,Beasley:2008kw,Donagi:2008ca,Donagi:2008kj}), $\OGroup(10)$ \cite{Chen:2010ts} and $E_6$ \cite{Chen:2010tg,Callaghan:2012rv,Callaghan:2013kaa} have been studied in detail (see also the reviews \cite{Heckman:2010bq,Weigand:2010wm} for further references).
In the sequel we will illustrate the key ideas in the context of Georgi-Glashow $SU(5)$ GUT theories.

In Georgi-Glashow $SU(5)$ GUTs, the Standard Model gauge algebra is embedded into $SU(5)$,
\bea
SU(3) \times SU(2) \times U(1)_Y \subset SU(5) \,.
\eea
In an ${\cal N}=1$ supersymmetric framework,
the charged matter content of the MSSM organises into three generations of chiral multiplets transforming as the ${\bf \bar 5_m}$ and ${\bf 10}$ representations of $SU(5)$:
\bea
{\bf \bar 5_m} &\rightarrow& ({d^c}, L)   \\
{\bf 10} &\rightarrow& ({Q},{u^c},{e^+}) \,.
\eea
The MSSM Higgs doublet is part of an additional vector-like pair of chiral multiplets, $H_u$ and $H_d$, which arise from the decomposition 
\bea \label{HuHd}
{\bf 5}_H \rightarrow (T_u,H_u) \,, \qquad {\bf \bar 5}_H \rightarrow (T_d,H_d)  \,.
\eea
The triplets $T_u$ and $T_d$ do not exist in the MSSM and must be sufficiently heavy by the process of doublet-triplet splitting such that they are not only unobservable at the massless level but also do not induce dangerous dimension-five proton decay operators in the low-energy effective theory. 
Finally, extensions of the MSSM by right-handed neutrinos $\nu^c$ contain extra $SU(5)$ singlets.

The MSSM Yukawa couplings are inherited from the two possible Yukawa couplings
of the $SU(5)$ theory:
\bea
{\bf 10} \,  {\bf 10} \,  {\bf 5}_H &\rightarrow & {Q} \, {u^c} \, H_u         \\
{\bf 10} \, {\bf \bar 5_m} \, {\bf \bar 5}_H &\rightarrow&  {Q} \,  {d^c} \, H_d + {e^+} \, L \, H_d   \,.  
\eea
By contrast, Yukawa couplings of the form ${\bf 10} \,   {\bf \bar 5_m} \,  {\bf \bar 5_m}$ or ${\bf 10} \,   {\bf \bar 5}_H \,  {\bf \bar 5}_H$
would lead to phenomenologically unacceptable dimension-four proton decay operators and must hence be suppressed by additional selection rules which distinguish between the ${\bf \bar 5_m}$ and the ${\bf \bar 5}_H$ representation.

This general framework can be embedded into F-theory as follows: \vspace{4mm}

\noindent {\bf GUT group, matter, Yukawas}

\vspace{2mm}

The GUT group $SU(5)$ can be realised on a stack of 7-branes encoded in a Kodaira Type $I_5$ singularity in the elliptic fiber over a divisor $\Sigma_{SU(5)}$ on the base $B_3$. 
Alternatively, the $SU(5)$ GUT group can also by itself be embedded into a higher gauge group which is broken accordingly by a gauge background, as studied systematically in \cite{Li:2021eyn,Li:2022aek}.

Additional non-Cartan $U(1)$
or discrete gauge symmetries may be realised via the geometric mechanisms described in the previous section. At the geometric level, such extra symmetries require a further tuning  of the Weierstrass model. The extra $U(1)$ gauge symmetries, if present, would have to be massive by a flux induced St\"uckelberg mechanism.

For definiteness, we will focus in the sequel on constructions with a geometrically tuned $SU(5)$ symmetry associated with a Kodaira Type $I_5$ fiber.
Such a singularity can be conveniently engineered by applying Tate's algorithm \cite{Bershadsky:1996nh,Katz:2011qp}, as reviewed in the $SU(5)$ GUT context in \cite{Weigand:2010wm}. 
The $SU(5)$ charged matter multiplets are localised on curves on $\Sigma_{SU(5)}$ where the GUT brane stack intersects other 7-branes. The two relevant representations, ${\bf R} ={\bf 10}$ and ${\bf R}={\bf 5}$, arise on curves $C_{\bf R}$ where the Kodaira type of the fibers enhance as follows:
\bea
I_5 \,  (A_4) &\rightarrow& I_6 \, (A_5):     \qquad C_{\bf 5}   \\
I_5 \, (A_4) &\rightarrow& I_1^\ast  \, (D_5): \qquad  C_{\bf 10}    \,. 
\eea
In the parantheses we display the the symmetry groups associated with these enhanced Kodaira singularities (see Table \ref{Kodtable}). 
In a Weierstrass model with an $I_5$ singularity over a divisor, both types of higher enhancements occur generically without further tuning. From a Type IIB orientifold point of view, $C_{\bf 5}$ is the intersection of the GUT brane stack with another 7-brane away from an orientifold plane, while $C_{\bf 10}$ represents the intersection of the GUT brane stack with its orientifold image on top of an O7-plane.
In absence of extra massive $U(1)$ or discrete gauge groups differentiating between ${\bf 5_m}$ and  ${\bf 5}_H$, both representations generically reside on the same matter curve, while extra such symmetries lead to a corresponding splitting of the matter curves. See also the discussion of Class $(iii)$ models in Section \ref{sec:4classes}.

At the intersection points of the matter curves, the overlap of the matter wavefunctions gives rise to the Yukawa couplings allowed by gauge symmetry. The intersection points are characterised by further singularity enhancements:
\bea
(I_6, I_1^\ast)&\rightarrow& {\rm IV} \,  (E_6): \qquad  {\bf 10} \,  {\bf 10} \,  {\bf 5}   \\
(I_6, I_1^\ast)&\rightarrow& I_2^\ast \, (D_6):  \qquad \, \, {\bf 10} \, {\bf \bar 5} \, {\bf \bar 5}  
\eea
The second type of enhancement corresponds to a symmetry group $\OGroup(12)$ and the associated couplings are present also in perturbative Type II orientifolds;
the enhancement to a symmetry group $E_6$, on the other hand, cannot be achieved perturbatively. The existence of such couplings is a trademark of mutually non-local $[p,q]$ 7-branes in F-theory.

Note that without extra $U(1)$ or discrete gauge symmetries, the ${\bf \bar 5_m}$ and the ${\bf 5}_H$ representations
are also indistinguishable 
at the level of the Yukawa couplings. In particular, this means that the phenomenologically unacceptable coupling ${\bf 10} \, {\bf \bar 5_m} \, {\bf \bar 5_m}$
or ${\bf 10} \,   {\bf \bar 5}_H \,  {\bf \bar 5}_H$ cannot be avoided without such selection rules \cite{Hayashi:2010zp}.

\vspace{4mm}

\noindent{\bf GUT breaking} \vspace{2mm}

There are two different ways to break the $SU(5)$ GUT group to the Standard Model gauge group:
Either via a dynamically generated vacuum expectation value for a GUT Higgs or by a topological gauge background.
The first mechanism faces the challenge of explaining the origin of the Higgs potential required for the GUT breaking. The smallest representation in which the GUT Higgs
field can occur is the ${\bf 24}$ representation, and there are two 
candidates for such a GUT Higgs:
If the  divisor $\Sigma_{\rm SU(5)}$ is non-rigid inside $B_3$, there arise $h^0(\Sigma_{\rm SU(5)}, K_{\Sigma_{\rm SU(5)}})$ massless chiral multiplets in the ${\bf 24}$, whose bosonic components represent geometric deformation moduli of the brane stack. The Higgsing of $SU(5) \to SU(3) \times SU(2) \times U(1)_Y$ amounts to a geometric deformation of the brane stack into an intersecting brane model.  The end result can equivalently be interpreted as a direct Standard Model construction in which the $SU(3)$ and $SU(2)$ brane stacks lie in the same homology class.
Alternatively, if $h^1(\Sigma_{\rm SU(5)}) > 0$ there exists a corresponding number of continuous Wilson line moduli, whose VEV can likewise break the $SU(5)$ gauge group.

In view of the difficulty of accounting for the symmetry breaking potential dynamically, the second, topological mechanism of GUT breaking is particularly attractive.
The $SU(5)$ group can be broken
to the Standard Model gauge group by an internal gauge background for the hypercharge $U(1)_Y$, the abelian subgroup of $SU(5)$ associated with a generator $T_{Y} ={\rm diag}(\frac{1}{3},\frac{1}{3},\frac{1}{3},-\frac{1}{2},-\frac{1}{2})$. This results in the following breaking pattern:
\be
\begin{aligned} 
\label{decompSU5GUT}
SU(5) &\rightarrow   SU(3) \times SU(2) \times U(1)_Y   \cr
{\bf 24} &\rightarrow ({ \bf 8}, { \bf 1},0) \oplus ({ \bf 1}, {\bf 3},0)  \oplus ({\bf 1},{\bf 1},0)  \oplus ({\bf 3},{\bf 2},\tfrac56) \oplus  ({\bf \bar 3},{\bf 2},-\tfrac56)    \cr
{ \bf 10} &\rightarrow ({\bf 3}, {\bf 2},\tfrac16) \oplus   ({ \3c},{\bf 1},-\tfrac23)   +   ({\bf 1}, {\bf 1},1) \cr
{\bf 5} &\rightarrow ({\bf 3}, {\bf 1},-\tfrac13) \oplus   ({\bf 1}, {\bf 2},\tfrac12)  \,. 
\end{aligned}
\ee
The ${\bf 24}$ here refers to matter fields propagating on the 7-brane stack along $\Sigma_{SU(5)}$, so-called bulk matter.
If $\pi_1(\Sigma_{SU(5)})$ is non-trivial, the $U(1)_Y$ gauge background can be taken to be flat, corresponding to a discrete Wilson line. This discrete version of the continuous GUT breaking via Wilson line moduli is studied in \cite{Marsano:2012yc}.

Another possibility which is available more generally is to consider a hypercharge gauge background \cite{Heckman:2008qa,Donagi:2008kj}
characterised by a non-trivial line bundle $L_Y$ on 
$\Sigma_{SU(5)}$ with $c_1(L_Y) = \frac{1}{2 \pi} F_Y \in H^2(\Sigma_{SU(5)})$.
The hypercharge flux is subject to a number of conditions in order for the gauge breaking mechanism to meet some basic phenomenological criteria:
\begin{itemize}
\item[-]
A hypercharge gauge background induces a St\"uckelberg mass for the hypercharge gauge boson unless
$F_Y$ is cohomologically trivial on $B_3$ \cite{Buican:2006sn}. This condition can be written, at the level of cohomology, as
\bea \label{pushforwardLY}
\iota_{!} c_1(L_Y) = 0  \,,
\eea 
where $\iota: \Sigma_{\rm SU(5)} \to B_3$ denotes the embedding of the divisor $\Sigma_{\rm SU(5)}$ into the base and the Gysin map $\iota_{!}: H^2(\Sigma_{\rm SU(5)}) \to H^4(\Sigma_{\rm SU(5)})$ is defined by first taking the Poincar\'e dual, then pushing forward at the level of homology and finally taking the Poincar\'e dual again.
As follows from \eqref{Stmassmatrix}, the 4-form flux $G^{Y}_4$ associated with such a line bundle must take
value in the so-called remainder piece $H^{2,2}(\hat Y_4)_{\rm rem}$ \cite{Braun:2014xka} in the orthogonal decomposition 
\bea
H^{2,2}(\hat Y_4) = H^{2,2}(\hat Y_4)_{\rm hor} \oplus H^{2,2}(\hat Y_4)_{\rm vert} \oplus H^{2,2}(\hat Y_4)_{\rm rem} \,,
\eea
where the first two summands refer to the primary horizontal and primary vertical subspaces of $H^{2,2}(\hat Y_4)$, respectively.

\item[-]
The second condition on the gauge background arises by demanding that there occur no exotic massless states in the representation $({\bf 3},{\bf 2},\frac56) + c.c$
from the decomposition \eqref{decompSU5GUT} of the  ${\bf 24}$ representation.
These so-called lepto-quarks are absent only if all cohomology groups of $L_Y^{\pm 5/6}$ on $\Sigma_{SU(5)}$ vanish,
\bea
H^i(L^{\pm 5/6}_Y) = (0,0,0)  \,,
\eea
which is a strong constraint. The power of $\tfrac56$ reflects the $U(1)_Y$ charge of the lepto-quarks. 
It turns out \cite{Beasley:2008kw,Donagi:2008kj} that rather than an integral line bundle $L_Y$, one can consider a fractionally quantised line bundle ${\cal L}_Y$ with the property that 
the lepto-quarks are counted by  the cohomology group $H^i({\cal L}^{\pm 1}_Y)$, $i=0,1,2$. The vanishing of these cohomology groups can be achieved for instance 
by taking ${\cal L}_Y = E_i - E_j$ on a del Pezzo surface dP$_n$ with exceptional curve classes $E_i$, $i=1, \ldots n$.

\item[-]
The Standard Model matter should appear in complete GUT multiplets. This means that 
 \bea \label{trivialrestr}
 \int_{C_{\bf 10}} F_Y = 0 =  \int_{C_{\bf \bar 5_m}} F_Y \,.
 \eea
 The requirement \eqref{trivialrestr} is only a necessary condition which guarantees that the chiral index of the MSSM representations do not differ within a GUT multiplet.
This must in fact be ensured also at the vector-like level.
\end{itemize}

Apart from being topological, one of the benefits of hypercharge GUT breaking is that the doublet-triplet splitting problem can be solved by a suitable restriction of the hypercharge gauge background to the Higgs curve. This will be discussed momentarily.

On the other hand, the hypercharge flux breaking affects precision gauge coupling unification via flux dependent subleading corrections to the gauge kinetic function, as pointed out first in \cite{Donagi:2008kj,Blumenhagen:2008aw}. 
This effect may be counter-balanced by the appearance vector-like exotics for instance at intermediate scales \cite{Dolan:2011aq}.

\vspace{4mm}
\noindent{\bf Proton decay and selection rules}
\vspace{2mm}

A closer analysis of this and related effects and more generally of the detailed realisation of the matter spectrum requires specifying the additional selection rules invoked to prevent the phenomenologically excluded
proton decay operators by which GUT models are typically inflicted.
As a minimal requirement, the extra symmetry must distinguish the ${\bf 5_m}$ and ${\bf 5}_H$ representations and forbid the dangerous dimension four operators 
${\bf 10} \, {\bf \bar 5_m} \, {\bf \bar 5_m}$ and ${\bf 10} \, {\bf \bar 5}_H \, {\bf \bar 5}_H$ while allowing for the Yukawa couplings 
${\bf 10} \,  {\bf \bar 5_m} \, {\bf \bar 5}_H$ and ${\bf 10} \, {\bf 10} \, {\bf 5}_H$.
Up to an overall normalisation, and assuming for now that the charge assignments do not distinguish between the three matter families, these conditions are met
by an extra (massive) $U(1)$ gauge symmetry with charge assignments
\bea \label{U1charges}
{\bf 10}_1 \,, \qquad  ({\bf \bar 5_m})_{-q} \,,  \qquad ({\bf 5}_H)_{-2} \,, \qquad  ({\bf \bar 5}_H)_{q-1} \,,   \qquad q \neq \frac{1}{2} \,,
\eea  
or a suitable discrete subgroup thereof. 

There are then two general possibilities:
The assignment $q=3$ corresponds to the so-called $U(1)_X$ symmetry, under which the charges of $H_u$ and $H_d$ merely differ by a sign.
In a geometric realisation of such an additional symmetry (and no additional symmetries on top), $H_u$ and $H_d$ form a  vector-like pair localised on a single
curve $C_{{\bf 5}_{H}}$, and altogether such models have three different types of localised matter curves, $C_{{\bf 10}_{1}}$, $C_{({\bf 5_m})_{-3}}$ and $C_{({\bf 5}_{H})_{-2}}$ \cite{Hayashi:2010zp}. 
The first globally consistent realisations of this model with three chiral generations of MSSM matter have been constructed in \cite{Krause:2011xj}.

The $U(1)_X$ charge assignment has the disadvantage that it does not forbid dimension-five proton decay operators.
To prevent these from being generated, the additional $U(1)$ (or, for that matter, discrete $\mathbb Z_k$) selection rule must distinguish also between
$H_u$ and $H_d$ other than just by an overall sign of the charge, which fixes $q \neq 3$.
A $U(1)$ gauge symmetry with this property is said to be of Peccei-Quinn (PQ) type \cite{Beasley:2008kw,Donagi:2008kj}. The different charge assignments have been studied intensively in the F-theory literature \cite{Marsano:2009wr,Dolan:2011iu,Dolan:2011aq,Mayrhofer:2012zy,Lawrie:2015hia}, to which we refer for details and further references.

Note that the above discussion assumes that the extra selection rule does not distinguish between the three different families of matter within the same MSSM representation. Without this requirement a plethora of new symmetry patterns opens up. At a phenomenological level, distinguishing between families can explain the hierarchical structure of Yukawa couplings via the Froggatt--Nielsen mechanism \cite{Dudas:2010zb,Dudas:2012wi}.

\vspace{6mm}
\noindent{\bf Gauge background}
\vspace{2mm}

The hypercharge flux must be complemented by additional gauge background which controls the multiplicities of charged matter, in particular the chiral index \eqref{chi-def}.
At the level of 4-form flux, the flux background takes the form
\bea
G_4 = G_4^{\rm c} + G_4^{Y} \,,
\eea 
where $G_4^{Y} \in H^{2,2}_{\rm rem}(\hat Y_4)$ denotes the hypercharge flux background and $G_4^{\rm c}$ is the part of the flux background which is blind to the GUT group breaking.
To obtain the correct chiral index of MSSM matter, the flux must satisfy the following conditions:
\bea \label{3gencond1}
\int_{S_{{\bf R}_{\vec{q}}}} G_4^{Y} = 0    \,, \qquad \sum_{\vec{q}} \int_{S_{{\bf R}_{\vec{q}}}} G_4^{\rm c} = 3     \,, \qquad {\bf R} = {\bf 10}, \, {\bf \bar 5_m} \,.
\eea
Here ${S_{{\bf R}_{\vec{q}}}}$ is the matter surface associated with representation ${\bf R}_{\vec{q}}$ of $U(1)$ charge vector $\vec{q} = (q_1, \ldots, q_n)$, where we are allowing for $n$ additional massive abelian gauge groups. 
The first condition is essentially \eqref{trivialrestr} and ensures that all representations appear in complete GUT multiplets. 

As for the Higgs sector, let us specialise for definiteness to a single extra massive $U(1)$ gauge group with charges \eqref{U1charges}. 
If $q=3$, one requires that at the chiral level
\bea \label{HiggsU1X-splitting}
\int_{S_{{\bf 5}_H}} G_4^{Y} = 0   \,, \qquad \int_{S_{{\bf 5}_H}} G_4^{\rm c} = 0    \,,
\eea
but the gauge background must give rise to one vector-like pair of  massless $H_u$ and $H_d$ fields
while both triplets $T_u$ and $T_d$ must be absent at the massless level. 
This means that the restriction of $G_4^{\rm c}$ and $G_4^{Y}$ must describe a line bundle of trivial curvature whose cohomology groups are compatible with this vector-like spectrum.
If, on the other hand, the abelian symmetry is of PQ type ($q \neq 3$), then doublet triplet splitting can already be imposed at the chiral level:
\be
\begin{aligned}
\label{HiggsU1PQ-splitting}
& \int_{{S_{{T_u}}}} G_4^{Y} +G_4^{\rm c} = 0 = \int_{{S_{{T_d}}}} G_4^{Y} +G_4^{\rm c}  \,, \cr
& \int_{{S_{{H_u}}}}  G_4^{Y} +G_4^{\rm c} = 1 = \int_{{S_{{H_d}}}}  G_4^{Y} +G_4^{\rm c} \,.
\end{aligned}
\ee
Here we have split the matter surface into a surface associated with the triplets and the doublets in the decomposition (\ref{HuHd}). Both these surfaces share the same base curve, but differ in the fiber. 
On top of these conditions, no vector-like pairs of states must be generated by the flux background to generate the exact massless MSSM matter content.

The second condition of \eqref{3gencond1} controlling the chiral matter content was for the first time realised in globally consistent SU(5) GUTs in \cite{Krause:2011xj}.
The conditions \eqref{HiggsU1X-splitting} have been exemplified in simple toy models in the literature \cite{Mayrhofer:2013ara,Braun:2014pva}. Their counterpart in PQ-type models,
 \eqref{HiggsU1PQ-splitting}, has not yet been realised, as of this writing, in globally consistent F-theory models where the hypercharge flux satisfies in addition the important constraint \eqref{pushforwardLY}. At the same time, considerations in the weak coupling limit suggest that this should be achievable in principle \cite{Mayrhofer:2013ara}.

\vspace{4mm}
\noindent{\bf Yukawa couplings and Flavour hierarchies}
\vspace{2mm}

The Yukawa couplings can in principle be computed locally by evaluating the overlap of the matter wavefunctions
at the intersection points of the matter curves.
The local structure of the Yukawa couplings favours a mass hierarchy among the different families whose wavefunctions overlap at the same point.
For details and a guide to the literature, we refer to \cite{Hayashi:2009ge,Cecotti:2009zf,Marchesano:2009rz,Cecotti:2010bp,Font:2013ida,Marchesano:2015dfa}. A remaining challenge for the future, however, is to connect such local computations to the data of the globally defined F-theory model. 
An alternative approach to explaining flavour hierarchies is via  the Froggatt--Nielsen mechanism \cite{Dudas:2010zb,Dudas:2012wi}.

\vspace*{1cm}

\centerline{\bf  Acknowledgements}

\vspace*{.5cm}

We thank I.~Garc\'ia-Etxebarria, F.~Quevedo, A.~Sagnotti and A.~M.~Uranga for discussions. FM is supported through the grants  CEX2020-001007-S and PID2021-123017NB-I00, funded by MCIN/AEI/10.13039/501100011033 and by ERDF A way of making Europe. TW is supported in part by
Deutsche Forschungsgemeinschaft under Germany's Excellence Strategy EXC 2121  Quantum Universe 390833306 and by Deutsche Forschungsgemeinschaft through a German-Israeli Project Cooperation (DIP) grant ``Holography and the Swampland”.

\bibliography{biblio}{}

\def\hre#1#2{\href{http://arxiv.org/abs/#1/#2}{[ArXiv:#1/#2]}}
\def\hspi#1#2{\href{http://www.slac.stanford.edu/spires/find/hep/www?irn=#1}{#2}}

\end{document}